\documentclass[journal]{IEEEtran}
\usepackage{graphics} 
\usepackage{epsfig} 
\usepackage{times} 
\usepackage{amsmath} 
\usepackage{amssymb}  
\usepackage{cite}
\usepackage{epstopdf}
\usepackage{subfigure,color,balance}
\usepackage{verbatim}
\usepackage{cases}
\usepackage{enumerate}
\usepackage{booktabs}
\usepackage{balance}
\usepackage{mathbbol}
\usepackage{algorithm}
\usepackage{algorithmicx}
\usepackage{algpseudocode}

\newcommand{\tr}{{{\mathsf T}}}

\usepackage[colorlinks=true,      
linkcolor=black,      
citecolor=black,      
filecolor=black,      
urlcolor=blue]{hyperref}

\newtheorem{definition}{Definition}
\newtheorem{theorem}{Theorem}

\newtheorem{remark}{Remark}
\newtheorem{lemma}{Lemma}
\newtheorem{corollary}{Corollary}

\def\0{{\bf 0}}
\def\1{{\bf 1}}

\def\eg{{\em e.g.}}
\def\ie{{\em i.e.}}

\begin{document}
%
\title{
Leading Cruise Control in Mixed Traffic Flow: System Modeling, Controllability, \\ and String Stability
}

\author{Jiawei Wang,~\IEEEmembership{Student Member,~IEEE}, Yang Zheng,~\IEEEmembership{Member,~IEEE},\\ Chaoyi Chen, Qing Xu and Keqiang Li
	\thanks{This work is supported by National Key R\&D Program of China with 2018YFE0204302,  National Natural Science Foundation of China with 52072212, and Tsinghua University-Didi Joint Research Center for Future Mobility. All correspondence should be sent to Y.~Zheng and K.~Li.}
	\thanks{J.~Wang, C.~Chen, Q.~Xu and K.~Li are with the School of Vehicle and Mobility, Tsinghua University, Beijing, China, and with Tsinghua University-Didi Joint Research Center for Future Mobility, Beijing, China. (\{wang-jw18,chency19\}@mails.tsinghua.edu.cn, \{qingxu,likq\}@tsinghua.edu.cn).}%
	\thanks{Y. Zheng is with the Department of Electrical and Computer Engineering, University of California San Diego, CA, USA. ({zhengy@eng.ucsd.edu}).}%
}

\maketitle

\begin{abstract}
	
Connected and autonomous vehicles (CAVs) have great potential to improve road transportation systems. Most existing strategies for CAVs' longitudinal control focus on downstream traffic conditions, but neglect the impact of CAVs' behaviors on upstream traffic flow. In this paper, we introduce a notion of Leading Cruise Control (LCC), in which the CAV maintains car-following operations adapting to the states of its preceding vehicles, and also aims to lead the motion of its following vehicles. Specifically, by controlling the CAV, LCC aims to attenuate downstream traffic perturbations and smooth upstream traffic flow actively. We first present the dynamical modeling of LCC, with a focus on three fundamental scenarios: car-following, free-driving, and Connected Cruise Control. Then, the analysis of controllability, observability, and head-to-tail string stability reveals the feasibility and potential of LCC in improving mixed traffic flow performance. Extensive numerical studies validate that the capability of CAVs in dissipating traffic perturbations is further strengthened when incorporating the information of the vehicles behind into the CAV's control.
	
\end{abstract}


\section{Introduction}

\IEEEPARstart{I}{mproving} 
the driving behavior of individual vehicles via vehicular automation provides new opportunities for smooth traffic flow and efficient mobility~\cite{baskar2011traffic}. One typical technology is Adaptive Cruise Control (ACC), which adjusts its own motion by monitoring the preceding vehicle ahead~\cite{vahidi2003research}. High-accuracy on-board sensors and advanced control algorithms enable ACC-equipped vehicles to achieve a better car-following behavior than typical human drivers~\cite{sugiyama2008traffic}. The potential of ACC has been further enhanced thanks to the emergence of connected and autonomous vehicles (CAVs)~\cite{li2017dynamical}. By exploiting wireless communication, \eg, vehicle-to-vehicle (V2V) or vehicle-to-infrastructure (V2I), CAVs can utilize other vehicles' information within their communication range, thereby allowing for more sophisticated driving strategies than traditional ACC-equipped vehicles.

\begin{figure*}[t]
	\vspace{1mm}
	\hspace{1.5mm}
	\includegraphics[scale=0.45]{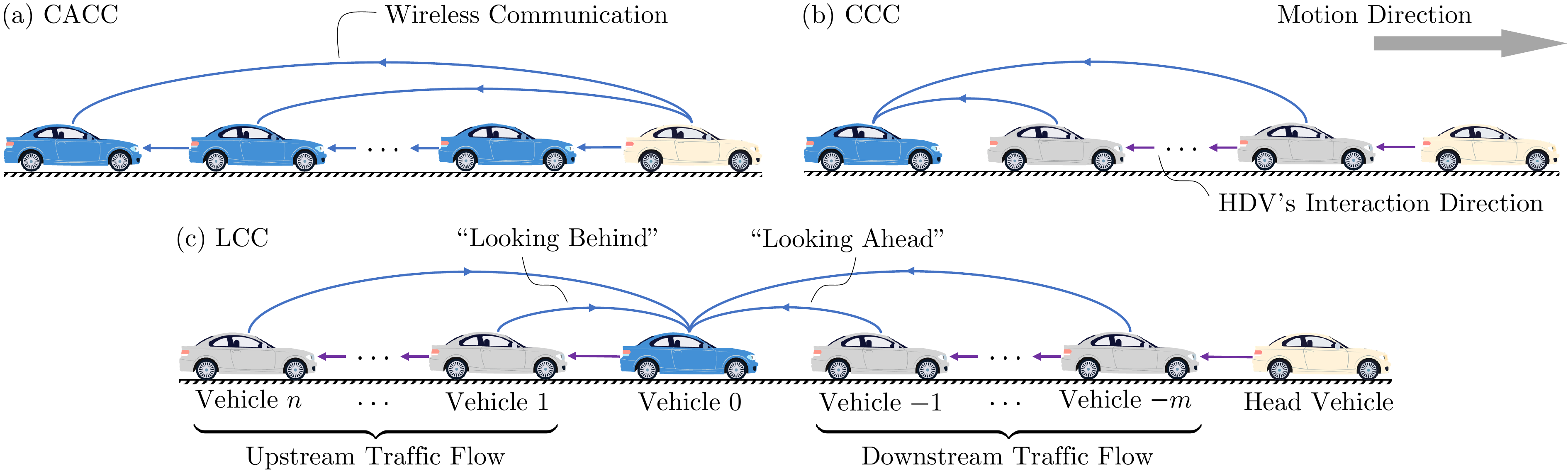}
	\vspace{-1mm}
	\caption{Schematic of different control frameworks for CAVs. The blue arrows represent the communication topology of the CAV, while the purple arrows illustrate the interaction direction in HDVs' dynamics. The blue vehicles, gray vehicles and yellow vehicles represent CAVs, HDVs and the head vehicle, respectively. (a) In a CACC platoon, multiple CAVs are controlled to follow a designated head vehicle. Here, we demonstrate a typical communication topology named predecessor-leader following (see~\cite{zheng2016stability} for other typical topologies). (b) CCC focuses on a mixed traffic scenario, where the CAV receives the information from multiple HDVs ahead~\cite{orosz2016connected}. (c) In LCC, the CAV explicitly considers the dynamics of both the vehicles ahead and the vehicles behind.}
	\label{Fig:SystemSchematic}
\end{figure*}

To coordinate multiple CAVs, Cooperative Adaptive Cruise Control (CACC) is a prevailing extension of ACC~\cite{milanes2013cooperative,li2017dynamical}. In CACC, a series of adjacent CAVs are organized into a platoon, following a designated head vehicle. Fig.~\ref{Fig:SystemSchematic}(a) demonstrates a typical CACC framework under a special communication topology, known as predecessor-leader following (PLF), where each CAV utilizes the information of its preceding vehicle and the head vehicle to determine its control input. Besides PLF, the potential of other communication topologies has also been explicitly investigated for improving the performance of CACC~\cite{zheng2016stability}. Extensive theoretical analysis and field experiments have revealed the capability of CACC in mitigating traffic perturbations while maintaining a small inter-vehicle distance, contributing to higher traffic capacity and throughput~\cite{zheng2016stability,sabuau2016optimal,besselink2017string}. 

Most CACC systems, especially in the platoon sense~\cite{milanes2013cooperative,li2017dynamical}, typically assume all the involved vehicles to have autonomous capabilities. Due to the long-period transition from human-driven vehicles (HDVs) to CAVs, the near future will have to meet mixed traffic scenarios with the coexistence of HDVs and CAVs~\cite{stern2018dissipation,zheng2020smoothing,di2020survey}, in which HDVs would still be the majority in traffic flow in the next few decades. 
In addition, since CACC systems do not explicitly consider other  HDVs' dynamics into their controller design, significant traffic performance improvement may be achieved only if the penetration rate of CAVs reaches a certain level~\cite{van2006impact,shladover2012impacts}. It is thus better to incorporate HDVs' dynamics for designing high-performance controllers. 
HDVs' car-following behavior has indeed been studied extensively since the last fifties~\cite{wilson2011car}, and many models have been developed, \eg,~optimal velocity model (OVM)~\cite{bando1995dynamical} and intelligent driver model (IDM)~\cite{treiber2000congested}.~These results facilitate the understanding of~human's driving behavior, which in turn contribute to the extension of ACC/CACC to mixed traffic scenarios. The notion of Connected Cruise Control (CCC) is a typical example~\cite{orosz2016connected}, which explicitly considers HDVs' dynamics and determines the CAV's control strategy at the tail by monitoring the motion of multiple HDVs ahead (Fig.~\ref{Fig:SystemSchematic}(b)). Many recent studies have revealed the potential of CCC in improving the performance of mixed traffic, and various topics have been addressed, such as scalability of the CCC framework~\cite{jin2014dynamics}, estimation of HDVs' dynamics~\cite{pandita2013preceding} and influence of communication delay~\cite{di2019cooperative}.

\subsection{Notion of Leading Cruise Control (LCC)}

Similar to human drivers' decision-making process~\cite{kesting2010enhanced}, CCC-type controllers monitor the downstream traffic conditions consisting of the vehicles ahead and aim to achieve a better car-following behavior~\cite{orosz2016connected,pandita2013preceding,jin2014dynamics,di2019cooperative,jin2017optimal}. Due to the front-to-rear reaction dynamics of human drivers, the behavior of one individual vehicle has a simultaneous influence on the upstream traffic flow containing its following vehicles behind. One example of such impact from the negative side is the  phenomenon of traffic waves, where a small perturbation of one vehicle persists and propagates upstream in a series of HDVs. In particular, if the involved vehicles have a poor string stability performance~\cite{stern2018dissipation,jin2014dynamics}, the perturbations will be amplified and might grow into a stop-and-go pattern, bringing a dramatic increase to travel time, fuel consumption and accident risks~\cite{sugiyama2008traffic}. The topic of enabling CAVs to dissipate perturbations from the front (\ie, downstream traffic) has gained significant attention, but to our knowledge, the influence of the perturbations on the upstream traffic flow behind the CAV has not been clearly addressed. By explicitly incorporating the motion of the vehicles behind into CAVs' control, the potential of CAVs in smoothing traffic flow could be further enhanced, compared to traditional CAV strategies.

In this paper, we introduce a new notion of Leading Cruise Control (LCC); see Fig.~\ref{Fig:SystemSchematic}(c) for illustration. In the spirit of ACC/CCC, an LCC-equipped vehicle retains car-following operations that adapt to the motion of its preceding vehicles. Meanwhile, LCC explicitly considers the impact of its own behavior on the upstream traffic, \ie, those HDVs following behind. Particularly, the LCC-equipped CAV acts as a leader to actively lead the motion of its following vehicles and aims to improve the performance of the entire upstream traffic flow. A related concept is the notion of Lagrangian control of traffic flow~\cite{stern2018dissipation,zheng2020smoothing}, where CAVs are utilized as mobile actuators for traffic control. Recent results have revealed the potential of one single CAV in stabilizing a closed ring-road traffic system from various aspects, including rigorous theoretical analysis~\cite{zheng2020smoothing,wang2020controllability,drummond2020impact,cui2017stabilizing}, small-scale field experiments~\cite{stern2018dissipation}, and large-scale numerical simulations~\cite{wu2017flow,vinitsky2018lagrangian}. Since the impact of the CAV's behavior propagates backwards, \ie, upstream in traffic flow, LCC generalizes these ring-road results and investigates traffic control and optimization via CAVs at an open straight road scenario. The possibility of controlling upstream traffic flow via CAVs has also been recently investigated based on macroscopic traffic models, by viewing CAVs as moving bottlenecks in traffic flow~\cite{piacentini2018traffic,vcivcic2018traffic} or establishing a closed-loop traffic system by enabling CAVs to respond to one vehicle behind~\cite{molnar2020open}.  Unlike~\cite{piacentini2018traffic,vcivcic2018traffic,molnar2020open}, LCC focuses on the microscopic car-following dynamics, which might be more feasible for practical use in individual vehicles, as individual vehicle control has been widely deployed in ACC or CCC.

Compared with existing strategies, \eg, ACC, CCC, and Lagrangian control of traffic flow, one distinctive feature of LCC is the explicit consideration of an individual CAV as both a leader and a follower in traffic flow. We note that there are two important topics in complex networks or multi-agent systems: 1) the control of follower agents, targeting at tracking a prescribed trajectory of leader agents~\cite{ni2010leader,ding2013network}; existing studies on ACC, CACC and CCC lie in this category, where one major objective of CAVs is to follow a designated leading vehicle; 2) the control of leader agents, acting as control inputs to achieve a desired performance for the entire system~\cite{clark2013supermodular,jafari2011leader}. Most existing research focuses on how to improve the behavior of the CAV as a follower agent, but neglects another role of the CAV as a leader agent with regard to the vehicles behind, especially in mixed traffic flow. In LCC, the potential of this dual identity of the CAV is explicitly addressed. Another distinction of LCC from existing ACC/CCC frameworks is the adequate employment of V2V connectivity --- the information of both the HDVs ahead and those behind is utilized in CAVs' control decision. Informally, such a general communication topology is called as both~``looking ahead'' and ``looking behind''~\cite{zheng2016stability} (see Fig.~\ref{Fig:SystemSchematic}(c)), while previous ACC/CCC frameworks fall in the ``looking ahead only" category. Note that such general communication topology is also motivated by the well-known bidirectional topology in CACC or vehicle platooning, which utilizes the information from one vehicle directly ahead and one directly behind~\cite{zheng2016stability,peppard1974string}. LCC generalizes bidirectional topology to the case where multiple vehicles are included in the communication network. Also, LCC 
explicitly takes the dynamics of surrounding HDVs into account.

\subsection{Contributions}

In this paper, we investigate fundamental properties of the proposed LCC framework, including dynamical modeling, controllability, observability, and head-to-tail string stability. One particular interest is to explore the potential of incorporating the motion of the vehicles behind into CAVs' control. Some preliminary results appeared in~\cite{wang2020leading}. Precisely, 
our contributions are as follows.

\begin{itemize}
	\item   We introduce a new notion of Leading Cruise Control (LCC). A general modeling framework of LCC is presented based on linearized car-following dynamics. Three special cases are discussed, which cover the notion of CCC~\cite{orosz2016connected} and two fundamental driving behaviors of individual vehicles in traffic flow: car-following and free-driving~\cite{treiber2013traffic}. This framework incorporates the motion of HDVs behind into controller design and exploits the role of the CAV as both a follower and a leader in mixed traffic flow.
	\item We prove that the motion of HDVs behind is controllable in LCC under a very mild condition, and also investigate the control energy of LCC at different system sizes. This result confirms the feasibility of the CAV for leading the motion of its following HDVs and achieving a desired traffic performance. Our controllability results generalize the previous stabilizability results in a closed ring-road system~\cite{zheng2020smoothing,wang2020controllability,cui2017stabilizing}, and verify the possibility of traffic control via CAVs on the common open road scenario. Moreover, our observability analysis reveals that when the CAV can directly measure the velocity error of one individual vehicle, the states of all the HDVs ahead of it are observable.  
	\item We finally investigate the head-to-tail string stability of the mixed traffic flow under the LCC framework. Compared to previous CCC-type strategies~\cite{di2019cooperative,jin2014dynamics }, we reveal that LCC enables the CAV to have more string-stable options in feedback policies and a bigger capability to dampen traffic perturbations coming from front. In addition, nonlinear traffic simulations demonstrate that LCC enables the CAV to respond actively to a perturbation that happens behind and reduce velocity fluctuations of the entire upstream traffic flow. These results confirm that LCC improves the capability of CAVs in suppressing traffic instabilities, after ``looking behind" appropriately compared with ``looking ahead" only~\cite{di2019cooperative,jin2014dynamics}.  
\end{itemize}

The rest of this paper is organized as follows. Section~\ref{Sec:2} introduces the modeling for the general LCC system and three special cases. Section~\ref{Sec:3} presents the controllability and observability analysis, and head-to-tail string stability is investigated in Section~\ref{Sec:4}. Traffic simulations are presented in Section~\ref{Sec:5}, and Section~\ref{Sec:6} concludes this paper.

\section{Theoretical Modeling Framework for LCC}
\label{Sec:2}

We focus on the longitudinal control of CAVs in mixed traffic flow. We first introduce the car-following dynamics of HDVs, and then present the modeling framework of the general LCC system. Three special cases are also discussed.

Consider an open single-lane setup, as shown in Fig.~\ref{Fig:SystemSchematic}(c). The CAV is indexed as vehicle $0$, and we define $\mathcal{F}=\{1,2,\ldots,n\}$ and $\mathcal{P}=\{-1,-2,\ldots,-m\}$  as the set of the following vehicles behind and the preceding vehicles ahead, respectively. The position, velocity and acceleration of vehicle $i$ ($i \in \{0\}\cup \mathcal{F} \cup \mathcal{P}$) is denoted as $p_i$, $v_i$ and $a_i$, respectively. The spacing of vehicle $i$ from its preceding vehicle, \ie, its relative distance from vehicle $i-1$, is defined as $s_i=p_{i-1}-p_i$. Without loss of generality, the vehicle length is ignored. There exists a head vehicle in the front of this series of vehicles, whose information is not received by the CAV via V2V, and its velocity is represented as $v_\mathrm{h}$. Note that all the HDVs ahead or behind, contained in $\mathcal{F} \cup \mathcal{P}$, are incorporated in the system modeling framework; however, not all of them are required to have V2V connections, and the CAV does not need to respond to the motion of all the HDVs either. More discussions can be found in Sections~\ref{Sec:3} and~\ref{Sec:4}.

\subsection{Nonlinear Dynamics of Individual Vehicles}

We now present the nonlinear car-following dynamics of individual HDVs. Many continuous-time models exist in the literature, \eg, optimal velocity model (OVM)~\cite{bando1995dynamical}, intelligent driver model (IDM)~\cite{treiber2000congested} and their variants. Most of them can be written in the following form~\cite{orosz2010traffic} ($i \in \mathcal{F} \cup \mathcal{P}$)
\begin{equation}\label{Eq:HDVModel}
\dot{v}_i(t)=F\left(s_i(t),\dot{s}_i(t),v_i(t)\right),
\end{equation}
where $\dot{s}_{i}(t)=v_{i-1}(t)-v_{i}(t)$. Function $F(\cdot)$ means that the acceleration of vehicle $i$ depends on the relative distance, relative velocity and its own velocity; see Fig. 2(a) for illustration. 

\begin{figure}[t]
	\vspace{1mm}
	\centering
	\subfigure[]
	{\includegraphics[scale=0.45]{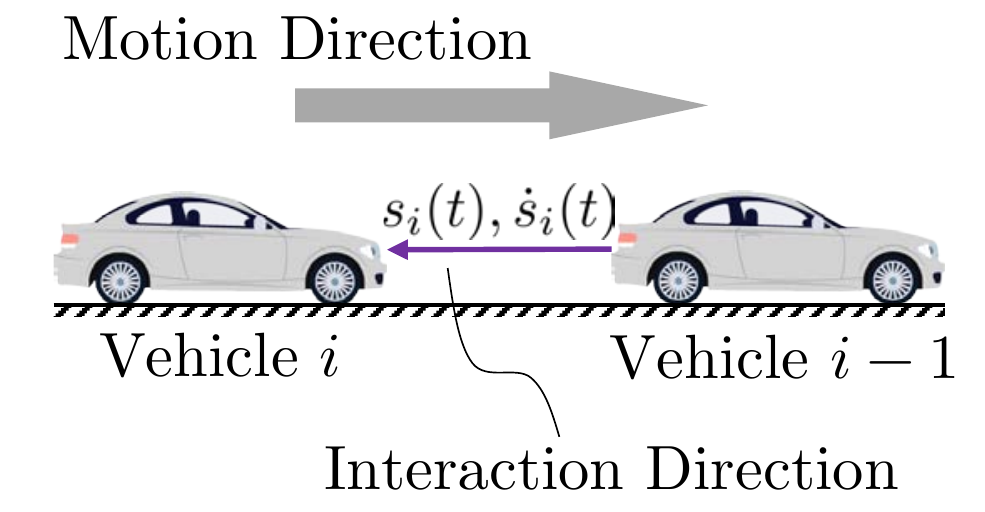}
		\label{Fig:HDVDynamicsA}}
	\hspace{3mm}
	\subfigure[]
	{\includegraphics[scale=0.55]{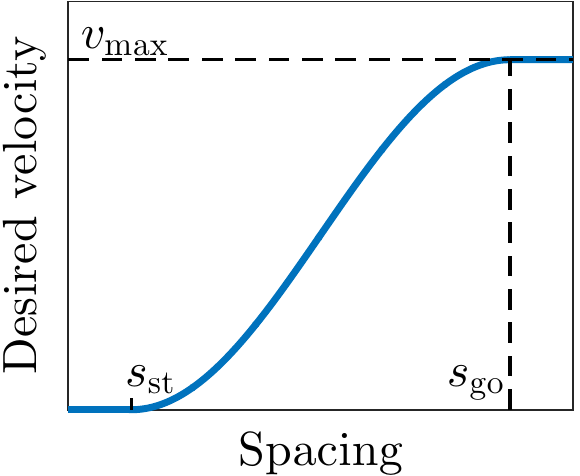}}\\
	\vspace{-1mm}
	\caption{Front-to-rear reaction dynamics of HDVs. (a) The driver's car-following action usually depends on the motion of the vehicle immediately ahead. (b) The typical relationship between the desired velocity of HDVs and the spacing, as shown in~\eqref{Eq:OVMDesiredVelocity} and~\eqref{Eq:DesiredVelocityPolicy}.}
	\label{Fig:HDVDynamics}
\end{figure}

One prevailing HDV model is the nonlinear OVM model~\cite{bando1995dynamical}, where the explicit expression of~\eqref{Eq:HDVModel} is
\begin{equation} \label{Eq:OVMmodel}
	F(\cdot)=\alpha\left(V\left(s_i(t)\right)-v_i(t)\right)+\beta\dot{s}_i(t),
\end{equation}
with $V(s)$ denoting the spacing-dependent desired velocity of the human driver, typically given by a continuous piecewise function
\begin{equation} \label{Eq:OVMDesiredVelocity}
	V(s)=\begin{cases}
		0, &s\le s_{\mathrm{st}};\\
		f_v(s), &s_{\mathrm{st}}<s<s_{\mathrm{go}};\\
		v_{\max}, &s\ge s_{\mathrm{go}}.
	\end{cases}
\end{equation}
In~\eqref{Eq:OVMDesiredVelocity}, the desired velocity $V(s)$ becomes zero for a small spacing $s_{\mathrm{st}}$, and reaches a maximum value $v_{\max}$ for a large spacing $s_{\mathrm{go}}$. When $s_{\mathrm{st}}<s<s_{\mathrm{go}}$, the desired velocity is given by a monotonically increasing function $f_v (s)$. A typical form of $f_v (s)$ is as follows
\begin{equation}
	\label{Eq:DesiredVelocityPolicy}
	f_v(s) = \frac{v_{\max }}{2}\left(1-\cos (\pi\frac{s-s_{\mathrm{st}}}{s_{\mathrm{go}}-s_{\mathrm{st}}})\right).
\end{equation}
Fig. 2(b) illustrates a profile of $V(s)$ under~\eqref{Eq:DesiredVelocityPolicy}. 

Regarding the CAV, indexed as $0$, its acceleration signal is utilized as the control input $u(t)$, shown as follows
\begin{equation}
	\dot{v}_0(t)=u(t).
\end{equation}
Note that the acceleration signal of the CAV also acts as the external control input of the entire LCC system, which can be directly designed, while all the HDVs are under human control.

\subsection{Linearized Dynamics of General LCC Systems}

It is known that in equilibrium traffic state, each vehicle moves with the same equilibrium velocity $v^{*}$ and corresponding equilibrium spacing $s^{*}$. According to~\eqref{Eq:HDVModel}, we have
\begin{equation} \label{Eq:Equilibrium}
	F\left(s^*,0,v^*\right)=0.
\end{equation}
In practice, the equilibrium velocity $v^*$ can be obtained from the steady value of the velocity of the head vehicle $v_\mathrm{h}$~\cite{jin2018connected}, and our main control objective is to stabilize the LCC system at this equilibrium state. Define the error state between actual and equilibrium state of vehicle $i$ as
\begin{equation*}
	\tilde{s}_i(t)=s_i(t)-s^*,\quad \tilde{v}_i(t)=v_i(t)-v^*,
\end{equation*}
where $\tilde{s}_{i}$, $\tilde{v}_{i}$ represent the spacing error and velocity error of vehicle $i$, respectively. Then a linearized second-order model for each HDV ($i \in \mathcal{F} \cup \mathcal{P}$) can be derived by using~\eqref{Eq:Equilibrium} and applying the first-order Taylor expansion to~\eqref{Eq:HDVModel}
\begin{equation}\label{Eq:LinearHDVModel}
	\begin{cases}
		\dot{\tilde{s}}_i(t)=\tilde{v}_{i-1}(t)-\tilde{v}_i(t),\\
		\dot{\tilde{v}}_i(t)=\alpha_{1}\tilde{s}_i(t)-\alpha_{2}\tilde{v}_i(t)+\alpha_{3}\tilde{v}_{i-1}(t),\\
	\end{cases}
\end{equation}
with $\alpha_{1} = \frac{\partial F}{\partial s}, \alpha_{2} = \frac{\partial F}{\partial \dot{s}} - \frac{\partial F}{\partial v}, \alpha_{3} = \frac{\partial F}{\partial \dot{s}}$ evaluated at the equilibrium state ($s^*,v^*$). To reflect the real driving behavior, we have $\alpha_{1}>0$, $\alpha_{2}>\alpha_{3}>0$ \cite{cui2017stabilizing,jin2017optimal}. Upon using the OVM model~\eqref{Eq:OVMmodel}, the coefficients in~\eqref{Eq:LinearHDVModel} become
\begin{equation*}
	\alpha_1 = \alpha \dot{V}(s^*),\; \alpha_2=\alpha+\beta, \; \alpha_3=\beta,
\end{equation*}
where $\dot{V}(s^*)$ denotes the derivative of $V(s)$ at $s^*$.

As for the CAV, we assume that it has the same equilibrium spacing as the other HDVs under the same equilibrium velocity, and then the longitudinal dynamics of the CAV around equilibrium can be written in a second-order form
\begin{equation}\label{Eq:LinearCAVModel}
	\begin{cases}
		\dot{\tilde{s}}_0(t)=\tilde{v}_{-1}(t)-\tilde{v}_0(t),\\
		\dot{\tilde{v}}_0(t)=u(t).\\
	\end{cases}
\end{equation}

We can now obtain the linearized dynamics model of the LCC system around the prescribed equilibrium state ($s^*,v^*$). Define the global state of LCC as
\begin{equation}\label{Eq:LCCstate}
x(t)=\begin{bmatrix}\tilde{s}_{-m}(t),\tilde{v}_{-m}(t),\ldots,\tilde{s}_{0}(t),\tilde{v}_{0}(t),\ldots,\tilde{s}_{n}(t),\tilde{v}_{n}(t)\end{bmatrix}^{\tr}.
\end{equation}
Based on the linearized HDVs' car-following model~\eqref{Eq:LinearHDVModel} and the CAV's dynamics~\eqref{Eq:LinearCAVModel}, the linearized state-space model for the LCC system is shown as follows
\begin{equation} \label{Eq:LCCSystemModel}
\dot{x}(t)=Ax(t)+Bu(t)+H\tilde{v}_{\mathrm{h}}(t),
\end{equation}
where $\tilde{v}_{\mathrm{h}}(t)$ denotes the velocity error of the head vehicle. The coefficient matrices $A\in \mathbb{R}^{(2n+2m+2)\times(2n+2m+2)}$, $B,H\in \mathbb{R}^{(2n+2m+2)\times 1}$ are given by
\begin{align*}
A&=\begin{bmatrix} P_1 & & & & &  \\
P_2 & P_1 & &  & &  \\
& \ddots& \ddots&  &  &\\
& & P_2& P_1 &  &\\
& & &  S_2 & S_1\\
& & & & P_2 &P_1\\
&&&&&\ddots&\ddots\\
&&&&&&P_2&P_1
\end{bmatrix},\\
B &= \begin{bmatrix}
b_{-m}^{\tr},\ldots,b_{-1}^{\tr},b_0^{\tr},b_1^{\tr},\ldots,b_n^{\tr}
\end{bmatrix}^{\tr},\\
H &= \begin{bmatrix}
h_{-m}^{\tr},\ldots,h_{-1}^{\tr},h_0^{\tr},h_1^{\tr},\ldots,h_n^{\tr}
\end{bmatrix}^{\tr},
\end{align*}
with block entries as ($i\in \mathcal{F}\cup \mathcal{P}$, $j\in \{0\}\cup\mathcal{F}\cup \mathcal{P} \backslash \{-m\}$)
\begin{align*}
&P_{1} = \begin{bmatrix} 0 & -1 \\ \alpha_{1} & -\alpha_{2} \end{bmatrix},
P_{2} = \begin{bmatrix} 0 & 1 \\ 0 & \alpha_{3} \end{bmatrix},\,
b_0 = \begin{bmatrix} 0  \\ 1 \end{bmatrix},
b_i = \begin{bmatrix} 0 \\ 0 \end{bmatrix},\\
&	S_1 = \begin{bmatrix} 0 & -1 \\ 0 & 0 \end{bmatrix},
S_2 = \begin{bmatrix} 0 & 1 \\ 0 & 0 \end{bmatrix},	\,
h_{-m} = \begin{bmatrix} 1 \\ \alpha_{3}\end{bmatrix},
h_j = \begin{bmatrix} 0 \\ 0 \end{bmatrix}.
\end{align*}

As shown in~\eqref{Eq:HDVModel} and Fig.~\ref{Fig:HDVDynamicsA}, the longitudinal motion of HDVs is influenced by its preceding vehicle immediately ahead. While V2V communication greatly enlarges the perception range for individual vehicles, most existing frameworks for CAV control in mixed traffic flow are limited to utilizing the information from multiple HDVs ahead --- CCC is a particular example. LCC generalizes these existing frameworks, and a straightforward distinction is the explicit consideration of those vehicles behind, as shown in~\eqref{Eq:LCCSystemModel}. To highlight this distinction, we present three special cases of the general LCC system below. In particular, we show that the CCC-type framework can be viewed as a special case in LCC as well.

\begin{figure}[t]
	\vspace{1mm}
	\centering
	\includegraphics[scale=0.45]{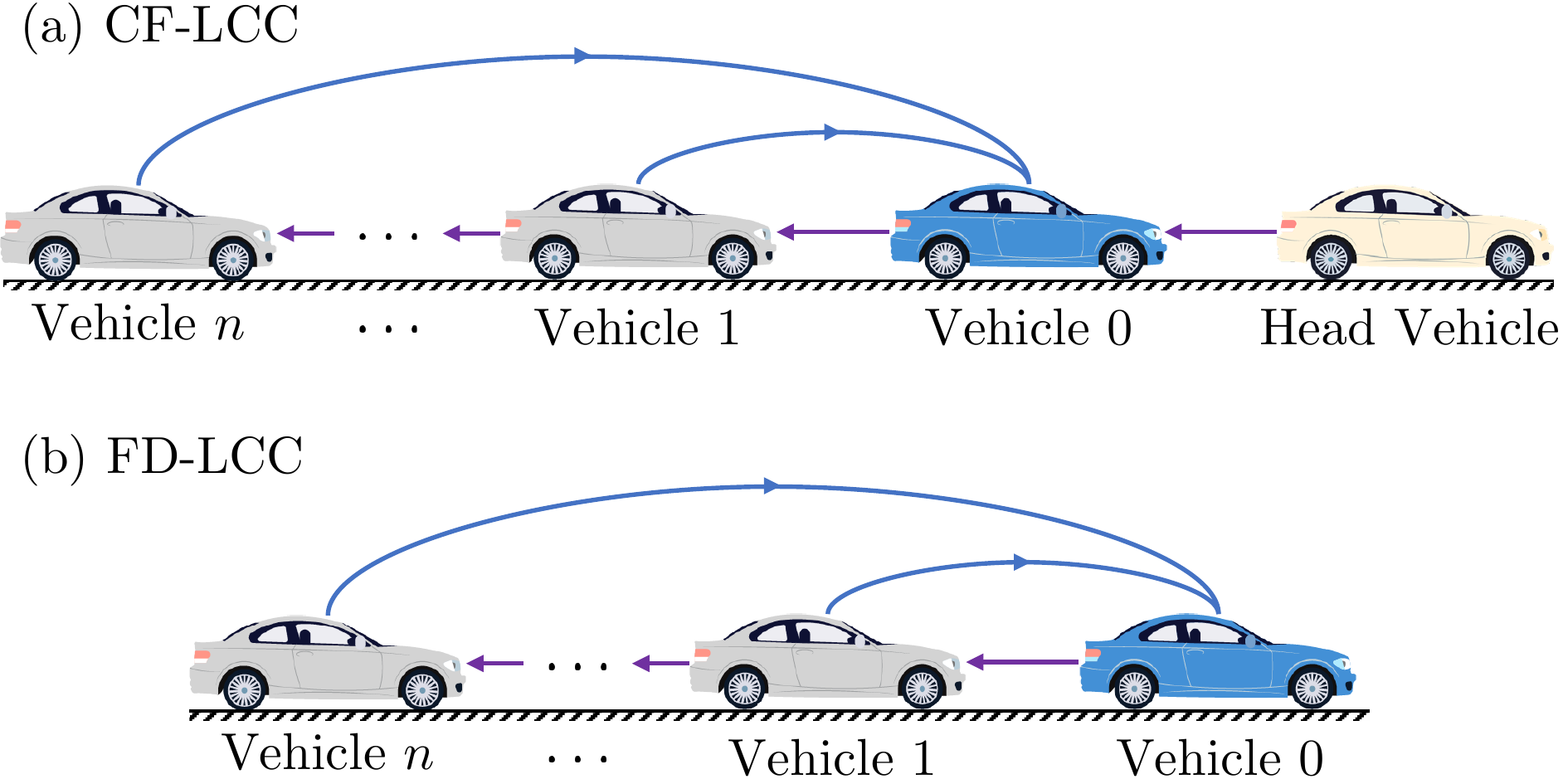}
	\vspace{-1mm}
	\caption{Two special cases of LCC when $\mathcal{P}=\emptyset$. (a) In CF-LCC, the CAV adopts the HDV's strategy \eqref{Eq:LinearHDVModel} to follow one single preceding vehicle, \ie, the head vehicle. (b) In FD-LCC, there are no preceding vehicles.}
	\label{Fig:FDLCC_CFLCC}
\end{figure}

\subsection{Special Cases of LCC}

The longitudinal behavior of individual vehicles includes two fundamental categories: car-following and free-driving~\cite{treiber2013traffic}. Accordingly, we here present two special LCC systems, Car-Following LCC and Free-Driving LCC, as demonstrated in Fig.~\ref{Fig:FDLCC_CFLCC}. In both cases, we assume that the CAV has no V2V communication with its preceding HDVs, \ie, $m=0,\mathcal{P}=\emptyset$, and we exclusively focus on ``looking behind" characteristics in the LCC framework. Finally, we also show CCC as a special case in LCC.

\emph{Special Case 1:} Car-Following LCC System (CF-LCC)

In the first case, we assume that the CAV adopts the HDVs' car-following strategy~\eqref{Eq:LinearHDVModel} to follow one single preceding vehicle, \ie, the head vehicle. Meanwhile, we also apply an additional control input $\hat{u} (t)$ into the CAV, which is determined by the state of the vehicles behind. Then, the longitudinal dynamics of the CAV can be expressed by
\begin{equation}\label{Eq:CFLCC_CAVModel}
\begin{cases}
\dot{\tilde{s}}_0(t)=\tilde{v}_{\mathrm{h}}(t)-\tilde{v}_0(t),\\
\dot{\tilde{v}}_0(t)=\alpha_{1}\tilde{s}_0(t)-\alpha_{2}\tilde{v}_0(t)+\alpha_{3}\tilde{v}_{\mathrm{h}}(t)+\hat{u}(t).\\
\end{cases}
\end{equation}
In this case, the global state of the CF-LCC system reduces from~\eqref{Eq:LCCstate} to
\begin{equation*}
\label{Eq:CFLCCstate}
x_{\mathrm{c}}(t)=[\tilde{s}_{0}(t),\tilde{v}_{0}(t),\tilde{s}_{1}(t),\tilde{v}_{1}(t),\ldots,\tilde{s}_{n}(t),\tilde{v}_{n}(t)]^{\tr}.
\end{equation*}
Then, the linearized state-space model for the CF-LCC system becomes
\begin{equation} \label{Eq:CFLCCSystemModel}
\dot{x}_{\mathrm{c}}(t)=A_{\mathrm{c}}x_\mathrm{c}(t)+B_1 \hat{u}(t)+H_1 \tilde{v}_{\mathrm{h}}(t),
\end{equation}
where $A_{\mathrm{c}}\in \mathbb{R}^{(2n+2)\times(2n+2)}$, $B_1,H_1\in \mathbb{R}^{(2n+2)\times 1}$ are given by
\begin{equation*}
A_{\mathrm{c}}=\begin{bmatrix} P_1 & & & \\
P_2 & P_1 & &    \\
& \ddots& \ddots&  \\
& & P_2& P_1 \\
\end{bmatrix},\,
B_1=\begin{bmatrix}
0\\1\\0\\ \vdots \\0
\end{bmatrix},\,
H_1=\begin{bmatrix}
1\\ \alpha_3 \\0\\ \vdots \\0
\end{bmatrix}.
\end{equation*}

\emph{Special Case 2:} Free-Driving LCC System (FD-LCC)

The second case is to assume that the CAV is driving freely with no vehicles ahead. Then the CAV's spacing $s_0$ has no real-world meaning and we consider its position instead as a state variable. Then, the longitudinal dynamics of the CAV can be given by a simple second-order model as 
\begin{equation*}
\begin{cases}
\dot{{p}}_0(t)=\tilde{v}_0(t),\\
\dot{\tilde{v}}_0(t)=u(t).\\
\end{cases}
\end{equation*}
Defining the global state of the FD-LCC system as
\begin{equation*} \label{Eq:FDLCCstate}
x_{\mathrm{f}}(t)=\begin{bmatrix}-{p}_{0}(t),\tilde{v}_{0}(t),\tilde{s}_{1}(t),\tilde{v}_{1}(t)\ldots,\tilde{s}_{n}(t),\tilde{v}_{n}(t)\end{bmatrix}^{\tr},
\end{equation*}
where the negative sign exits for consistency with \eqref{Eq:LinearCAVModel} and \eqref{Eq:CFLCC_CAVModel}, then the linearized state-space model for the FD-LCC system is consequently obtained
\begin{equation} \label{Eq:FDLCCSystemModel}
\dot{x}_{\mathrm{f}}(t)=A_{\mathrm{f}}x_\mathrm{f}(t)+B_1 u(t),
\end{equation}
where $A_{\mathrm{f}}\in \mathbb{R}^{(2n+2)\times(2n+2)}$ is given by
\begin{equation*} \label{Eq:AfExpression}
A_{\mathrm{f}}=\begin{bmatrix} S_1 & & & \\
P_2 & P_1 & &    \\
& \ddots& \ddots&  \\
& & P_2& P_1 \\
\end{bmatrix}.
\end{equation*}

\emph{Special Case 3:} Connected Cruise Control (CCC)

In the existing CCC framework~\cite{orosz2016connected,jin2014dynamics,di2019cooperative }, the CAV only utilizes the information from multiple preceding HDVs to determine its control input, without considering the motion of the vehicle behind. By letting $n=0,\mathcal{F}=\emptyset$, the proposed LCC system~\eqref{Eq:LCCSystemModel} is naturally reduced to the CCC system. Specifically, the state of CCC is chosen as
\begin{equation*}
\label{Eq:CCCstate}
x_{\mathrm{p}}(t)=[\tilde{s}_{-m}(t),\tilde{v}_{-m}(t),\ldots,\tilde{s}_{-1}(t),\tilde{v}_{-1}(t),\tilde{s}_{0}(t),\tilde{v}_{0}(t)]^{\tr}.
\end{equation*}
The linearized state-space model for the CCC system is 
\begin{equation} \label{Eq:CCCSystemModel}
\dot{x}_{\mathrm{p}}(t)=A_{\mathrm{p}}x_\mathrm{p}(t)+B_2 u(t)+H_2 \tilde{v}_{\mathrm{h}}(t),
\end{equation}
where $A_{\mathrm{p}}\in \mathbb{R}^{(2m+2)\times(2m+2)}$, $B_2,H_2\in \mathbb{R}^{(2m+2)\times 1}$ are given by
\begin{equation*}
A_{\mathrm{p}}=\begin{bmatrix} P_1 & & & \\
P_2 & P_1 & &    \\
& \ddots& \ddots&  \\
& & P_2& P_1 \\
&&&S_2&S_1
\end{bmatrix},\,
B_2=\begin{bmatrix}
0\\0\\0\\ \vdots \\0\\1
\end{bmatrix},\,
H_2=\begin{bmatrix}
1\\ \alpha_3 \\0\\ \vdots \\0\\0
\end{bmatrix}.
\end{equation*}

\begin{remark}
	In the general LCC system~\eqref{Eq:LCCSystemModel}, the CF-LCC system~\eqref{Eq:CFLCCSystemModel} and the CCC system~\eqref{Eq:CCCSystemModel}, the velocity error of the head vehicle $\tilde{v}_{\mathrm{h}}(t)$ can be viewed as an external disturbance signal into the system. Note that the models \eqref{Eq:LCCSystemModel}, \eqref{Eq:CFLCCSystemModel}, \eqref{Eq:FDLCCSystemModel}, and \eqref{Eq:CCCSystemModel} are all based on an linearization around an equilibrium velocity $v^*$ that needs to be chosen or designed carefully. For LCC, CF-LCC and CCC, the equilibrium velocity $v^*$ is determined by the equilibrium velocity of the head vehicle $v_\mathrm{h}$, and this information can be estimated from its historical trajectory data or current traffic condition. In contrast, for FD-LCC system, the equilibrium velocity can be designed according to the CAV's own desired velocity. We also refer the interested reader to~\cite[Section IV.B]{zheng2020smoothing} and~\cite[Section IV.C]{wang2020controllability} for more discussions on the equilibrium velocity $v^*$ in ring-road mixed traffic systems. 
\end{remark}

\section{Controllability and Observability \\ of LCC Systems}
\label{Sec:3}

In this section, we analyze the controllability and observability of the LCC systems. This allows us to reveal the theoretical potential of the CAV on controlling the mixed traffic flow consisting of its following HDVs behind and preceding HDVs ahead.

\subsection{Controllability Analysis}

According to the front-to-rear reaction dynamics of human drivers (see the purple arrows in Fig.~\ref{Fig:SystemSchematic}(c) and Fig. 2(a)), it is easy to see that the CAV's action will have certain influence on its following HDVs. The notion of controllability is an essential metric to quantify such influence. If a linear system is controllable, it can be driven from any initial state to arbitrary desired state within arbitrary finite time.

\begin{definition}[Controllability \rm{\cite{skogestad2007multivariable}}]
	The dynamical system $\dot{x}=Ax+Bu$ is controllable, if for any initial state $x(0)=x_0$, any time $t_{\text{f}}>0$ and any final state $x_{\text{f}}$, there exists an input $u(t)$ such that $x(t_{\text{f}})=x_{\text{f}}$.
\end{definition}

\begin{lemma}[PBH controllability test \rm{\cite{skogestad2007multivariable}}] \label{Lemma:PBH}
	System $(A,B)$ is controllable, if and only if  $\left[  \lambda I-A,B \right]$  is of full row rank for all  $\lambda$  being an eigenvalue of  $ A$, where $I$ denotes an identity matrix with compatible dimension.
\end{lemma}

\begin{lemma}[Controllability invariance \rm{\cite{skogestad2007multivariable}}] \label{Lemma:ControllabilityInvariance}
The controllability is invariant under state feedback. Precisely, ($A,B$) is controllable if and only if ($A-BK,B$) is controllable for any matrix $K$ with compatible dimensions.
\end{lemma}

We first consider the FD-LCC system in which there exist several HDVs following the CAV (see Fig.~\ref{Fig:FDLCC_CFLCC}(b)). We have the following result.

\begin{theorem} \label{Theorem:FDLCCControllability}
	The FD-LCC system with no vehicle ahead and  $n$  HDVs behind, given by \eqref{Eq:FDLCCSystemModel}, is controllable, if the following condition holds
	\begin{equation}
	\vspace{2mm}
	\label{Eq:ControllabilityCondition}
	\alpha _{1}- \alpha _{2} \alpha _{3}+ \alpha _{3}^{2} \neq 0.
	\end{equation}
\end{theorem}

\begin{IEEEproof}
	We prove this result by contradiction. Suppose that the FD-LCC system~\eqref{Eq:FDLCCSystemModel} is not controllable. By Lemma~\ref{Lemma:PBH}, there exists an eigenvalue \(  \lambda  \)  of  \( A_{\mathrm{f}} \) such that \(  \left[  \lambda I-A_{\mathrm{f}},B_{1} \right]  \) is not of full rank. Hence, there exists a nonzero vector \(  \rho  \) such that $\rho^{\tr} \left[  \lambda I-A_{\mathrm{f}},B_{1} \right] =0$, which leads to
 \begin{subequations}
		\begin{align}
		\rho ^{\tr} \left(  \lambda I-A_{\mathrm{f}} \right) &=0; \label{Eq:RhoEquationA}\\
		\rho ^{\tr}B_{1}&=0.  \label{Eq:RhoEquationB}
		\end{align}
	\end{subequations}
	Denote  $\rho$  as
	\begin{equation*}
			\rho = \begin{bmatrix}  \rho _{0}^{\tr}, \rho _{1}^{\tr}, \rho _{2}^{\tr}, \ldots , \rho _{n}^{\tr} \end{bmatrix}^{\tr},
	\end{equation*}
	where $\rho _{i}= \begin{bmatrix}  \rho _{i1}, \rho _{i2} \end{bmatrix}^{\tr} \in \mathbb{R}^{2 \times 1}$, $i=0,1, \ldots ,n $. Since only the second element in  $B_{1}$  is nonzero, \eqref{Eq:RhoEquationB} leads to  $\rho _{02}=0$. Substituting the expression \eqref{Eq:FDLCCSystemModel} of  $A_{\mathrm{f}}$ into \eqref{Eq:RhoEquationA}, we have ($i \in \{1, \ldots ,n-1\}$)
	\begin{subequations} \label{Eq:LittleRhoEquation}
		\begin{align}
		\rho _{0}^{\tr} \left( S_{1}- \lambda I \right) + \rho _{1}^{\tr}P_{2}&=0; \label{Eq:LittleRhoEquationA}\\
		\rho _{i}^{\tr} \left( P_{1}- \lambda I \right) + \rho _{i+1}^{\tr}P_{2}&=0;
		\label{Eq:LittleRhoEquationB}\\
		\rho _{n}^{\tr} \left( P_{1}- \lambda I \right) &=0. \label{Eq:LittleRhoEquationC}
		\end{align}
	\end{subequations}
	
	In the following, we solve the equations~\eqref{Eq:LittleRhoEquation} by discussing two cases separately.
	
	\emph{Case 1:} $\lambda ^{2}+ \alpha _{2} \lambda + \alpha _{1} \neq 0 $. 
	
	In this case,  \( P_{1}- \lambda I\)  is nonsingular. According to \eqref{Eq:LittleRhoEquationC}, we have  \(  \rho _{n}^{\tr}=0 \). Substituting  \(  \rho _{n}^{\tr}=0 \)  into \eqref{Eq:LittleRhoEquationB}, we have  \(  \rho _{n-1}^{\tr}=0 \). Using \eqref{Eq:LittleRhoEquationB} recursively, we can obtain that  \(  \rho _{i}^{\tr}=0,\,i=1, \ldots ,n \), which also leads to  \(  \rho _{0}^{\tr} \left( S_{1}- \lambda I \right) =0 \). Expanding this equation, we have  \(  \rho _{01}+ \lambda  \rho _{02}=0 \). Since  \(  \rho _{02}=0\), it is obtained that  \(  \rho _{01}=0 \). Consequently, we arrive at  \(  \rho =0 \), which contradicts the fact that  \(  \rho  \)  is nonzero.
	
	\emph{Case 2:} $\lambda ^{2}+ \alpha _{2} \lambda + \alpha _{1} = 0 $. 
	
	In this case, we have  \(  \lambda  \neq 0 \)  since  \(  \alpha _{1}>0 \). Also, it can be obtained that  \(  \alpha _{3} \lambda + \alpha _{1} \neq 0 \); otherwise, condition \eqref{Eq:ControllabilityCondition} will be contradicted. Expanding \eqref{Eq:LittleRhoEquationA} leads to  \(  \lambda  \rho _{01}=0 \)  and  \(  \rho _{01}+ \lambda  \rho _{02}+ \rho _{11}+ \alpha _{3} \rho _{12}=0 \). Hence, we have  \(  \rho _{01}=0 \)  and  \(  \rho _{11}+ \alpha _{3} \rho _{12}=0 \). Meanwhile, letting  \( i=1 \)  and expanding \eqref{Eq:LittleRhoEquationB} yields  \(  \rho _{12}=\frac{ \lambda }{ \alpha _{1}} \rho _{11} \), which, combined with  \(  \rho _{11}+ \alpha _{3} \rho _{12}=0 \), leads to  \(  \rho _{11}= \rho _{12}=0 \). Letting  \( i=2, \ldots ,n \)  and expanding \eqref{Eq:LittleRhoEquationB} and \eqref{Eq:LittleRhoEquationC}, we can obtain the following results (\( i=2, \ldots ,n \))
	\begin{subequations}
		\begin{align}
		\lambda  \rho _{i1}- \alpha _{1} \rho _{i2}&=0, \label{Eq:ControllabilityCase2EquationA}\\
		\left(  \lambda ^{2}+ \alpha _{2} \lambda + \alpha _{1} \right)  \rho _{ \left( i-1 \right) 1}&=\left(  \alpha _{3} \lambda + \alpha _{1} \right)  \rho _{i1} . \label{Eq:ControllabilityCase2EquationB}
		\end{align}
	\end{subequations}
	Since  \(  \alpha _{3} \lambda + \alpha _{1} \neq 0\), substituting  \(  \lambda ^{2}+ \alpha _{2} \lambda + \alpha _{1}=0 \) into~\eqref{Eq:ControllabilityCase2EquationB} and then \eqref{Eq:ControllabilityCase2EquationA} yields  \(  \rho _{i1}= \rho _{i2}=0,\,i=2, \ldots ,n \). Accordingly, we arrive at  \(  \rho =0 \), which contradicts  \(  \rho  \neq 0\).
	
	To summarize, the assumption does not hold. We now conclude that the FD-LCC system \eqref{Eq:FDLCCSystemModel} is controllable.
\end{IEEEproof}

Theorem~\ref{Theorem:FDLCCControllability} allows us to establish the controllability of the CF-LCC system~\eqref{Eq:CFLCCSystemModel}, as summarized in the following corollary.

\begin{corollary}
	 The CF-LCC system where the CAV adopts the HDVs' dynamics~\eqref{Eq:LinearHDVModel} to follow one vehicle ahead with $n$ HDVs following behind, given by~\eqref{Eq:CFLCCSystemModel}, is controllable if the condition~\eqref{Eq:ControllabilityCondition} holds.
\end{corollary}

\begin{IEEEproof}
	The proof is immediate by noting that the system matrix $A_\mathrm{c}$ in the CF-LCC system~\eqref{Eq:CFLCCSystemModel} can be derived from $A_\mathrm{f}$ in the FD-LCC system~\eqref{Eq:FDLCCSystemModel} under state feedback. Specifically, we have
	\begin{equation*}
		A_{\mathrm{c}}=A_{\mathrm{f}}-B_{1} K_{1},
	\end{equation*}
with $K_{1}=\left[-\alpha_{1}, \alpha_{2}, 0,0, \ldots, 0,0\right] \in \mathbb{R}^{1 \times(2 n+2)}$. Therefore, the proof follows from Lemma~\ref{Lemma:ControllabilityInvariance} and Theorem~\ref{Theorem:FDLCCControllability}.
\end{IEEEproof}

\begin{remark}
	Theorem~\ref{Theorem:FDLCCControllability} generalizes the stabilizability results in the closed ring-road mixed traffic system~\cite{zheng2020smoothing,wang2020controllability}, where it is shown that one single CAV can stabilize the entire traffic system. Due to the existence of the ring-road structure, there always exists an uncontrollable mode in the ring-road system~\cite{zheng2020smoothing}, while in the open straight road scenario, the FD-LCC or CF-LCC system is completely controllable under condition~\eqref{Eq:ControllabilityCondition}. We note that~\eqref{Eq:ControllabilityCondition} is a sufficient condition, and with random choices of $\alpha_{1}, \alpha_{2}, \alpha_{3}$, condition~\eqref{Eq:ControllabilityCondition} holds with probability one. Accordingly, FD-LCC and CF-LCC are controllable with probability one, meaning that the closed-loop poles of the linearized FD-LCC or CF-LCC system can be placed arbitrarily. 
\end{remark}

The following theorem characterizes the controllability of the general LCC system. 

\begin{theorem} \label{Theorem:LCCControllability}
	Consider the general LCC system with $m$ vehicles ahead and $n$ HDVs behind, given by \eqref{Eq:LCCSystemModel}. The following statements hold:
	\begin{enumerate}
		\vspace{-1.5mm}
		\item The subsystem consisting of the states of the vehicles ahead, \ie,  \( \tilde{s}_{-m} \left( t \right) ,\tilde{v}_{-m} \left( t \right) , \ldots ,\tilde{s}_{-1} \left( t \right) ,\tilde{v}_{-1} \left( t \right)  \)  is uncontrollable.
		\item The subsystem consisting of the states of the CAV and the vehicles behind, \ie,  $ \tilde{s}_{0} \left( t \right) ,\tilde{v}_{0} \left( t \right) ,$   $\tilde{s}_{1} \left( t \right) ,\tilde{v}_{1} \left( t \right) , \ldots ,\tilde{s}_{n} \left( t \right) ,\tilde{v}_{n} \left( t \right)$, is controllable, if the condition \eqref{Eq:ControllabilityCondition} holds.
	\end{enumerate}
\end{theorem}

\begin{IEEEproof}
	We first decompose the LCC system~\eqref{Eq:LCCSystemModel} into two subsystems. Consider a nonsingular matrix $T$ given by
	\begin{equation*}
		T^{-1}=\begin{bmatrix}0 & I_{2 n+2} \\ I_{2 m} & 0\end{bmatrix},
	\end{equation*}
	where $I_r$ denotes an identity matrix of dimension $r$. Then, the LCC system~\eqref{Eq:LCCSystemModel} can be represented by a different basis, defined as
	\begin{equation*}
		\tilde{x}=T^{-1} x=\begin{bmatrix}x_{1} \\ x_{2}\end{bmatrix},
	\end{equation*}
	where
	\begin{equation*}
		\begin{aligned}
		x_1&=[\tilde{s}_{0}(t),\tilde{v}_{0}(t),\tilde{s}_{1}(t),\tilde{v}_{1}(t),\ldots,\tilde{s}_{n}(t),\tilde{v}_{n}(t)]^{\tr},\\
		x_2&=[\tilde{s}_{-m}(t),\tilde{v}_{-m}(t),\ldots,\tilde{s}_{-1}(t),\tilde{v}_{-1}(t)]^{\tr}.
		\end{aligned}
	\end{equation*}
	The linear dynamics~\eqref{Eq:LCCSystemModel} can thus be transformed into 
	\begin{equation}\label{Eq:LCCSystemModel_Decomposed}
		\dot{\tilde{x}}(t)=\widetilde{A} \tilde{x}(t)+\widetilde{B} u(t)+\widetilde{H} \tilde{v}_{\mathrm{h}}(t).
	\end{equation}
	where
	\begin{align*}
		\widetilde{A}&=T^{-1} A T=\begin{bmatrix}A_{\mathrm{f}} & A_{12} \\ 0 & A_{22}\end{bmatrix},\\
		\widetilde{B}&=T^{-1} B=\begin{bmatrix}B_{1} \\ 0\end{bmatrix},\\
		\widetilde{H}&=T^{-1} H=\left[h_{0}^{\tr}, h_{1}^{\tr}, \ldots, h_{n}^{\tr}, h_{-m}^{\tr}, \ldots, h_{-1}^{\tr}\right]^{\tr},
	\end{align*}
	with
	\begin{equation*}
		\begin{aligned}
				A_{12}&=\begin{bmatrix} 0 & \ldots & 0&S_2 \\
		 &\ddots & & 0   \\
		& &\ddots & \vdots  \\
		& & & 0
		\end{bmatrix}\in \mathbb{R}^{(2n+2) \times (2n+2)},\\
		A_{22}&=\begin{bmatrix} P_1 & & & \\
		P_2 & P_1 & &    \\
		& \ddots& \ddots&  \\
		& & P_2& P_1
		\end{bmatrix}\in \mathbb{R}^{2m \times 2m}.
		\end{aligned}
	\end{equation*}
	The linear model~\eqref{Eq:LCCSystemModel_Decomposed} is equivalent to~\eqref{Eq:LCCSystemModel} for the general LCC system. This is written as
	\begin{equation} \label{Eq:ControllabilityDecomposition}
		\begin{bmatrix}\dot{x}_{1}(t) \\ \dot{x}_{2}(t)\end{bmatrix}=\begin{bmatrix}A_{\mathrm{f}} & A_{12} \\ 0 & A_{22}\end{bmatrix}\begin{bmatrix}x_{1}(t) \\ x_{2}(t)\end{bmatrix}+\begin{bmatrix}B_{1} \\ 0\end{bmatrix} u(t)+\widetilde{H} \tilde{v}_{\mathrm{h}}(t).
	\end{equation}
	Then, it can be clearly observed that $x_2 (t)$ constitutes an uncontrollable subspace of the LCC system, thus leading to the first statement in Theorem~\ref{Theorem:LCCControllability}. As for $x_1 (t)$, it has been revealed from Theorem~\ref{Theorem:FDLCCControllability} that when the condition~\eqref{Eq:ControllabilityCondition} holds, ($A_\mathrm{f},B_1$) is controllable. Therefore, we can obtain the second statement in Theorem~\ref{Theorem:LCCControllability}.
\end{IEEEproof}

In CCC-type frameworks, it has been pointed out that the CCC system is not controllable~\cite{jin2017optimal}. This is a direct corollary of Theorem~\ref{Theorem:LCCControllability}.

\begin{corollary}
	The CCC system with no vehicles behind and $m$ HDVs ahead given by~\eqref{Eq:CCCSystemModel} is not completely controllable. The controllable subspace consists of the state of the CAV itself only, \ie, $\tilde{s}_{0}(t), \tilde{v}_{0}(t)$.
\end{corollary}

In~\eqref{Eq:ControllabilityDecomposition}, we present a controllability decomposition of the general LCC system~\eqref{Eq:LCCSystemModel}. The physical interpretation of Theorem~\ref{Theorem:LCCControllability} is that the control input of the CAV has no influence on the state of the preceding HDVs, but has full control of the motion of the following HDVs. In the CCC system, the only controllable part is the state of the CAV itself, \ie, $\tilde{s}_{0}(t), \tilde{v}_{0}(t)$; consequently, the control objective of CCC is limited to improving the performance of the CAV's own car-following behavior. By contrast, the controllability of the state of the vehicles behind allows the CAV to act as a leader with a global consideration. Precisely, the CAV has the potential to improve the performance of the entire upstream traffic flow.

\subsection{Control Difficulty at Different System Sizes}

The controllability condition~\eqref{Eq:ControllabilityCondition} has no bound on the system size $n$, indicating that independent to the number of HDVs behind, the system remains controllable using the input of a single CAV. This result is consistent with the mechanism of traffic waves, where the perturbation of one individual vehicle may cause persistent velocity fluctuations propagating upstream the entire traffic flow. In LCC systems, however, it might be impractical for one single CAV to lead the motion of a large number of HDVs, which might cause unnecessary control actions for the CAV or even raise safety concerns given that the CAV still needs to adapt to the traffic ahead. Accordingly, it is important to determine how many CAVs behind should be incorporated into the LCC framework before designing specific controllers. We here provide discussions from the perspective of control difficulty.

In fact, controllability is a qualitative criterion but fails to quantify the difficulty of a control task for a controllable system. If a controllable dynamical system is ``hard" to control, a large amount of input energy might be required to reach a target state from an initial state~\cite{muller1972analysis}. Indeed, a class of energy-related metrics has been proposed to further quantify the control difficulty besides controllability. They have found many applications in complex networks and multi-agent systems~\cite{pasqualetti2014controllability,yan2012controlling,clark2012leader}. Here, we discuss the influence of the size $n$ of the following HDVs in the LCC systems using a control energy-related index, defined as $\int_0^t u(\tau)^{\tr} u(\tau) d \tau$. In vehicular systems, this index is also closely related to fuel consumption, driving comfort and safety~\cite{li2015effect}. 

\begin{definition}[Controllability Gramian \rm{\cite{muller1972analysis}}] \label{Def:ControllabilityGramian}
For a controllable dynamical system $\dot{x}(t)=Ax(t)+Bu(t)$, its Controllability Gramian at time $t$ is defined as
\begin{equation} \label{Eq:ControllabilityGramian}
	W(t)=\int_{0}^{t} e^{A \tau} B B^{\tr} e^{A^{\tr} \tau} d \tau,
\end{equation}
which is always positive definitive.
\end{definition}

\begin{lemma}[Minimum control energy \rm{\cite{muller1972analysis}}] \label{Lemma:MinimumEnergy}
	For a controllable dynamical system $\dot{x}(t)=Ax(t)+Bu(t)$, the minimum control energy required to move the system from the initial state $x(0)=x_0$ to the target state $x(t)=x_{\mathrm{tar}}$ is given by
	\begin{equation} \label{Eq:Definition_MinimumEnergy}
	\begin{aligned}
	&\min \int_{0}^t u(\tau)^{\tr} u(\tau) d \tau\\
	=\;&\left(x_{\mathrm{tar}}-e^{A t} x_{0}\right)^{\tr} W\left(t\right)^{-1}\left(x_{\mathrm{tar}}-e^{A t} x_{0}\right).
	\end{aligned}
	\end{equation}
\end{lemma}

\begin{figure}[t]
	\vspace{1mm}
	\centering
	\includegraphics[scale=0.45]{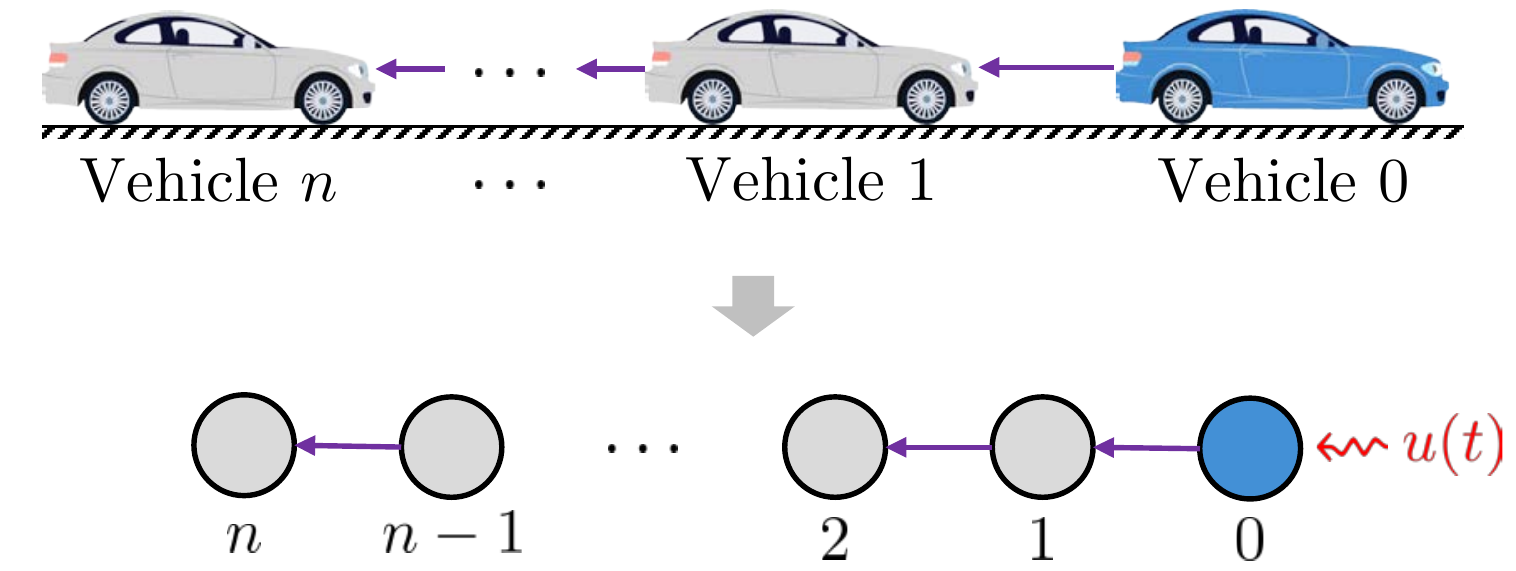}
	\vspace{-1mm}
	\caption{A network schematic of the FD-LCC system consisting of one leading CAV and $n$ HDVs. The blue node with control input $u(t)$ represents the CAV, while the gray uncontrolled nodes represent the HDVs.}
	\label{Fig:NetworkSchematic}
\end{figure}

According to Lemma~\ref{Lemma:MinimumEnergy}, the state space corresponding to larger eigenvalues of $W(t)^{-1}$, \ie, smaller eigenvalues of $W(t)$, requires higher input energy to reach, \ie, is ``harder'' to control. 
This interpretation inspires using specific metrics related to $W(t)$ or $W(t)^{-1}$ to quantify control difficulty~\cite{muller1972analysis}. Two typical examples are $\lambda_{\mathrm{min}}(W(t))$ and $\mathrm{Tr}(W(t)^{-1} )$, where $\lambda_{\mathrm{min}}(\cdot)$ and $\mathrm{Tr}(\cdot)$ denotes the smallest eigenvalue and the trace of a matrix, respectively. In particular, $\lambda_{\mathrm{min}}(W(t))$ represents a worst-case metric inversely related to the energy required to move the system in the direction that is the most difficult to move, while $\mathrm{Tr}(W(t)^{-1} )$ measures the average control energy over random target states~\cite{pasqualetti2014controllability}.

\begin{figure}[t]
	\vspace{1mm}
	\centering
		\subfigure[]
	{\includegraphics[scale=0.36]{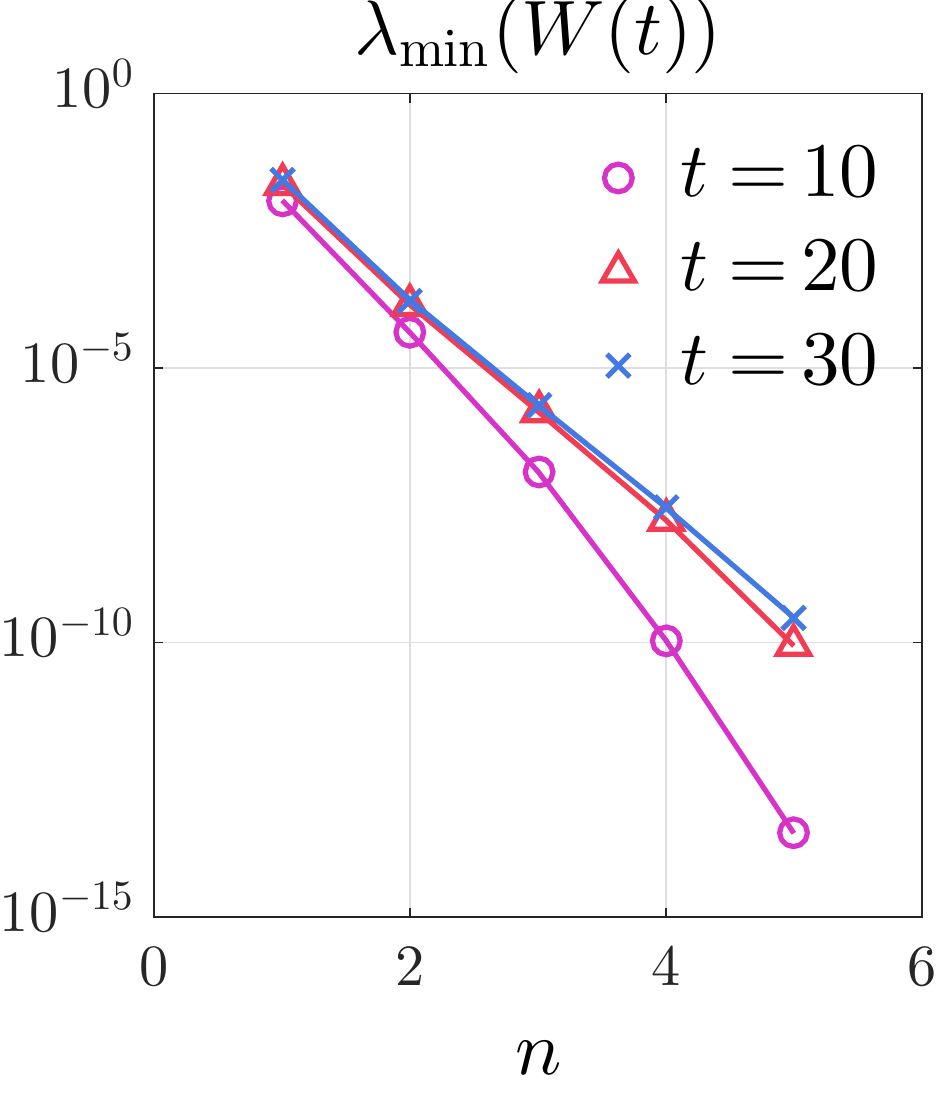}
		\label{Fig:MinimumEigenvalue}}
	\hspace{4mm}
	\subfigure[]
	{\includegraphics[scale=0.36]{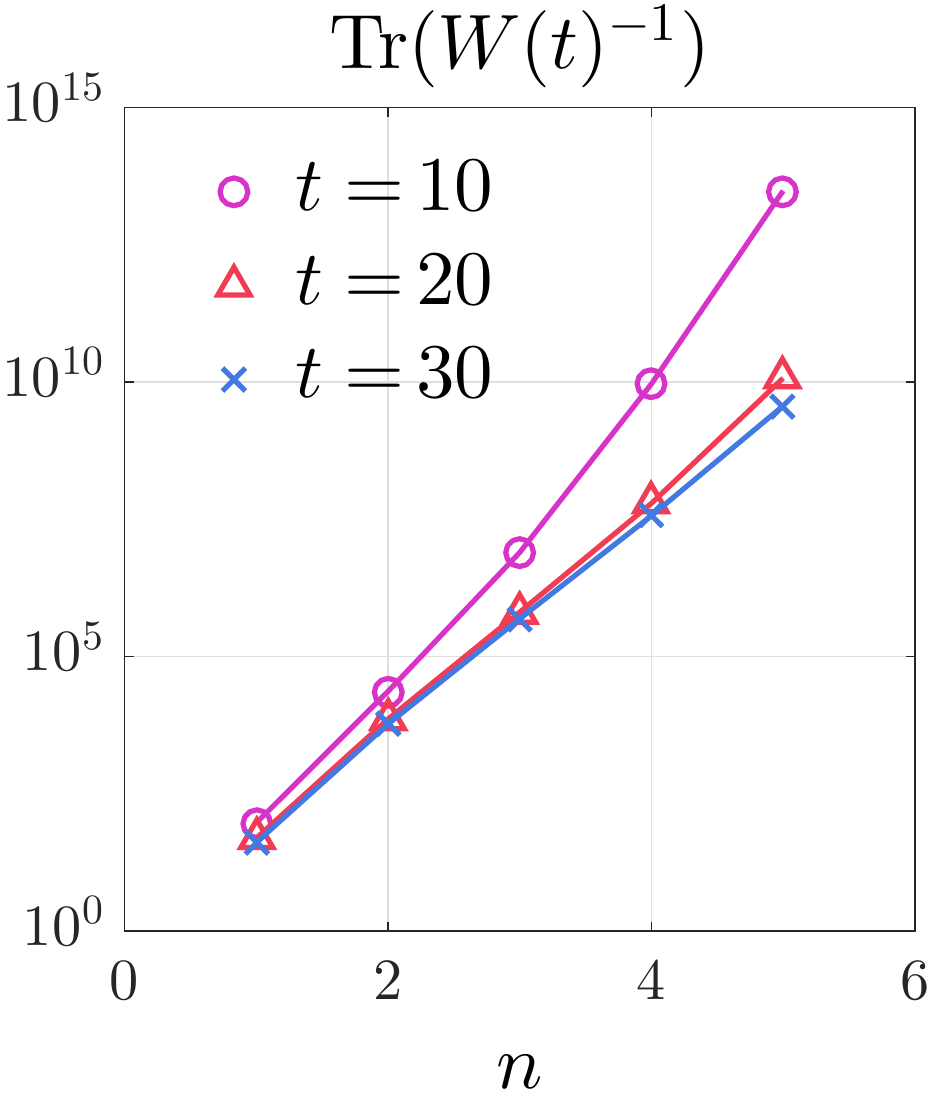}
		\label{Fig:AverageEnergy}}
	\vspace{-1mm}
	\caption{Results of control energy at different system sizes. (a) and (b) demonstrate the smallest eigenvalue of $W(t)$ and the trace of $W(t)^{-1}$, respectively. At a relatively large value of $n$, $W(t)^{-1}$ is computationally intractable, since $W(t)$ is close to singularity. 
	}
	\label{Fig:ControlEnergy}
\end{figure}

We utilize these two metrics to measure the energy-related control difficulty of the LCC systems at different system sizes. We focus on the FD-LCC system~\eqref{Eq:FDLCCSystemModel}, which can be abstracted as a chain network with one single input node and $n$ uncontrolled nodes, as shown in Fig.~\ref{Fig:NetworkSchematic}. The results below can be extended to the CF-LCC system or general LCC system. 
Then, we numerically calculate the value of $\lambda_{\mathrm{min}}(W(t))$, $\mathrm{Tr}(W(t)^{-1} )$ and $E_{\mathrm{min}}(t)$ at different system sizes $n$. The OVM model~\eqref{Eq:OVMmodel} is utilized to derive the value of the system matrix $A_\mathrm{f}$ in a typical parameter setup~\cite{jin2017optimal,zheng2020smoothing,wang2020controllability}: $\alpha=0.6, \beta=0.9, v_{\max }=30, s_{\mathrm{st}}=5, s_{\mathrm{go}}=35, v^* = 15$. The results under three different time lengths $t=10\,\mathrm{s}, 20\,\mathrm{s}, 30\,\mathrm{s}$ are shown in Fig.~\ref{Fig:ControlEnergy}. It can be clearly observed that as the system size $n$  increases, $\lambda_{\mathrm{min}}(W(t))$ drops down rapidly, while $\mathrm{Tr}(W(t)^{-1} )$ grows up dramatically. 

Based on  Fig.~\ref{Fig:ControlEnergy}, one can deduce that the FD-LCC system requires an exponential control energy with respect to system size $n$. 
As expected, the control energy is reduced when the time length $t$ increases, indicating that it becomes easier to control the FD-LCC system to a target state given a longer time period. Although the FD-LCC system remains controllable at arbitrary system size $n$, our numerical results suggest that it is more feasible for the CAV to respond to the motion of a moderate number of the HDVs behind in the LCC framework, considering practical fuel consumption, driving comfort and safety of the CAV~\cite{li2015effect}.

\begin{remark}
Note that network control with a few input nodes has attracted extensive interest, and controllability and control energy are two fundamental metrics; see, \eg, \cite{summers2015submodularity,liu2011controllability,yan2012controlling}. For asymptotically stable systems, the Controllability Gramian~\eqref{Eq:ControllabilityGramian} is finite when $t \rightarrow \infty$, which can be computed by solving a Lyapunov equation~\cite{muller1972analysis}. However, there exists a zero eigenvalue in open-loop LCC systems, and thus the Controllability Gramian can only be calculated at a finite time horizon in~\eqref{Eq:ControllabilityGramian}. It is non-trivial to get an analytical expression for~\eqref{Eq:ControllabilityGramian}. We also note that some network systems with first-order dynamics are shown to require an exponentially growing control energy; see, \eg, \cite[Example 1]{pasqualetti2014controllability}. For the CAV control in mixed traffic, it will be interesting to quantify the control difficulty more precisely. We left this in the future work.
\end{remark}

\subsection{Observability Analysis}

In practice, not all the HDVs might have V2V capabilities, or the CAV might only have access to the information of its neighboring HDVs due to limit of communication range. Here, we are interested in whether the CAV can estimate the states of all the HDVs in the LCC system with limited information available. In particular, we analyze the observability property of the LCC system. As a dual concept of controllability, observability quantifies the ability of estimating the system's state from its output. 

\begin{definition}[Observability \rm{\cite{skogestad2007multivariable}}] \label{Def:Observability}
	Consider a dynamical system $\dot{x}=A x+B u$, $y=C x$. The pair ($A,C$) is observable if, for any time $t_{\text{f}}>0,$ the initial state $x(0)=$ $x_{0}$ can be determined from the time history of the input $u(t)$ and the output $y(t)$ in the interval $\left[0, t_{\text{f}}\right]$.
\end{definition}

\begin{lemma}[PBH observability test \rm{\cite{skogestad2007multivariable}}] \label{Lemma:PBH_Observability}
($A, C$) is observable, if and only if $\left[(\lambda I-A)^{\tr}, C^{\tr}\right]$ is of full row rank for all $\lambda$ being a right eigenvalue of $A$.
\end{lemma}

We assume that the CAV can only receive the velocity error signal $\tilde{v}_{k}(t)$ of one HDV indexed as $k$ ($1 \leq k \leq n$). This signal might be easier to obtain compared to the spacing error signal $\tilde{s}_{i}(t)$, since the equilibrium spacing $s^{*}$ of one HDV is non-trivial to measure accurately. We now consider the FD-LCC system~\eqref{Eq:FDLCCSystemModel}, and the output is given by
\begin{equation} \label{Eq:FDLCC_Output}
	y_{\mathrm{f}}(t)=C_{\mathrm{f}}^{(k)} x(t),
\end{equation}
where $y_{\mathrm{f}}$ denotes the output and $C_{\mathrm{f}}^{(k)} \in \mathbb{R}^{1 \times(2 n+2)}$ denotes the output matrix, given by
\begin{equation*}
	C_{\mathrm{f}}^{(k)}=\left[c_{0}^{\tr}, c_{1}^{\tr}, \ldots, c_{n}^{\tr}\right],
\end{equation*}
with
\begin{equation*}
	c_{k}=\begin{bmatrix}
	0 \\
	1
	\end{bmatrix}, c_{i}=\begin{bmatrix}
	0 \\
	0
	\end{bmatrix}, i \in\{0\} \cup \mathcal{F} \backslash\{k\}.
\end{equation*}
Then the observability result is as follows.

\begin{theorem} \label{Theorem:Observability_FDLCC}
	Consider the FD-LCC system with no vehicle ahead and $n$ HDVs behind, given by~\eqref{Eq:FDLCCSystemModel}. Suppose the CAV can measure the velocity error of vehicle $k$ ($1 \leq k < n$), given by~\eqref{Eq:FDLCC_Output}. Then, we have the following statements:
	\begin{enumerate}
		\vspace{-0.5mm}
		\item The subsystem consisting of the states of the vehicles indexed from $1$ to $k$, \ie,  \( \tilde{s}_{1} \left( t \right) ,\tilde{v}_{1} \left( t \right) , \ldots ,\tilde{s}_{k} \left( t \right) ,\tilde{v}_{k} \left( t \right)  \)  is observable, if the condition~\eqref{Eq:ControllabilityCondition} holds.
		\item The subsystem consisting of the states of the vehicles indexed from $k+1$ to $n$, \ie,  $ \tilde{s}_{k+1} \left( t \right) ,\tilde{v}_{k+1} \left( t \right) ,\ldots ,$ $\tilde{s}_{n} \left( t \right) ,\tilde{v}_{n} \left( t \right)$, is unobservable.
	\end{enumerate}
\end{theorem}

This result is expected considering the dual relationship between controllability and observability, and its proof is similar to that of Theorem~\ref{Theorem:FDLCCControllability} and Theorem~\ref{Theorem:LCCControllability} based on the PBH observability test and Kalman decomposition. In addition, we also have the following corollary.

\begin{corollary} \label{Corollary:Observability_FDLCC}
Consider the FD-LCC system with no vehicle ahead and $n$ HDVs behind, given by~\eqref{Eq:FDLCCSystemModel}. The CAV can measure the velocity error signal of vehicle $n$. Then the states of all the HDVs, \ie, \( \tilde{s}_{1} \left( t \right) ,\tilde{v}_{1} \left( t \right) , \ldots ,\tilde{s}_{n} \left( t \right) ,\tilde{v}_{n} \left( t \right)  \), are observable, if the condition~\eqref{Eq:ControllabilityCondition} holds. 
\end{corollary}

Note that Theorem~\ref{Theorem:Observability_FDLCC} and Corollary~\ref{Corollary:Observability_FDLCC} are also applicable to the CF-LCC system. In the following, we proceed to consider the general LCC system~\eqref{Eq:LCCSystemModel}. In addition to velocity error signal of one HDV indexed as $k$, the CAV is naturally able to measure its own states, including the spacing error and the velocity error. Thus, the output of LCC can be represented by
\begin{equation} \label{Eq:LCC_Output}
	y(t)=C^{(k)} x(t),
\end{equation}
where $y$ denotes the output signal for LCC. The output matrix $C^{(k)} \in$ $\mathbb{R}^{3 \times(2 n+2 m+2)}$ is given by
\begin{equation*}
	C^{(k)} = \begin{bmatrix}
		\mathbb{e}_{2m+1},\mathbb{e}_{2m+2},\mathbb{e}_{2m+2+2k}
	\end{bmatrix}^\tr,
\end{equation*}
where the vector $\mathbb{e}_r$ denotes a $(2 n+2 m+2)\times 1$ unit vector, with the $r$-th entry being one and the others being zeros.

We are now ready to present the observability result for the general LCC system.

\begin{theorem} \label{Theorem:Observability_LCC}
	Consider the general LCC system with $m$ vehicles ahead and $n$ HDVs behind, given by~\eqref{Eq:LCCSystemModel}. Suppose that the CAV can measure its own states and the velocity error signal of vehicle $k$ ($1 \leq k < n$), given by~\eqref{Eq:LCC_Output}. Then, we have the following statements:
	\begin{enumerate}
		\vspace{-0.5mm}
		\item The subsystem consisting of the states of the vehicles indexed from $-m$ to $k$, \ie,  
		$\tilde{s}_{-m} \left( t \right) ,\tilde{v}_{-m} \left( t \right) , \ldots ,$ 
		$\tilde{s}_{k} \left( t \right) ,\tilde{v}_{k} \left( t \right) $  
		is observable, if the condition~\eqref{Eq:ControllabilityCondition} holds.
		\item The subsystem consisting of the states of the vehicles indexed from $k+1$ to $n$, \ie,  $ \tilde{s}_{k+1} \left( t \right) ,\tilde{v}_{k+1} \left( t \right) ,\ldots ,$ $\tilde{s}_{n} \left( t \right) ,\tilde{v}_{n} \left( t \right)$, is unobservable.
	\end{enumerate}
\end{theorem}

\begin{corollary} \label{Corollary:Observability_LCC}
Consider the general LCC system with $m$ vehicles ahead and $n$ HDVs behind, given by~\eqref{Eq:LCCSystemModel}. Suppose the CAV can measure its own states and the velocity error signal of vehicle $n$, \ie, the output is given by~\eqref{Eq:LCC_Output} with $k=n$. Then the LCC system is observable, if the condition~\eqref{Eq:ControllabilityCondition} holds.
\end{corollary}

The following result for the CCC system is immediate.

\begin{corollary} \label{Corollary:Observability_CCC}
	Consider the CCC system with no vehicles behind and $m$ HDVs ahead given by~\eqref{Eq:CCCSystemModel}. Suppose the CAV can measure its own states. Then, the CCC system is observable, if the condition~\eqref{Eq:ControllabilityCondition} holds. 
\end{corollary}


\begin{remark}
We note that the observability property for mixed traffic flow is less studied before, and the prevailing CCC system~\cite{jin2014dynamics,di2019cooperative} did not address this issue explicitly. 
When designing control strategies for CAVs, most existing work focuses on state feedback, especially a full-state feedback controller (see, \eg, \cite{zheng2020smoothing,di2019cooperative,jin2017optimal}), which might be impractical. To deal with limited information, one approach is to design a structured controller (see, \eg,~\cite{rotkowitz2005characterization,jovanovic2016controller,furieri2020sparsity} and the references therein), which has shown promising results in controller synthesis for ring-road mixed traffic systems~\cite{wang2020controllability}. Another approach is to design a dynamical output feedback controller. The observability results in Theorems~\ref{Theorem:Observability_FDLCC} and~\ref{Theorem:Observability_LCC} indicate that we can design an optimal state estimator given a dynamical model of the LCC systems, and the linear-quadratic-Gaussian control is applicable. However, it might be non-trivial to obtain an accurate model for LCC systems, and possible model mismatches should be handled explicitly to achieve good robustness performance~\cite{zhou1996robust}. 
It will be interesting for future work to design a robust controller for LCC systems. 
\end{remark}

\section{Head-to-Tail String Stability}
\label{Sec:4}

The controllability/observability analysis  confirms the potential of the CAV to actively lead the motion of the following HDVs and stabilize upstream traffic flow. Regarding the preceding vehicles ahead, it is important to address the CAV's capability in dampening the perturbations coming from front. This capability can be captured by the notion of string stability~\cite{swaroop1996string}. In this section, we study the string stability performance of mixed traffic flow under the proposed LCC framework.

\subsection{Head-to-Tail Transfer Function}

String stability depicts the ability of an individual vehicle in attenuating velocity fluctuations coming from the vehicle immediately ahead. For a series of vehicles, head-to-tail string stability is utilized more often, especially in a mixed traffic scenario. The definition is presented below.

\begin{definition}[Head-to-Tail String Stability \rm{\cite{jin2014dynamics}}]
	\label{Def:HeadtoTail}
	Given a series of consecutive vehicles, denote the velocity deviation of the vehicle at the head and the one at the tail as  $ \tilde{v}_\mathrm{h} \left( t \right) $  and  \( \tilde{v}_\mathrm{t} \left( t \right)  \) , respectively. The head-to-tail transfer function is defined as
	\begin{equation} \label{Eq:GeneralTransferFunction}
	\Gamma (s) = \frac{\widetilde{V}_\mathrm{t} (s)  }{\widetilde{V}_\mathrm{h} (s) },
	\end{equation}
	where $\widetilde{V}_\mathrm{h}(s), \widetilde{V}_\mathrm{t}(s)$ denote the Laplace transform of  $ \tilde{v}_\mathrm{h} (t) $  and  $ \tilde{v}_\mathrm{t} (t) $, respectively. Then the system is called head-to-tail string stable if and only if
	\begin{equation} \label{Eq:HeadtoTailDefinition}
	\vert  \Gamma  \left( j \omega  \right)  \vert ^{2}<1,  \forall  \omega >0,
	\end{equation}
	where $ j^2=-1 $, and $\vert \cdot \vert  $ denotes the modulus.
\end{definition}

Head-to-tail string stability describes a property in a series of vehicles where the perturbation signals are attenuated between the head and the tail vehicles for all excitation frequencies. When head-to-tail string stability is violated, a small perturbation in the head vehicle might cause severe stop-and-go behaviors in the following vehicles. 

We investigate the head-to-tail string stability property in the proposed LCC framework. As shown in Fig.~\ref{Fig:SystemSchematic}(c), we consider a general LCC system with $m$ preceding vehicles and $n$ following HDVs. Let the velocity deviation $\tilde{v}_\mathrm{h} (t)$ of the head vehicle (the vehicle colored yellow in Fig.~\ref{Fig:SystemSchematic}(c)) be the input of the LCC system, and the velocity deviation $\tilde{v}_{n} (t)$ of the HDV at the very tail be the output. We then derive the head-to-tail transfer function of the LCC system, defined as~\eqref{Eq:GeneralTransferFunction}. 
First, the Laplace transform of the linearized car-following model~\eqref{Eq:LinearHDVModel} of HDVs yields the local transfer function of HDVs' dynamics as follows ($i \in \mathcal{F} \cup \mathcal{P}$)

\begin{equation*}\label{Eq:HDVTransferFunction}
\frac{\widetilde{V}_{i} \left( s \right) }{\widetilde{V}_{i-1} \left( s \right) }=\frac{ \alpha _{3}s+ \alpha _{1}}{s^{2}+ \alpha _{2}s+ \alpha _{1}}=\frac{ \varphi  \left( s \right) }{ \gamma  \left( s \right) },
\end{equation*}
with
\begin{equation*}
\varphi  \left( s \right) = \alpha _{3}s+ \alpha _{1},  \gamma  \left( s \right) =s^{2}+ \alpha _{2}s+ \alpha _{1}.
\end{equation*}

\begin{figure*}[t]
	\vspace{1mm}
	\centering
	\hspace{1.5mm}
	\includegraphics[scale=0.45]{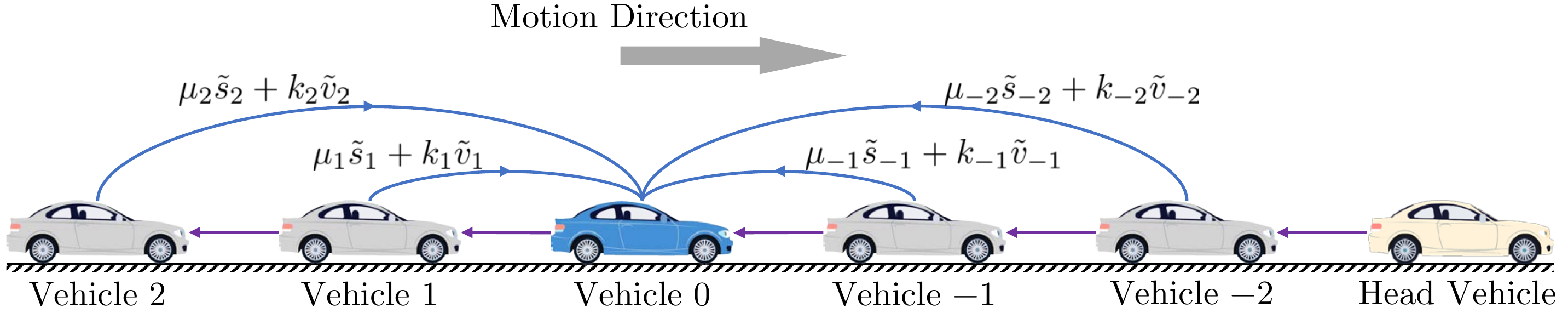}
	\vspace{-1mm}
	\caption{Schematic of the LCC system when $n=m=2$ and the CAV utilizes the controller~\eqref{Eq:ControlInputDefinition}. Note that the CAV adopts the HDV's strategy~\eqref{Eq:LinearHDVModel} to follow the vehicle immediately ahead, as depicted by the purple arrow pointing to vehicle 0. Meanwhile, $\mu_i,k_i$, $i=1,2$ represent ``looking behind'' feedback gains, which correspond to the following vehicles, while $\mu_i,k_i$, $i=-1,-2$ represent ``looking ahead'' feedback gains, which correspond to the preceding vehicles.}
	\label{Fig:SystemSchematic_Ahead2Behind2}
\end{figure*}

For the CAV, we assume that it adopts the HDVs' strategy~\eqref{Eq:LinearHDVModel} to follow the vehicle immediately ahead, while also exploiting the state of other HDVs for feedback control; see Fig.~\ref{Fig:SystemSchematic_Ahead2Behind2} for illustration when $n=m=2$. Denote $  \mu _{i},k_{i} $ as the static feedback gain of the spacing error and the velocity error of vehicle $ i $  ($ i \in \mathcal{F} \cup \mathcal{P} $), respectively. Then, the control input is given by
\begin{equation}\label{Eq:ControlInputDefinition}
	\small
u \left( t \right) = \alpha _{1}\tilde{s}_{0}(t)- \alpha _{2}\tilde{v}_{0}(t)+ \alpha _{3}\tilde{v}_{-1}(t)
+  \sum _{i\in \mathcal{F} \cup \mathcal{P}} \left(  \mu _{i}\tilde{s}_{i} \left( t \right) +k_{i}\tilde{v}_{i} \left( t \right)  \right).
\end{equation}
Note that if vehicle $i$ is not connected or the CAV does not need to respond to its motion, its corresponding feedback gain is naturally set to zeros, \ie, $\mu _{i}=k_{i}=0$. Particularly, when  $\mu _{i}=k_{i}=0$, $\forall i \in \mathcal{F} \cup \mathcal{P}$, this control strategy is reduced to the linearized car-following dynamics of HDVs.
Substituting~\eqref{Eq:ControlInputDefinition} into the CAV's longitudinal dynamics~\eqref{Eq:LinearCAVModel} and combining the Laplace transform of~\eqref{Eq:LinearHDVModel} and \eqref{Eq:LinearCAVModel}, we can obtain the head-to-tail transfer function of the LCC system as follows
\begin{equation}\label{Eq:LCCTransferFunction}
\Gamma  \left( s \right) =G(s) \cdot \left( \frac{ \varphi  \left( s \right) }{ \gamma  \left( s \right) } \right) ^{n+m},
\end{equation}
where
\begin{equation*}
	G(s)=\frac{ \varphi  \left( s \right) + \sum _{i\in \mathcal{P}}H_{i} \left( s \right)  ( \frac{ \varphi  \left( s \right) }{ \gamma  \left( s \right) } ) ^{i+1}}{ \gamma  \left( s \right) - \sum_{i \in \mathcal{F}} H_{i} \left( s \right)  ( \frac{ \varphi  \left( s \right) }{ \gamma  \left( s \right) } ) ^{i}},
\end{equation*}
with
\begin{equation*}
H_{i} \left( s \right) = \mu _{i}\left(\frac{\gamma (  s )}{\varphi  ( s )} - 1\right)+k_{i}s,\,i \in \mathcal{F} \cup \mathcal{P}.
\end{equation*}

We consider three special cases to provide further insights of the transfer function~\eqref{Eq:LCCTransferFunction}.

1) When  $\mu _{i}=k_{i}=0$, $i \in\mathcal{F} \cup \mathcal{P}$ , \ie, the CAV follows the same control strategy as human drivers, the head-to-tail transfer function then degrades to
\begin{equation}
\Gamma _{1} \left( s \right) = \left( \frac{ \varphi  \left( s \right) }{ \gamma  \left( s \right) } \right) ^{n+m+1},
\end{equation}
corresponding to a platoon of  $n+m+1$  HDVs.

2) When $\mu _{i}=k_{i}=0$, $i \in \mathcal{F} $, \ie, the CAV does not exploit the information of the following HDVs for feedback control, the controller~\eqref{Eq:ControlInputDefinition} becomes a typical CCC strategy. This strategy considers the motion of multiple vehicles ahead for longitudinal control. In this case, the head-to-tail transfer function for the LCC framework becomes
\begin{equation} \label{Eq:AheadTransferFunction}
\Gamma_2  \left( s \right) =\frac{ \varphi  \left( s \right) + \sum _{i\in \mathcal{P}}H_{i} \left( s \right)  ( \frac{ \varphi  \left( s \right) }{ \gamma  \left( s \right) } ) ^{i+1}}{ \gamma  \left( s \right)} \cdot \left( \frac{ \varphi  \left( s \right) }{ \gamma  \left( s \right) } \right) ^{n+m}.
\end{equation}

3) When  $ \mu _{i}=k_{i}=0$, $i \in \mathcal{P}$, the CAV adopts the same strategy as that of HDVs to follow the preceding vehicle, and meanwhile considers the information of multiple HDVs behind to adjust its own motion. Under this circumstance, the head-to-tail transfer function becomes
\begin{equation} \label{Eq:BehindTransferFunction}
\Gamma_3  \left( s \right) =\frac{ \varphi  \left( s \right)  }{ \gamma  \left( s \right)- \sum_{i \in \mathcal{F}} H_{i} \left( s \right)  ( \frac{ \varphi  \left( s \right) }{ \gamma  \left( s \right) } ) ^{i}} \cdot \left( \frac{ \varphi  \left( s \right) }{ \gamma  \left( s \right) } \right) ^{n+m}.
\end{equation}

As clearly observed from~\eqref{Eq:AheadTransferFunction} and~\eqref{Eq:BehindTransferFunction}, incorporating either the information of the preceding vehicles or the following vehicles brings a significant change to the head-to-tail transfer function of mixed traffic flow, but the changes of the two types work in different ways. Most existing studies addressed the head-to-tail string stability of mixed traffic flow when the CAV monitors the motion of the vehicles ahead, \ie, adopting a CCC-type framework~\cite{jin2014dynamics,di2019cooperative}. The influence of the incorporation of the vehicles behind remains unclear on the head-to-tail string stability of mixed traffic flow. 
\begin{remark}
    Previous research on CCC-type frameworks mostly focus on the transfer function from the head vehicle to the CAV itself, instead of a certain vehicle behind the CAV. It is worth noting that \emph{the perturbations continue to propagate upstream after reaching the CAV}: although the CAV can mitigate the perturbations coming from front, the perturbations might still be amplified behind the CAV. It is more desirable to incorporate the motion information of the HDVs behind into the CAV's control, thus improving the capability of CAVs in attenuating perturbations in the entire mixed traffic flow.
\end{remark}

\subsection{Head-to-Tail String Stable Region}

\begin{figure}[t]
	\vspace{1mm}
	\centering
	\subfigure[]
	{\includegraphics[scale=0.5,trim=5 2 17 10,clip]{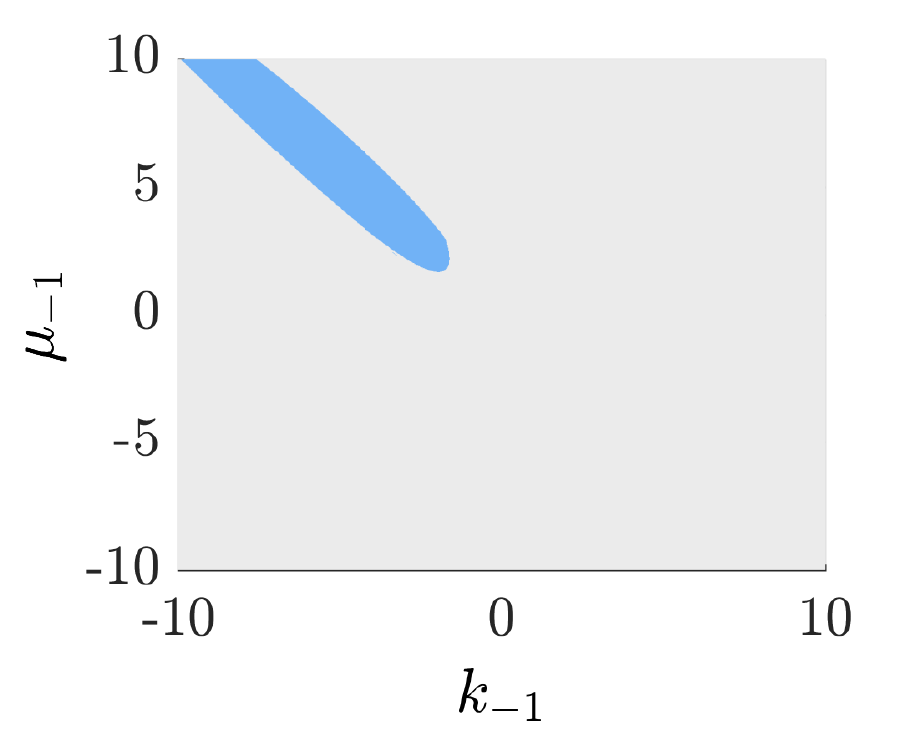}}
	\subfigure[]
	{\includegraphics[scale=0.5,trim=5 2 17 10,clip]{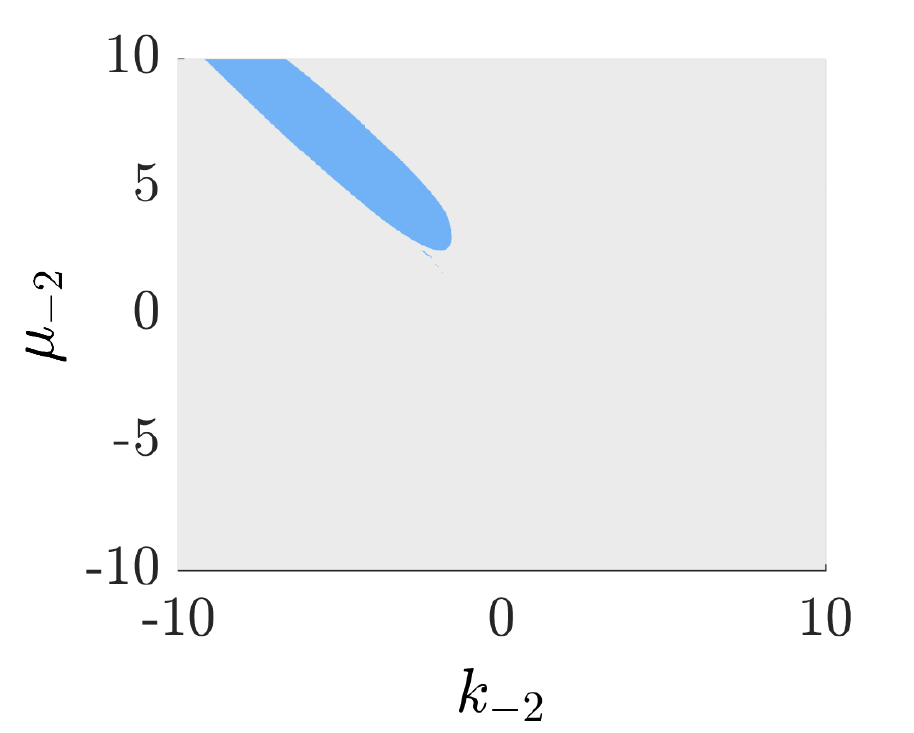}}\\
	\subfigure[]
	{\includegraphics[scale=0.5,trim=5 2 17 10,clip]{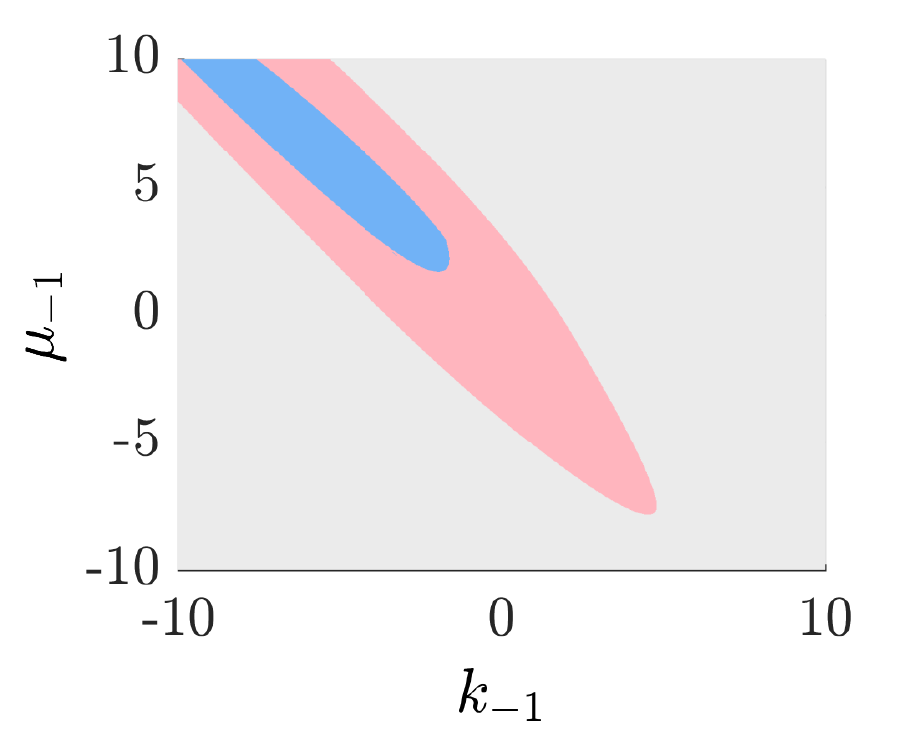}}
	\subfigure[]
	{\includegraphics[scale=0.5,trim=5 2 17 10,clip]{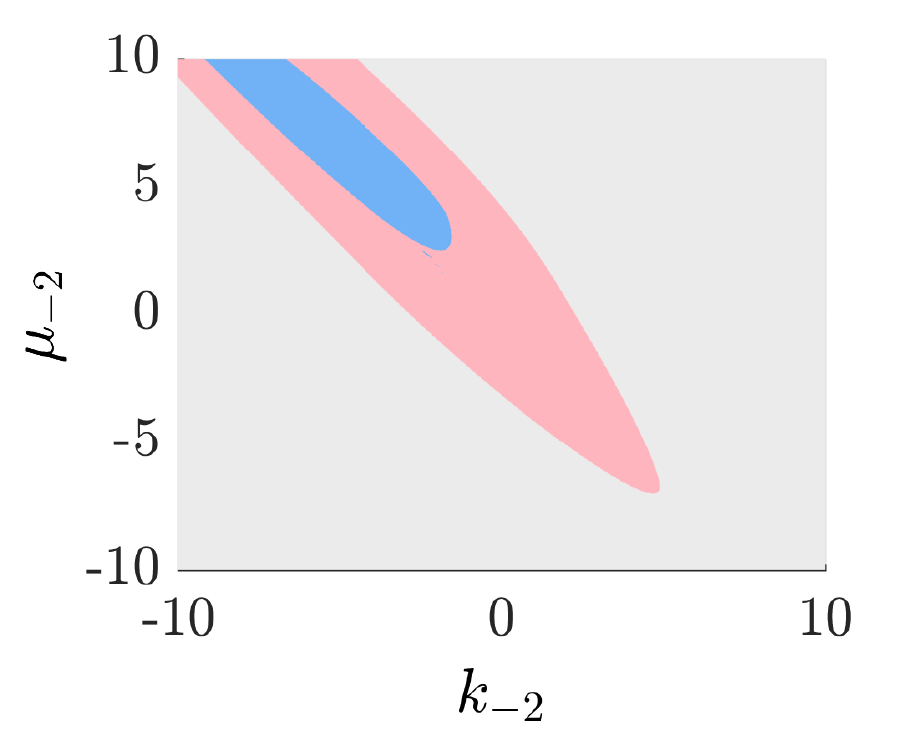}}\\
	\subfigure[]
	{\includegraphics[scale=0.5,trim=5 2 17 10,clip]{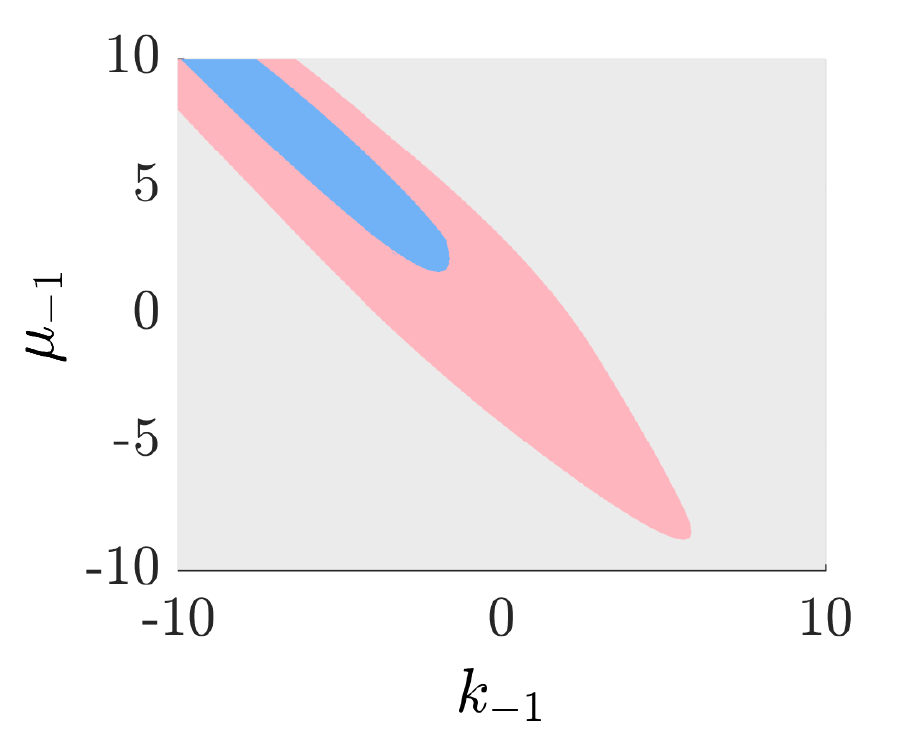}}
	\subfigure[]
	{\includegraphics[scale=0.5,trim=5 2 17 10,clip]{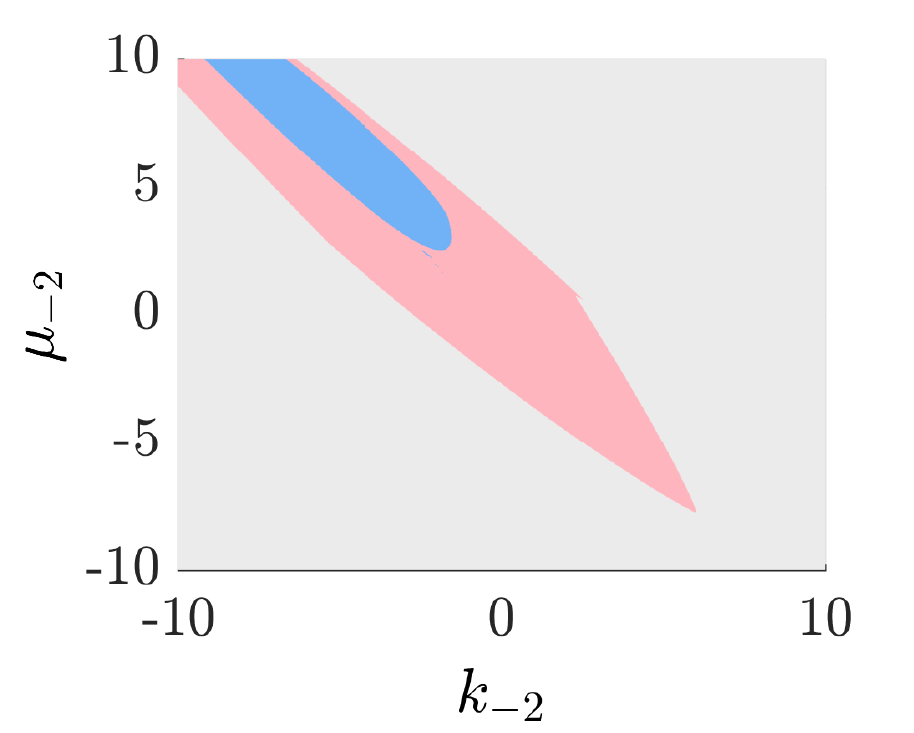}}
	\vspace{-1mm}
	\caption{``Looking ahead" head-to-tail string stable regions ($n=m=2$). (a)(c)(e) and (b)(d)(f) correspond to the feedback gains of vehicle $-1$ and vehicle $-2$ respectively. (a)(b) represent the original string stable regions (colored in blue) when monitoring one single preceding vehicle, while (c)-(f) represent the string stable regions after incorporation of one vehicle behind with the expanded region colored in red. Specifically, $\mu_1=k_1=-1$ in (c)(d), and $\mu_2=k_2=-1$ in (e)(f).} 
	\label{Fig:LookingAheadRegion}
\end{figure}

We proceed to numerically solve the specific head-to-tail string stable regions of the feedback gains in~\eqref{Eq:ControlInputDefinition}. We consider the LCC system in Figure~\ref{Fig:SystemSchematic_Ahead2Behind2} with a specific scenario $n=m=2$. The OVM model~\eqref{Eq:OVMmodel} is employed for the numerical solution of~\eqref{Eq:HeadtoTailDefinition} with a typical parameter setup~\cite{jin2017optimal,wang2020controllability}: $\alpha=0.6, \beta=0.9, v_{\max }=30, s_{\mathrm{st}}=5, s_{\mathrm{go}}=35, v^* = 15$. It is not difficult to verify that this parameter setup yields a string unstable behavior of HDVs. This indicates that when $\mu_i=k_i=0$ for $i=-2,-1,1,2$, \ie, the CAV adopts exactly the same driving strategy as HDVs, a perturbation of the head vehicle will be amplified along upstream traffic flow consisting of HDVs only.

\begin{figure}[t]
	\vspace{1mm}
	\centering
	\subfigure[]
	{\includegraphics[scale=0.5,trim=5 2 17 10,clip]{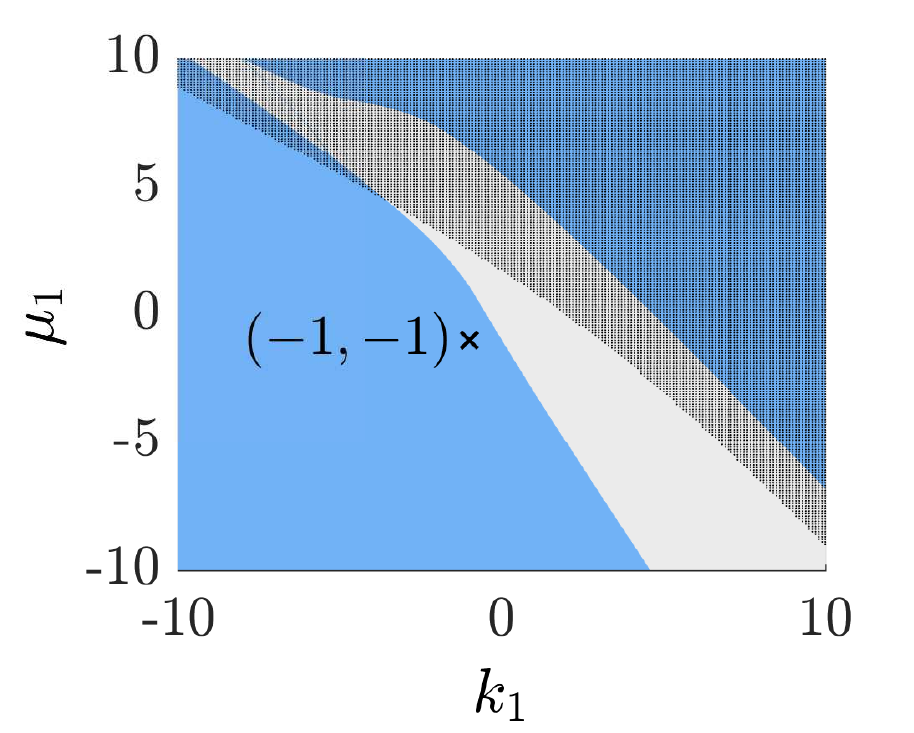}}
	\subfigure[]
	{\includegraphics[scale=0.5,trim=5 2 17 10,clip]{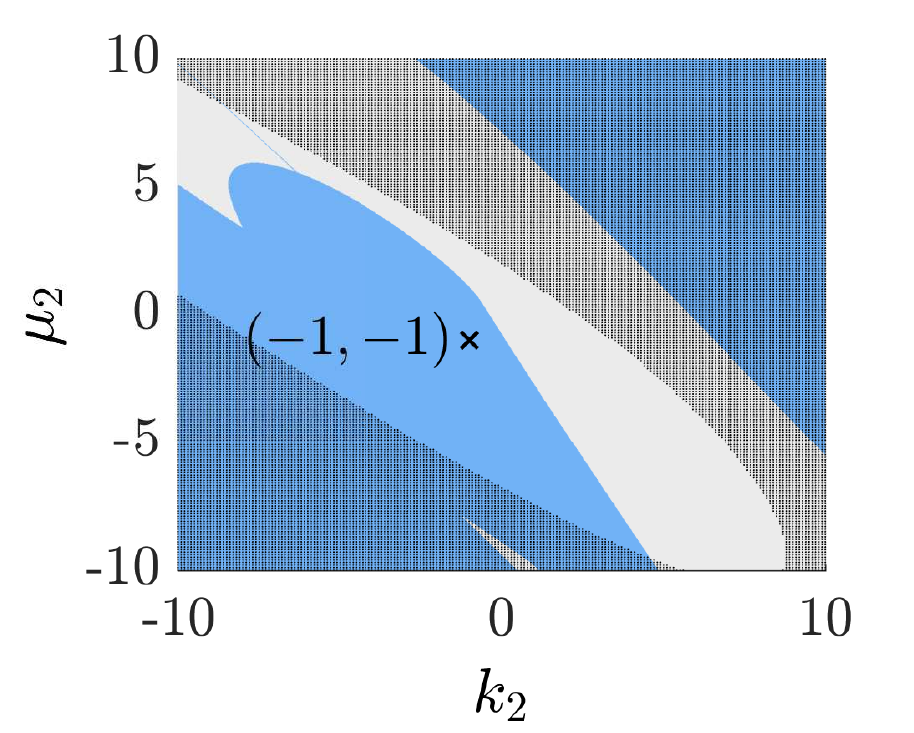}}
	\vspace{-1mm}
	\caption{``Looking behind" head-to-tail string stable regions ($n=m=2$) when the CAV monitors one following HDV. (a) and (b) correspond to the feedback gains of vehicle $1$ and vehicle $2$ respectively. The blue areas represent head-to-tail string stable regions, while the dashed areas represent asymptotically unstable regions.}
	\label{Fig:LookingBehindRegion}
\end{figure}

The first numerical study investigates the head-to-tail string stable region of the mixed traffic flow when the CAV only utilizes the information of one single HDV. The results are demonstrated in Fig.~\ref{Fig:LookingAheadRegion}(a)(b) and Fig.~\ref{Fig:LookingBehindRegion}(a)(b), where the head-to-tail string stable regions are colored blue. Comparing Fig.~\ref{Fig:LookingAheadRegion}(a)(b) and Fig.~\ref{Fig:LookingBehindRegion}(a)(b), we can clearly observe that ``looking behind" string stable areas of the feedback gains are apparently larger than ``looking ahead'' ones. This means that when the CAV monitors the motion of the vehicles behind, it has a larger feasible region in the feedback policies~\eqref{Eq:ControlInputDefinition} to dampen traffic perturbations, compared to considering the state of the vehicles ahead. 

Second, we investigate the change in the ``looking ahead" string stable regions after incorporating the motion of the vehicles behind. Precisely, we fix the feedback gains corresponding to one following vehicle to an appropriate value according to Fig.~\ref{Fig:LookingBehindRegion}, and then solve again the string stable regions of the feedback gains corresponding to one preceding vehicle. Two cases are chosen: $\mu_1=-1,k_1=-1$ or $\mu_2=-1,k_2=-1$ (highlighted in Fig.~\ref{Fig:LookingBehindRegion}), corresponding to additional consideration of vehicle $1$ or vehicle $2$, respectively. 

The new ``looking ahead" string stable results are demonstrated in Fig.~\ref{Fig:LookingAheadRegion}(c)-(f). The blue areas denote the original head-to-tail string stable regions when monitoring one single preceding vehicle, which remain the same as those in Fig. 7(a)(b). However, after the CAV considers the motion of one following vehicle for feedback control at the same time, the string stable regions of the ``looking ahead" feedback gains witness a significant expansion, as highlighted in red areas in Fig.~\ref{Fig:LookingAheadRegion}(c)-(f). This indicates that after incorporating the information of the following vehicles into CAV's controller design, the CAV has more string stable choices in the feedback gains corresponding to the preceding vehicles. This result reveals a remarkable potential of the LCC framework: combining the information from both the vehicles ahead and the vehicles behind provide  \emph{more possibilities in dampening traffic waves and smoothing traffic flow} than the traditional ``looking ahead only" strategies.

\section{Nonlinear Traffic Simulations}
\label{Sec:5}

Our theoretical results are obtained based on the linearized dynamics of the LCC systems. In this section, we present simulation results by employing nonlinear car-following models. In particular, the nonlinear OVM model~\eqref{Eq:OVMmodel} is utilized for modeling longitudinal behaviors of HDVs, where $\alpha=0.6,\beta=0.9,v_{\max }=30,s_{\mathrm{st}}=5,s_{\mathrm{go}}=35$. Two types of simulations, mitigating traffic perturbation ahead and leading the motion of HDVs behind, are conducted to validate the potential of LCC. All the simulations are carried out in MATLAB; the scripts can be downloaded from \url{https://github.com/wangjw18/LCC}.

Note that in all our nonlinear traffic simulations, we added a low-level emergency braking system to guarantee safety for the CAV, which is given by
\begin{equation} \label{Eq:AEB}
	\dot{v}_0(t)=a_{\min }, \;\mathrm { if } \, \frac{v_{0}^{2}(t)-v_{-1}^{2}(t)}{2 s_{0}(t)} \geq\left|a_{\min }\right|,
\end{equation}
where the maximum acceleration and deceleration of each vehicle are set as $a_{\max }=2 \,\mathrm{m} / \mathrm{s}^{2}$ and $a_{\min }=-5 \,\mathrm{m} / \mathrm{s}^{2}$, respectively.

\subsection{Mitigating Traffic Perturbations Ahead}
\label{Sec:SimulationAhead}

The head-to-tail string stability analysis in Section~\ref{Sec:4} has revealed that incorporating the vehicles behind contributes to larger string stable regions for CAV's feedback policies under the controller~\eqref{Eq:ControlInputDefinition}. We here demonstrate the improvement of the CAV's capability in mitigating perturbations after ``looking behind'' in both frequency domain and time domain. Consider the scenario where there are two vehicles ahead and two vehicle behind the CAV, \ie, $m=2,n=2$, as shown in Fig.~\ref{Fig:SystemSchematic_Ahead2Behind2}. The four cases in Table~\ref{Tb:FeedbackGainSetup} are under specific investigation. Note that the parameter setup in Table~\ref{Tb:FeedbackGainSetup} represents different V2V communication patterns, or different combinations of HDVs that the CAV respond to. Precisely, in Cases A or B, the CAV responds to one or two HDVs ahead, corresponding to typical ``looking ahead" only strategies, while in Cases C or D, the CAV not only responds to the two HDVs ahead, but also responds to one or two HDVs behind. The equilibrium velocity is chosen as $v^*=15$. 

\begin{figure}[t]
	\vspace{1mm}
	\centering
	\includegraphics[scale=0.4]{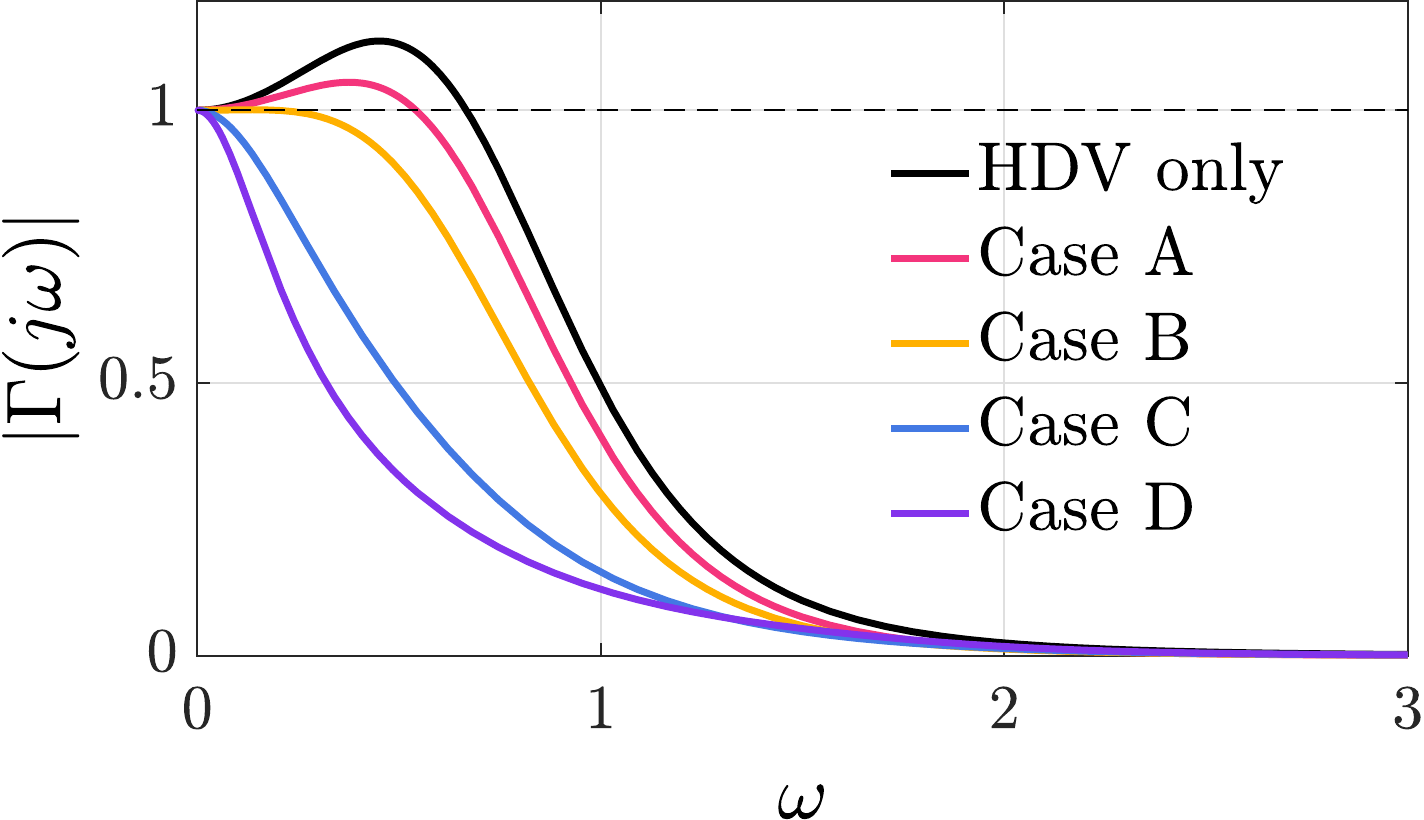}
	\vspace{-1mm}
	\caption{Frequency-domain response, \ie, magnitude of transfer function~\eqref{Eq:LCCTransferFunction}, when $n=m=2$. Parameter setups are shown in Table~\ref{Tb:FeedbackGainSetup}. The magnitude continues to decrease from HDV-only case to Case A to Case D especially at low frequencies, indicating better dissipation of perturbations in traffic flow. }
	\label{Fig:Simulation_TransferFunction}
\end{figure}

First, the numerical value of the magnitude of the transfer function~\eqref{Eq:LCCTransferFunction} of the LCC system at various excitation frequencies $\omega$, \ie, $|G(j\omega)|$ is illustrated in Fig.~\ref{Fig:Simulation_TransferFunction}. When all the vehicles are HDVs, the transfer function~\eqref{Eq:LCCTransferFunction} has magnitude larger than one among certain frequency range, indicating a string unstable performance. After an explicit consideration of surrounding HDVs in the LCC framework, the magnitude witnesses an apparent drop from Case A to Case D. In particular, after the extra incorporation of the vehicles behind (Cases C and D), the magnitude of the transfer function~\eqref{Eq:LCCTransferFunction} is smaller than those of ``looking ahead only" cases (Cases A and B), especially at low excitation frequencies, indicating {an even better attenuation of perturbations in the traffic flow.}

\begin{table}[t]
	\begin{center}
		\caption{Parameter Setup in Feedback Gains}\label{Tb:FeedbackGainSetup}
		\begin{tabular}{ccccccccc}
		\toprule
			&$\mu_{-2}$ & $k_{-2}$ & $\mu_{-1}$ & $k_{-1}$ & $\mu_{1}$ & $k_{1}$ & $\mu_{2}$ & $k_{2}$  \\\hline
			HDV only & 0 & 0 & 0  & 0 & 0 & 0 & 0 & 0\\
			Case A & 1 & -1 & 0  & 0 & 0 & 0 & 0 & 0\\
			Case B & 1 & -1 & 1  & -1 & 0 & 0 & 0 & 0\\
			Case C & 1 & -1 & 1  & -1 & -1 & -1 & 0 & 0\\
			Case D & 1 & -1 & 1  & -1 & -1 & -1 & -1 & -1\\
			\bottomrule
		\end{tabular}
	\end{center}
\end{table}

\begin{figure}[t]
	\vspace{1mm}
	\centering
	\subfigure[Case A]
	{\includegraphics[scale=0.36,height=3cm]{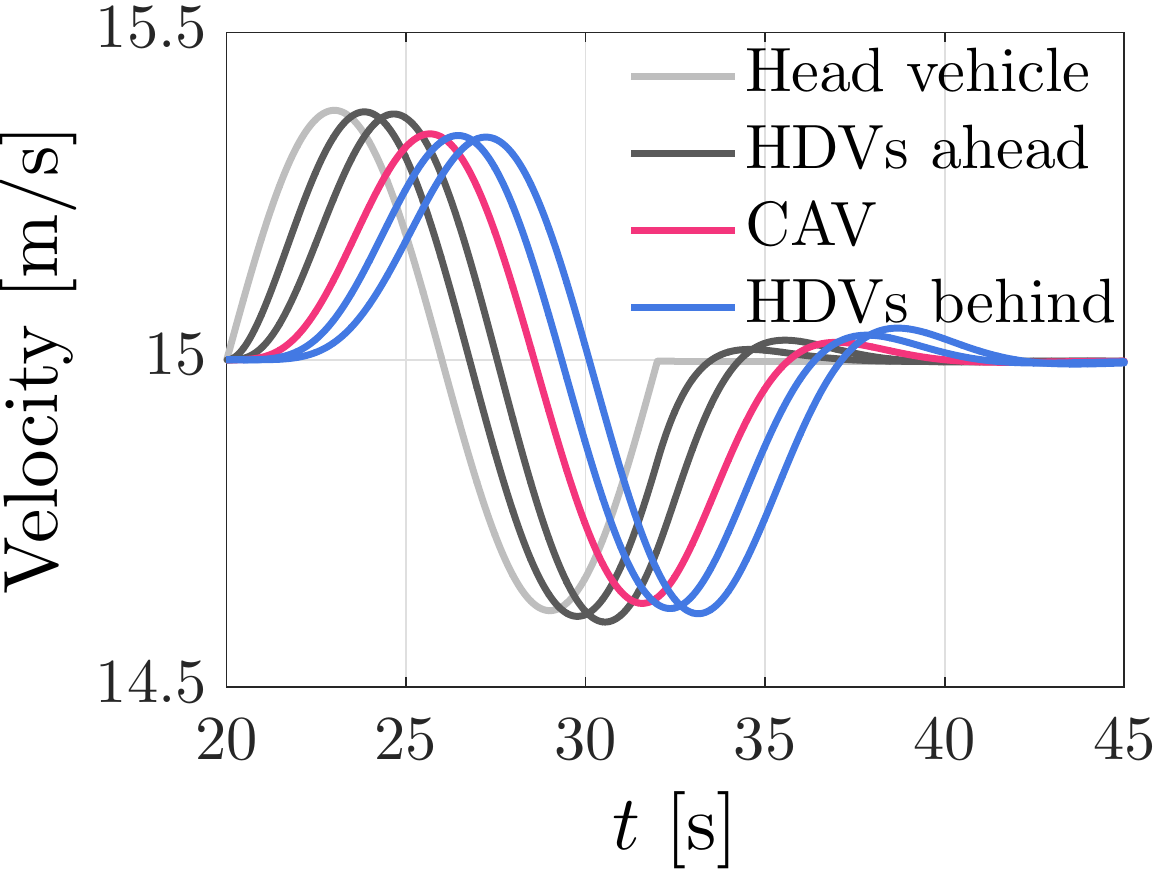}
		\label{Fig:Simulation_PerturbationAhead_1}}
	\subfigure[Case B]
	{\includegraphics[scale=0.36,height=3cm]{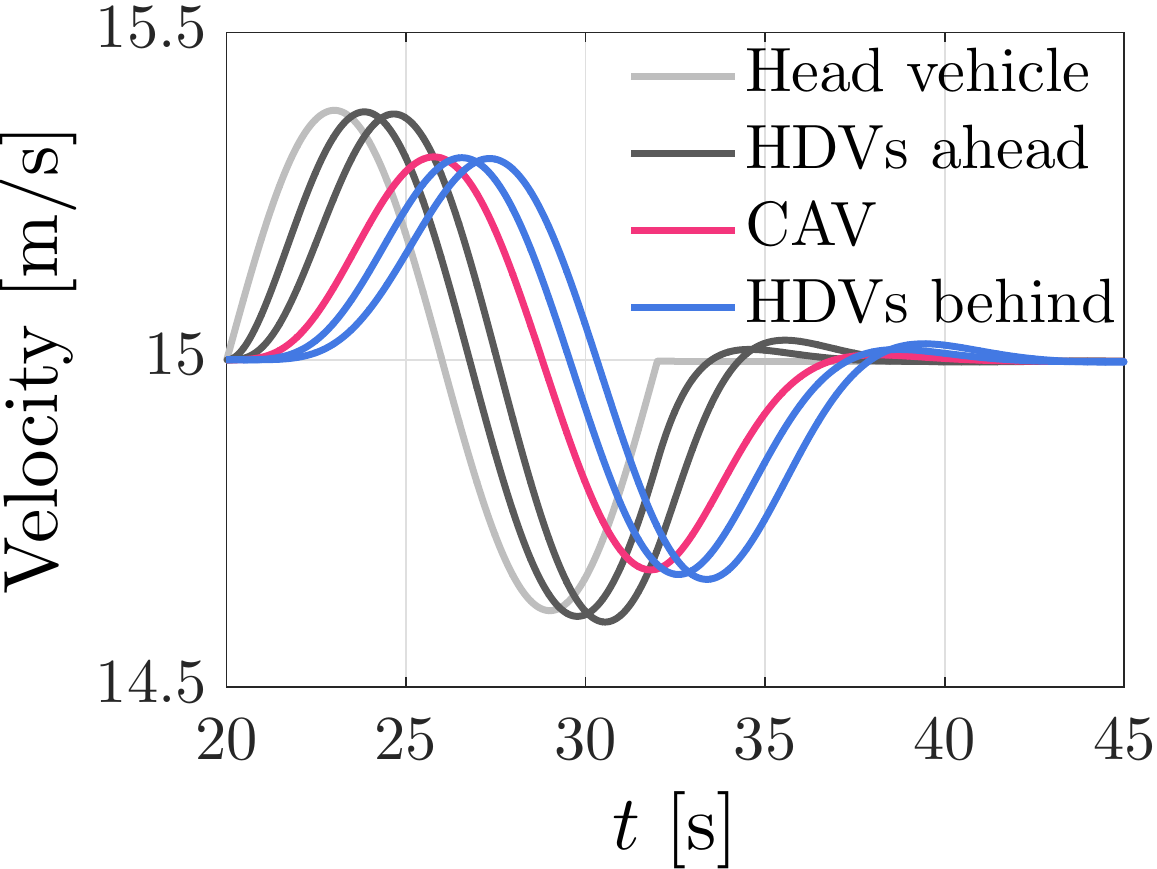}
		\label{Fig:Simulation_PerturbationAhead_2}}
	\subfigure[Case C]
	{\includegraphics[scale=0.36,height=3cm]{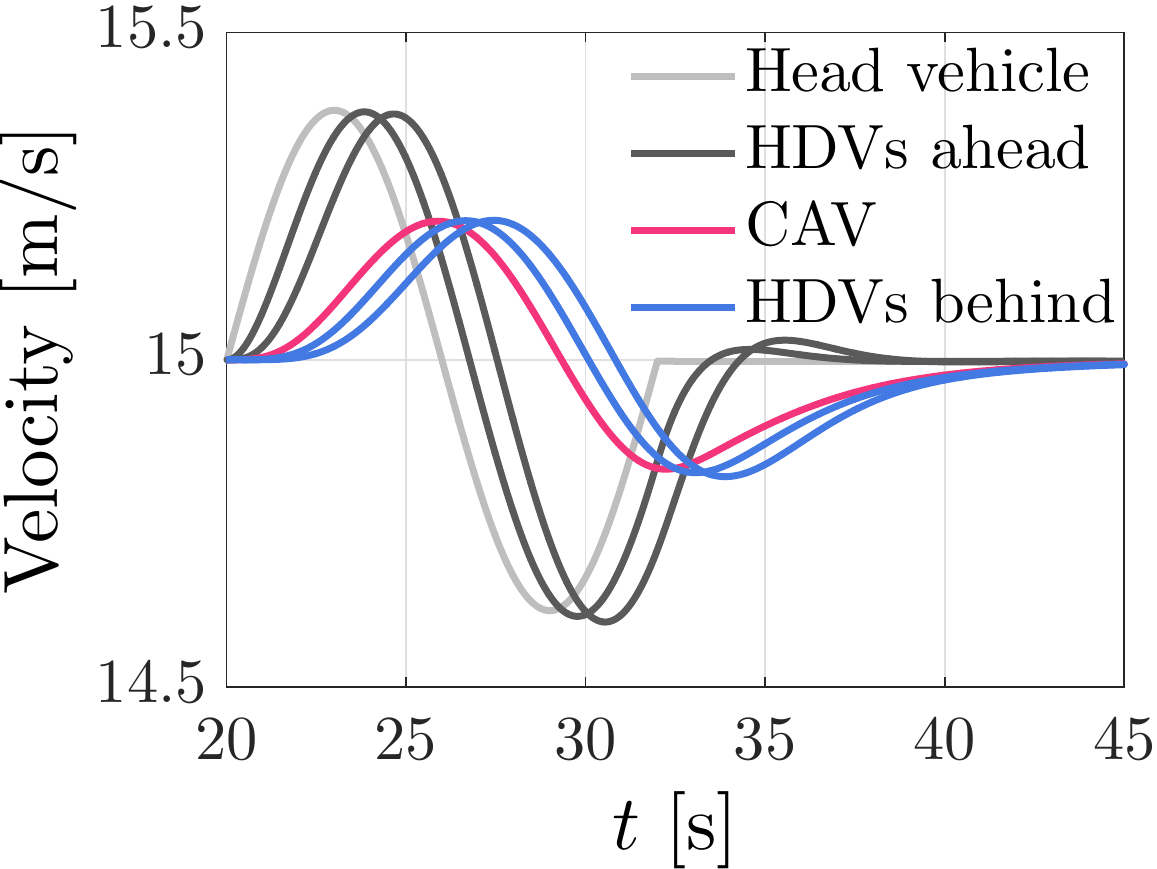}
		\label{Fig:Simulation_PerturbationAhead_3}}
	\subfigure[Case D]
	{\includegraphics[scale=0.36,height=3cm]{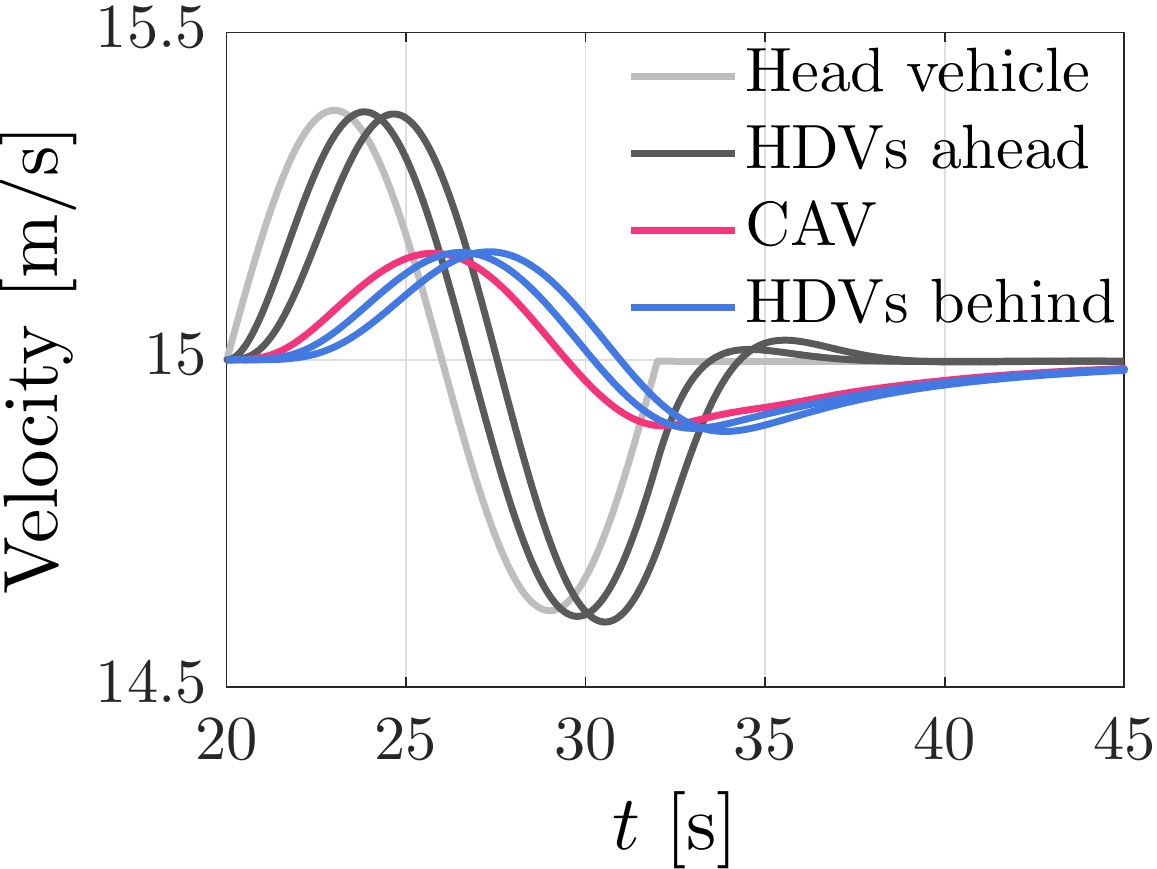}
		\label{Fig:Simulation_PerturbationAhead_4}}
	\vspace{-1mm}
	\caption{Time-domain response, \ie, velocity profile of each vehicle, in the nonlinear simulation when $n=m=2$. (a) - (d) correspond to Case A to Case D in Table~\ref{Tb:FeedbackGainSetup}. The amplitude of velocity perturbations of the CAV and two HDVs behind becomes smaller from (a) to (d), indicating that ``looking behind" appropriately improves the capability of the CAV in dampening perturbations ahead.}
	\label{Fig:Simulation_PerturbationAhead}
\end{figure}

Then, we conduct time-domain simulations utilizing the nonlinear OVM model~\eqref{Eq:OVMmodel} with four different connectivity patterns shown in Table~\ref{Tb:FeedbackGainSetup}. At the beginning of the simulation, the traffic flow is in equilibrium state with a velocity of $15\,\mathrm{ m/s}$. From $t=20\,\mathrm{ s}$, the velocity of the head vehicle is under a slight sinusoid perturbation. The velocity trajectories of all the vehicles are shown in Fig.~\ref{Fig:Simulation_PerturbationAhead}. As can be clearly observed, the amplitude of the velocity fluctuations of the following vehicles becomes smaller from Case A to Case D with more vehicles under incorporation in the LCC framework. Consequently, both the frequency-domain and the time-domain observations indicate that the degree to which the CAV mitigates the perturbation coming from ahead becomes higher after ``looking behind" appropriately compared with ``looking ahead" only.

\subsection{Leading the Motion of the Vehicles Behind}
\label{Sec:SimulationBehind}

\begin{figure*}[t]
	\vspace{1mm}
	\centering
	\subfigure{\includegraphics[scale=0.35]{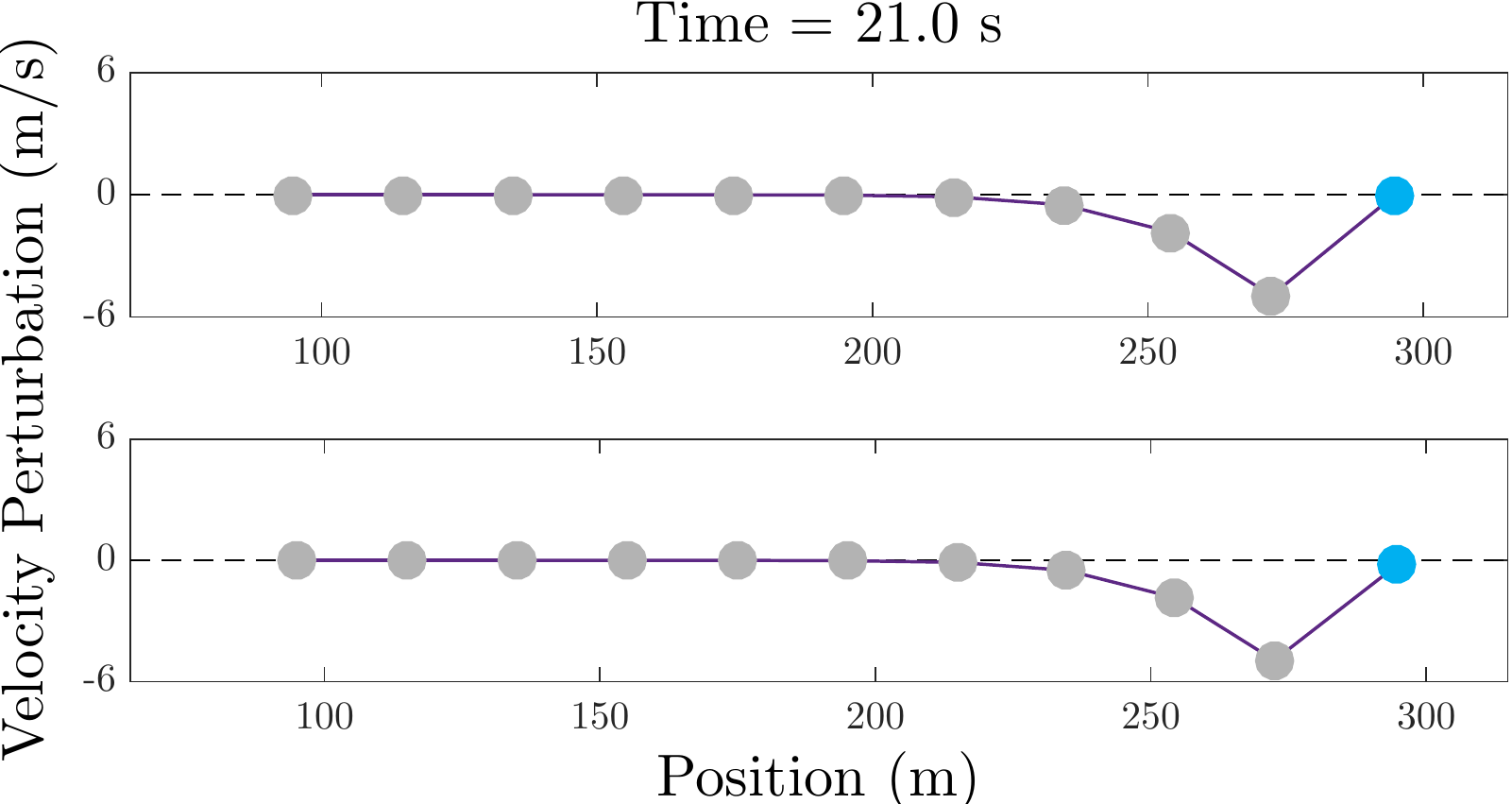}}
	\addtocounter{subfigure}{-1}
		\subfigure{\includegraphics[scale=0.35]{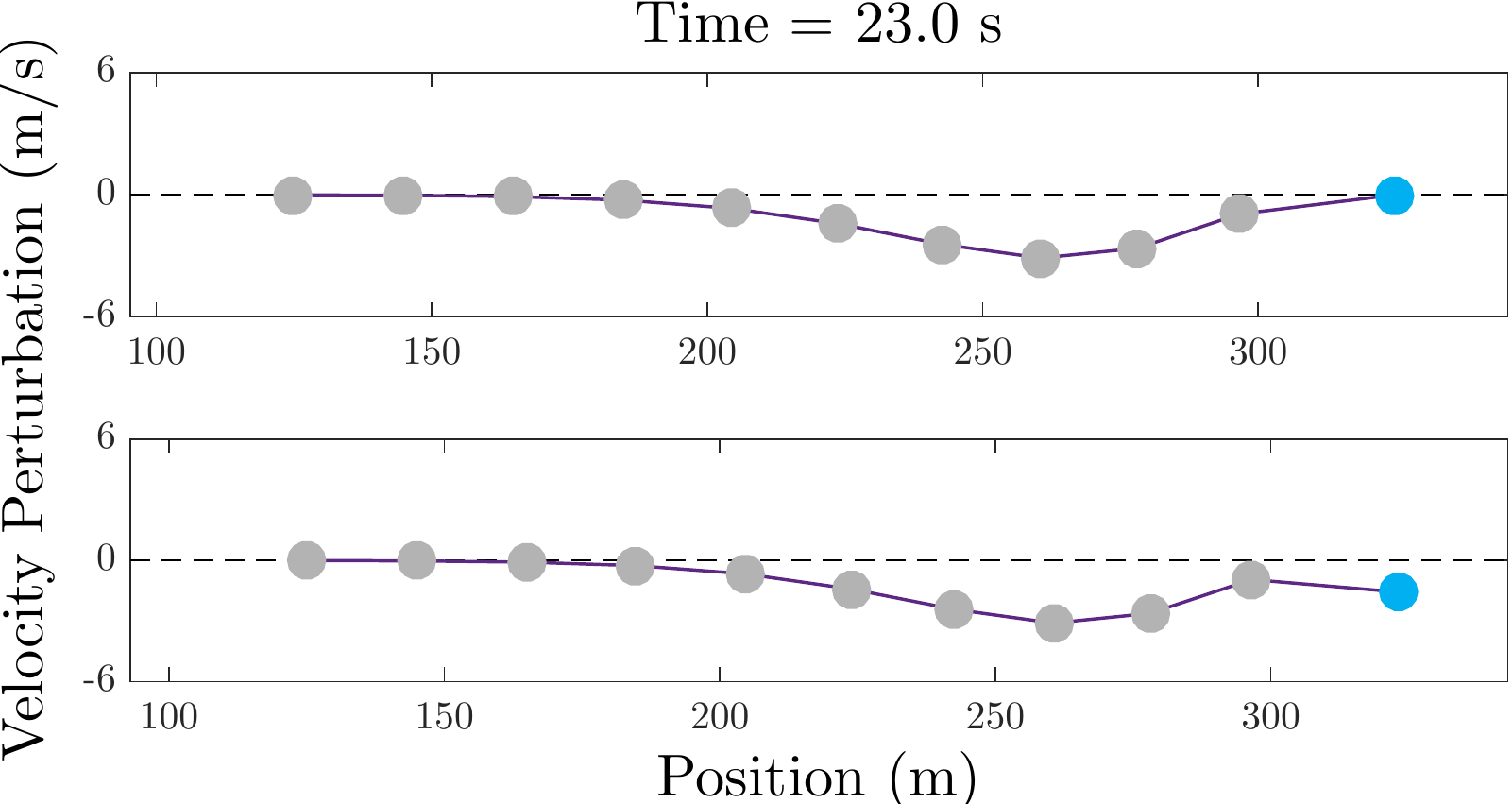}}
		\addtocounter{subfigure}{-1}
		\subfigure{\includegraphics[scale=0.35]{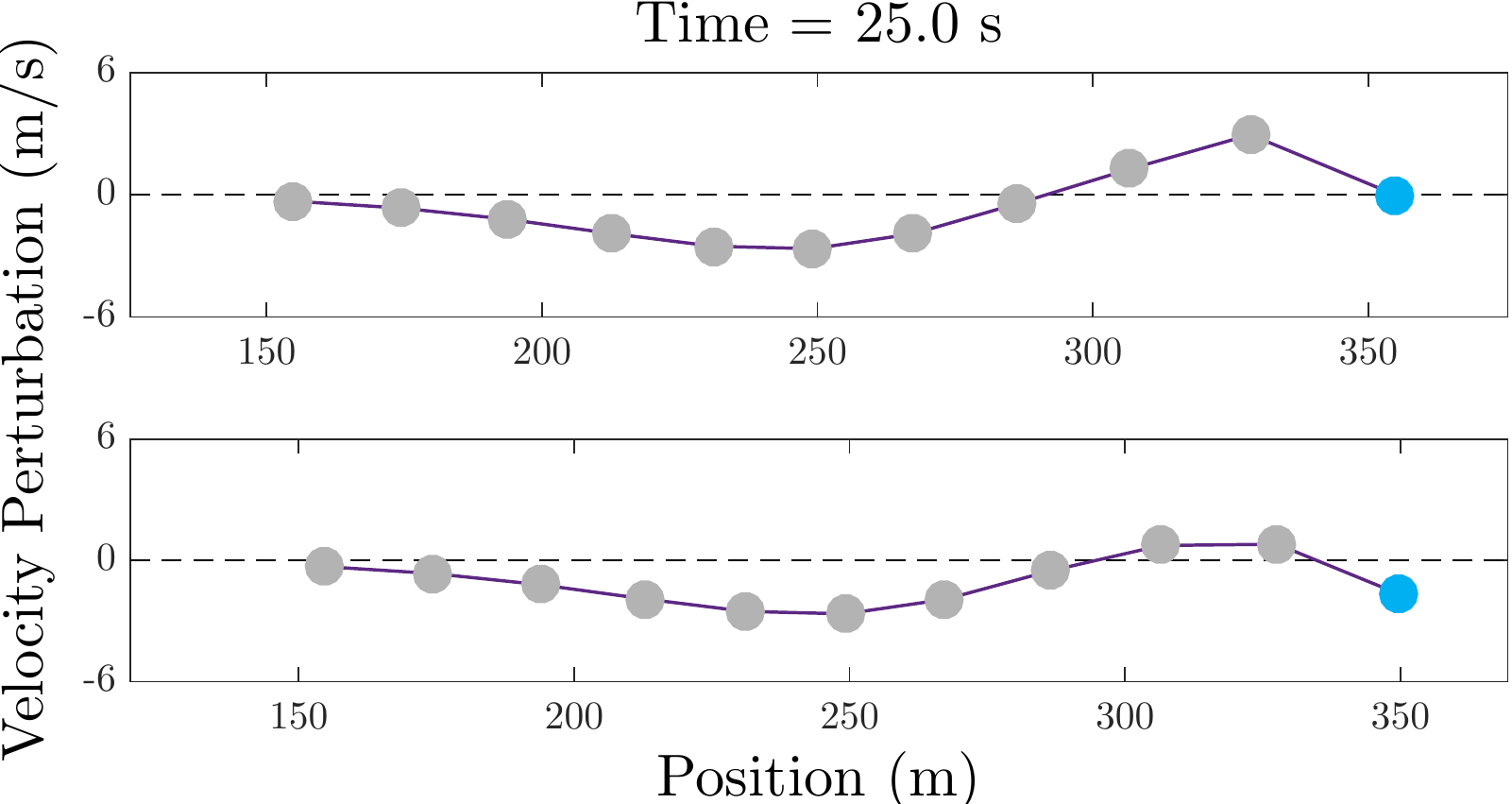}}
		\addtocounter{subfigure}{-1}
		\subfigure{\includegraphics[scale=0.35]{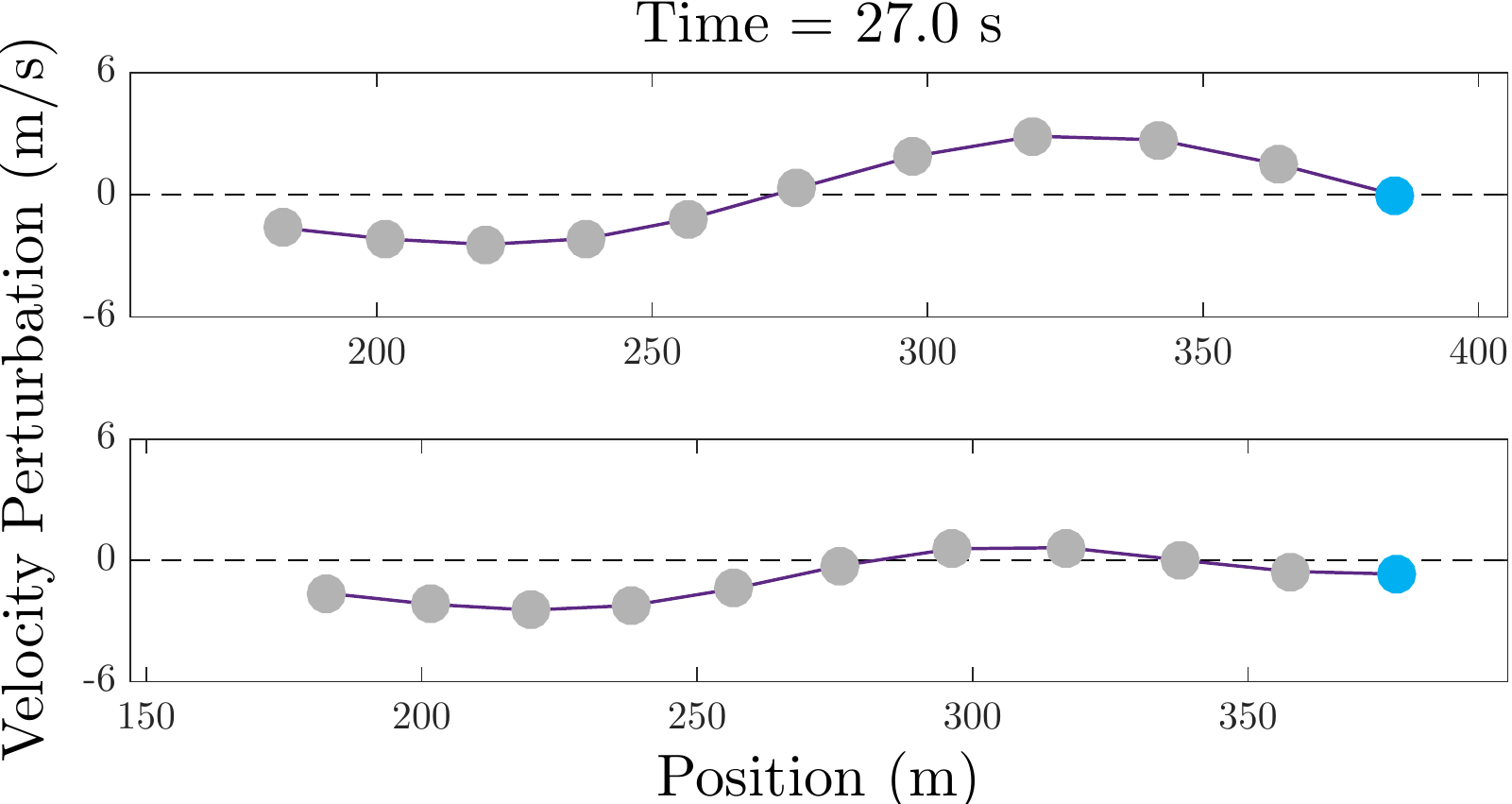}}
		\addtocounter{subfigure}{-1}
		\subfigure[FD-LCC]{\includegraphics[scale=0.35]{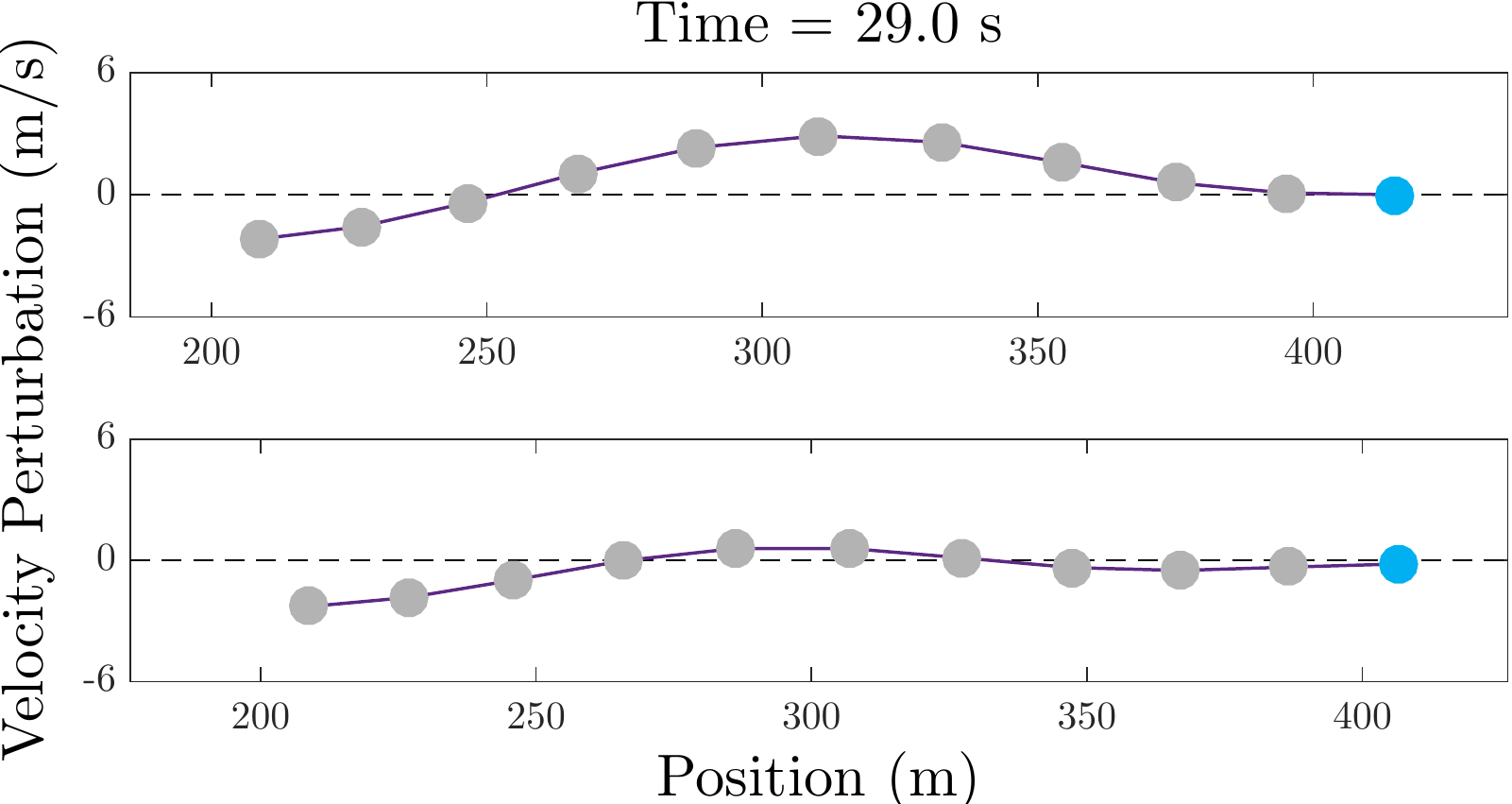}}
		\addtocounter{subfigure}{-1}
		\subfigure{\includegraphics[scale=0.35]{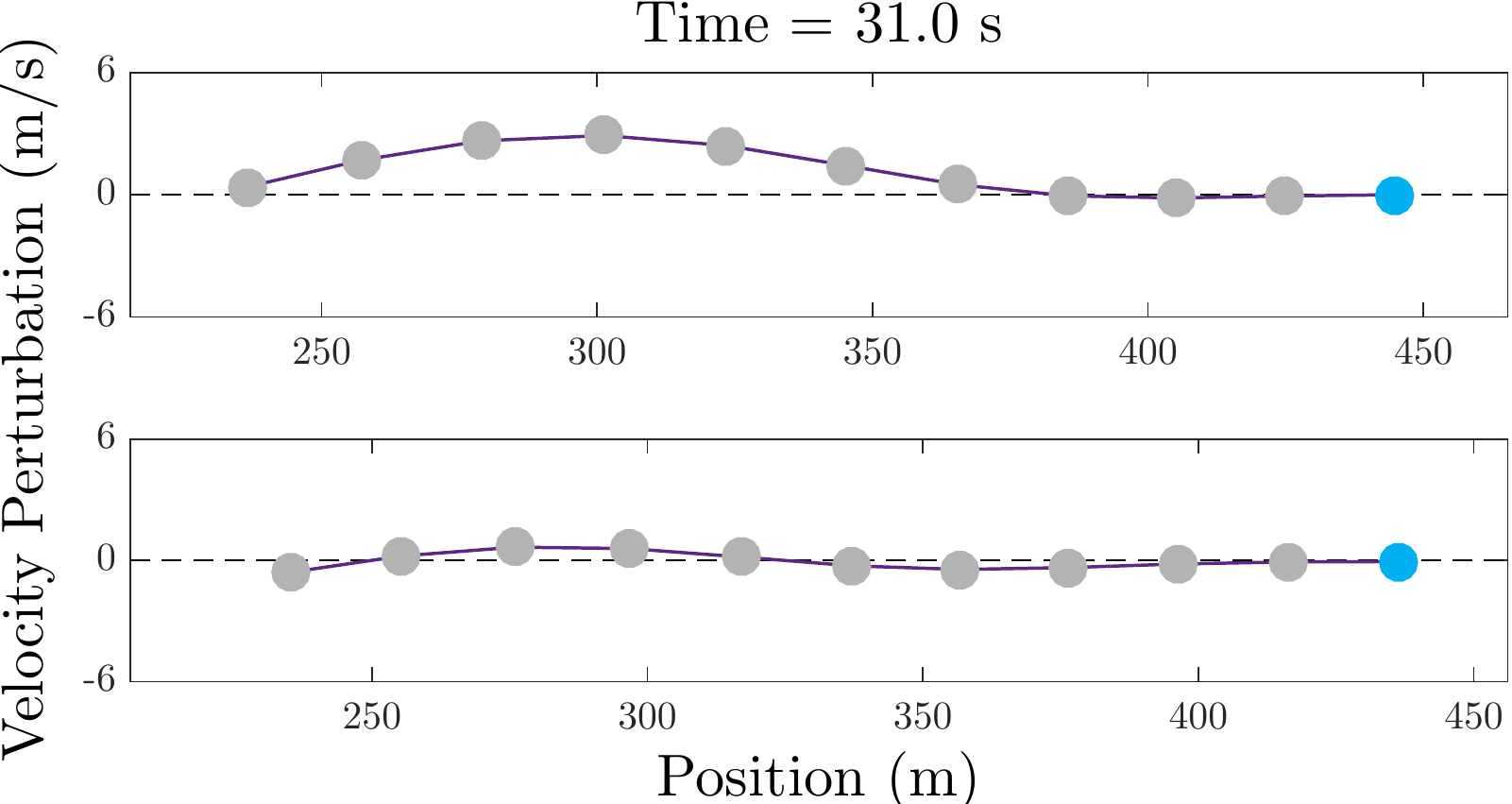}}
		\subfigure{\includegraphics[scale=0.35]{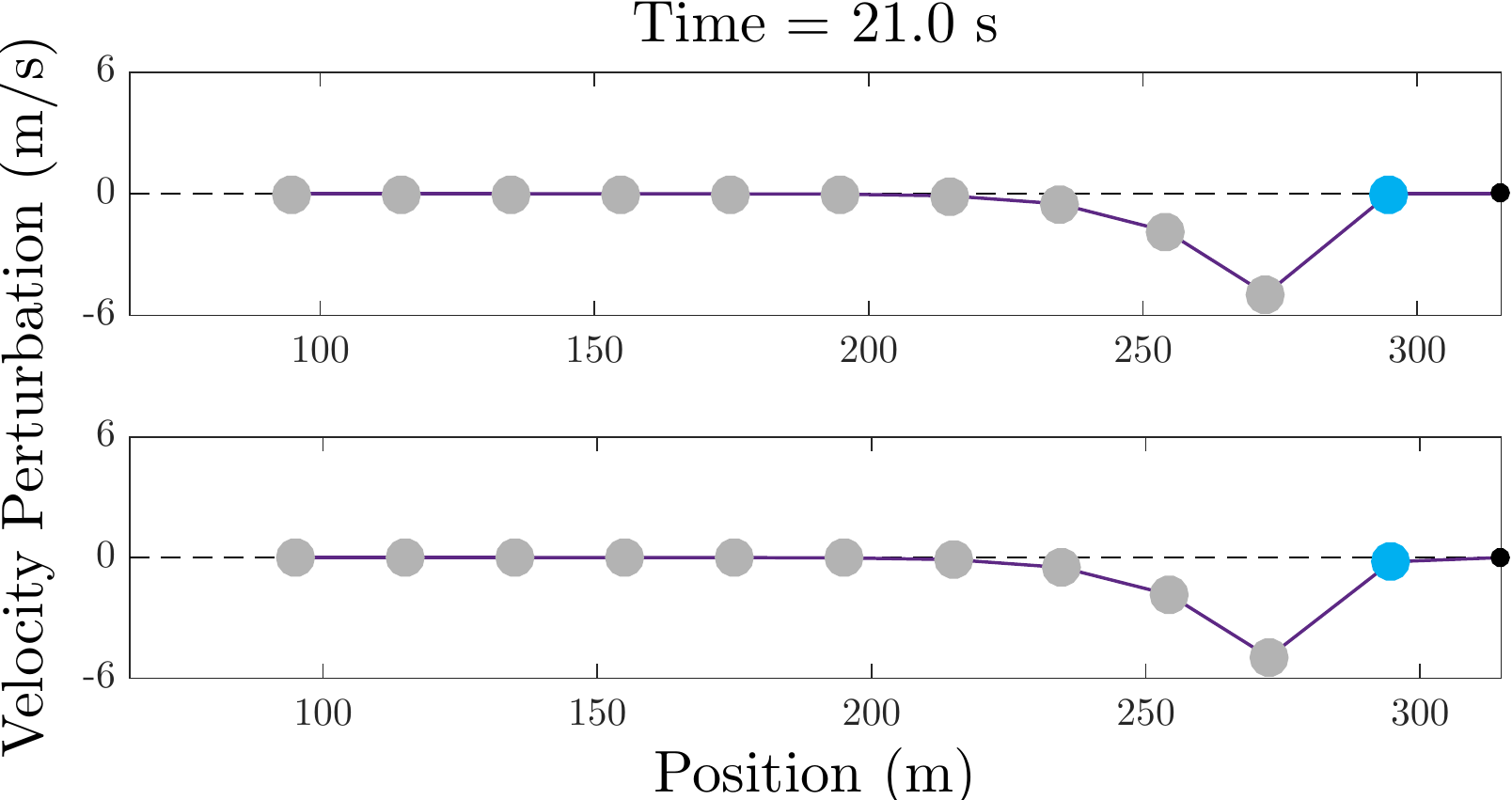}}
		\addtocounter{subfigure}{-1}
		\subfigure{\includegraphics[scale=0.35]{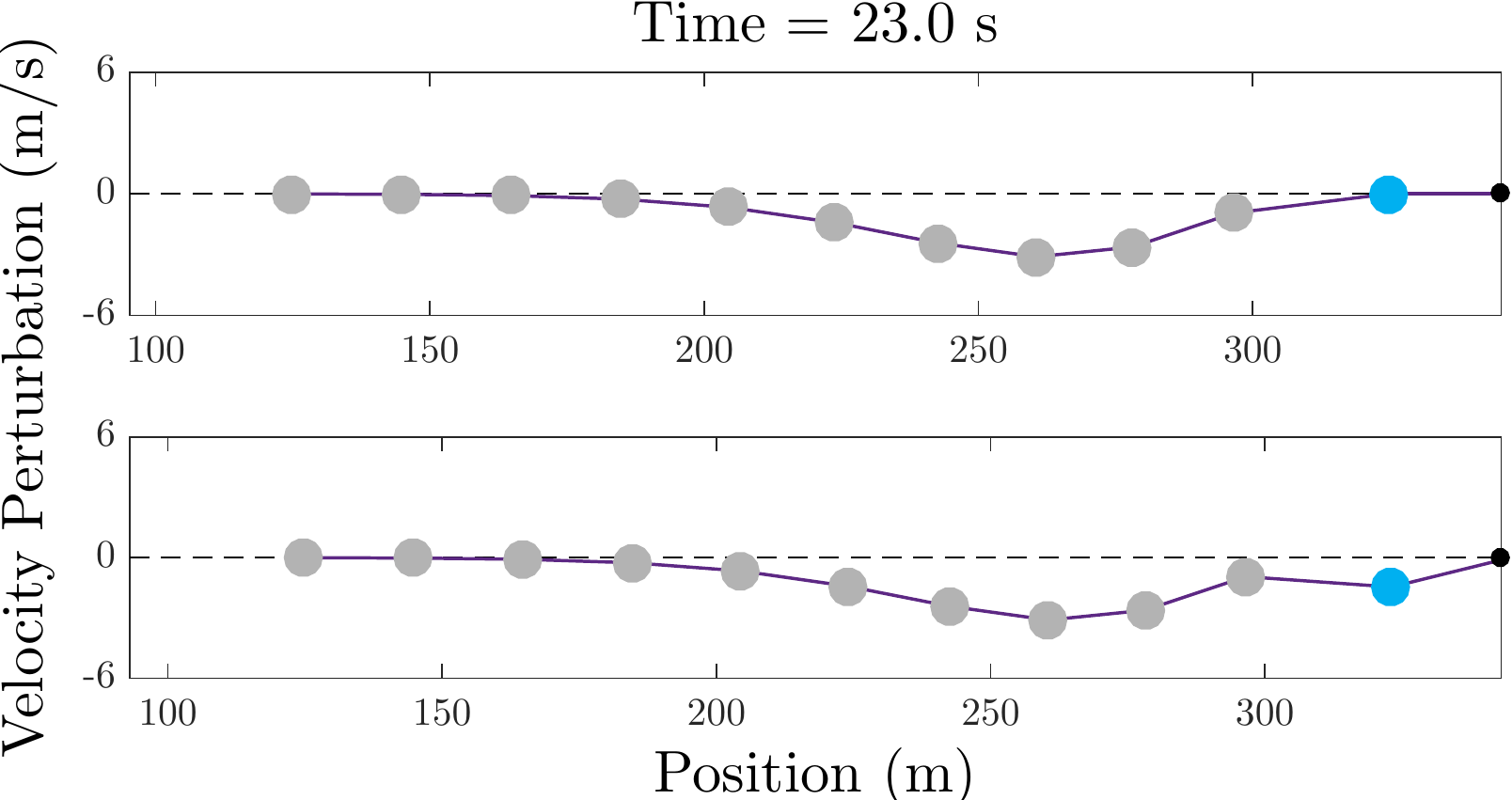}}
		\addtocounter{subfigure}{-1}
		\subfigure{\includegraphics[scale=0.35]{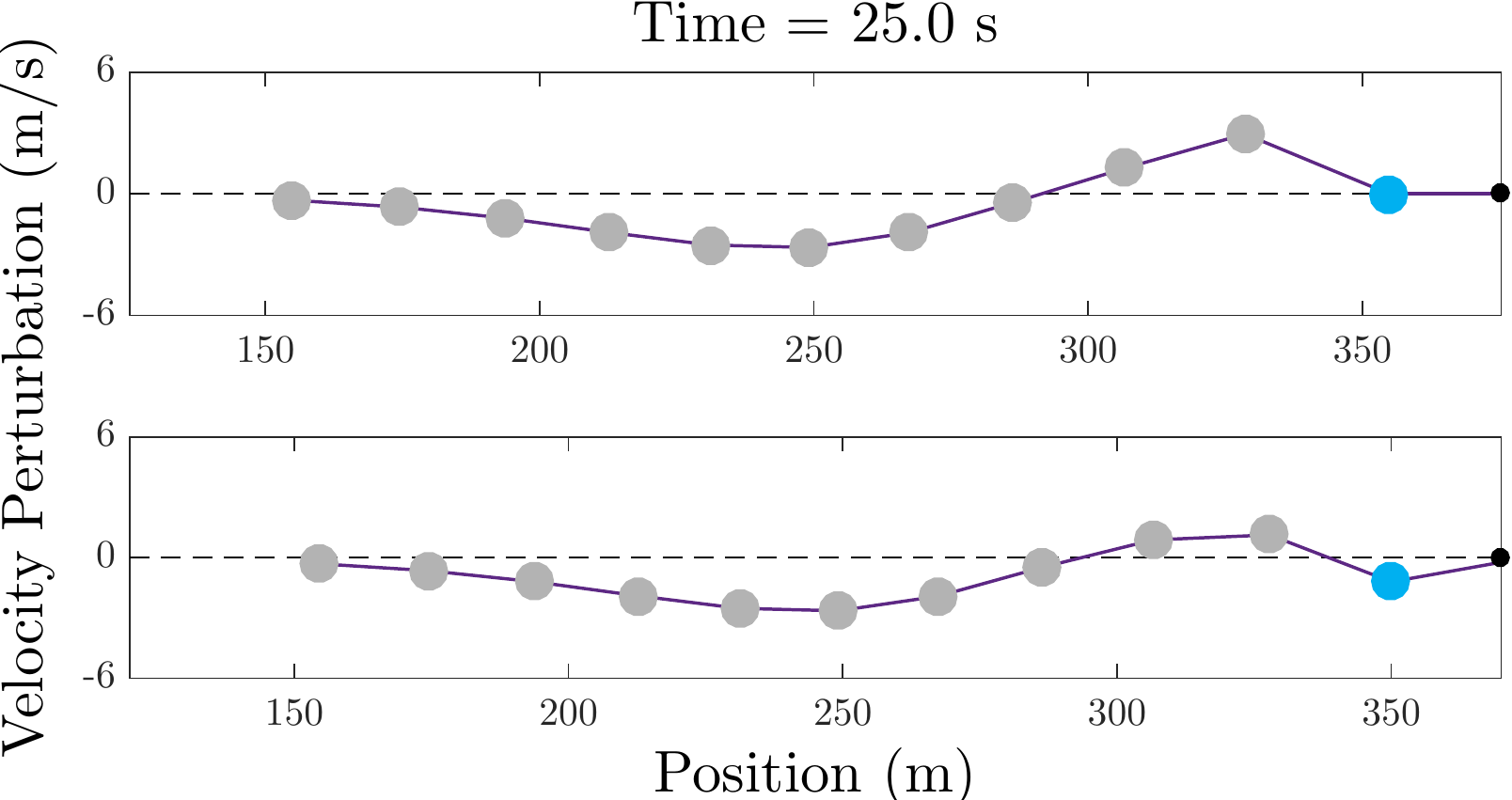}}
		\addtocounter{subfigure}{-1}
		\subfigure{\includegraphics[scale=0.35]{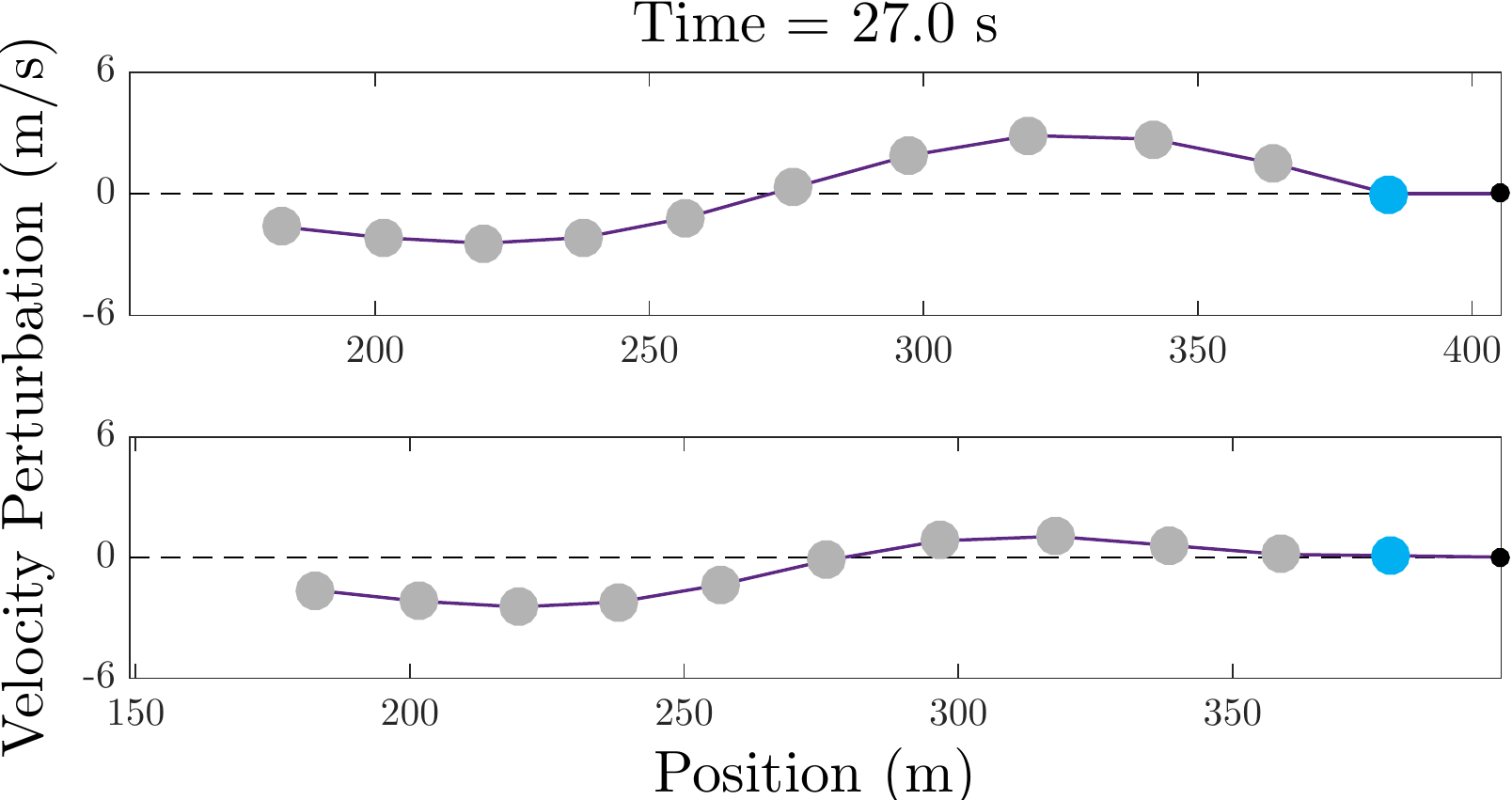}}
		\addtocounter{subfigure}{-1}
		\subfigure[CF-LCC]{\includegraphics[scale=0.35]{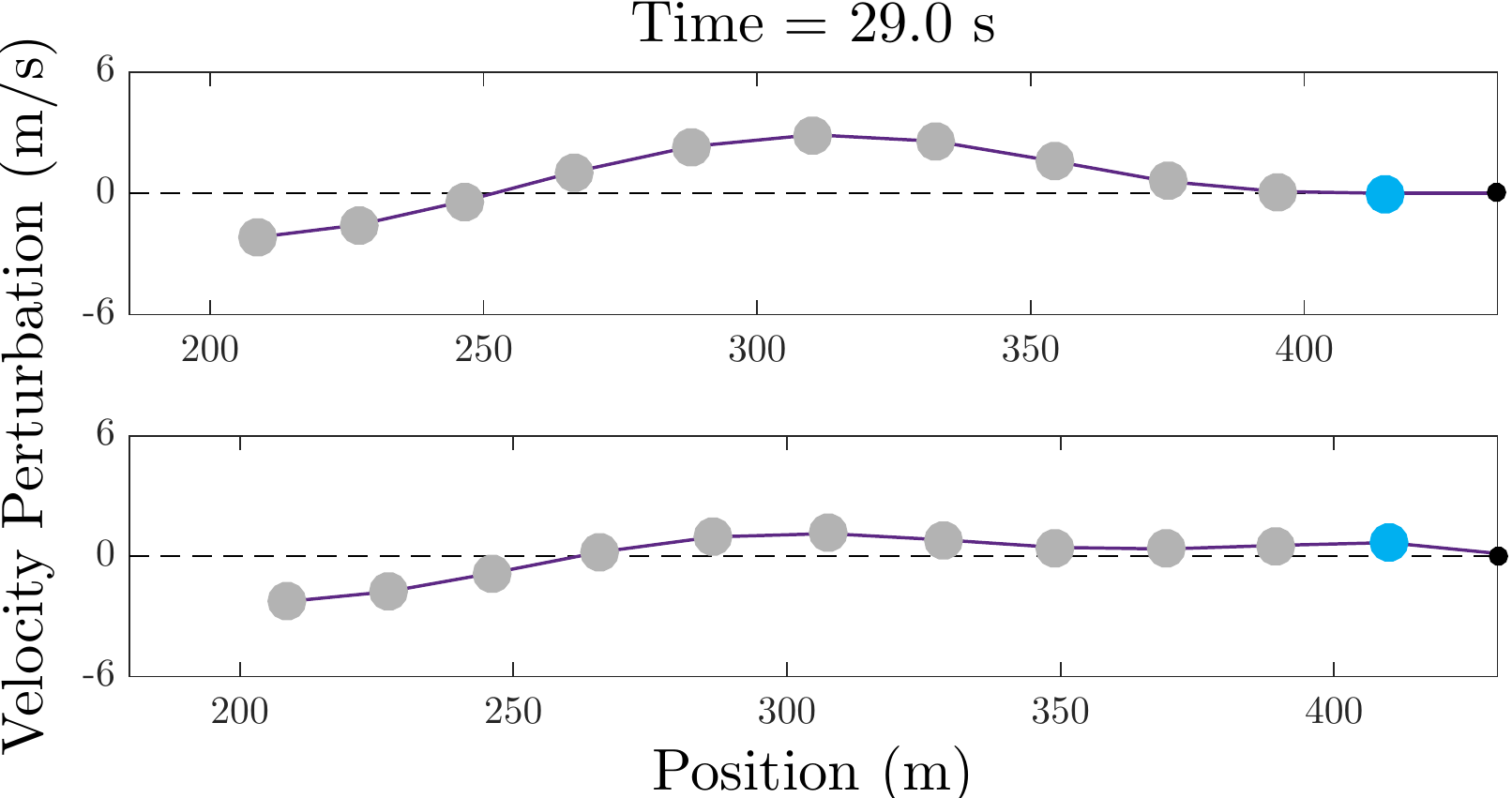}}
		\addtocounter{subfigure}{-1}
		\subfigure{\includegraphics[scale=0.35]{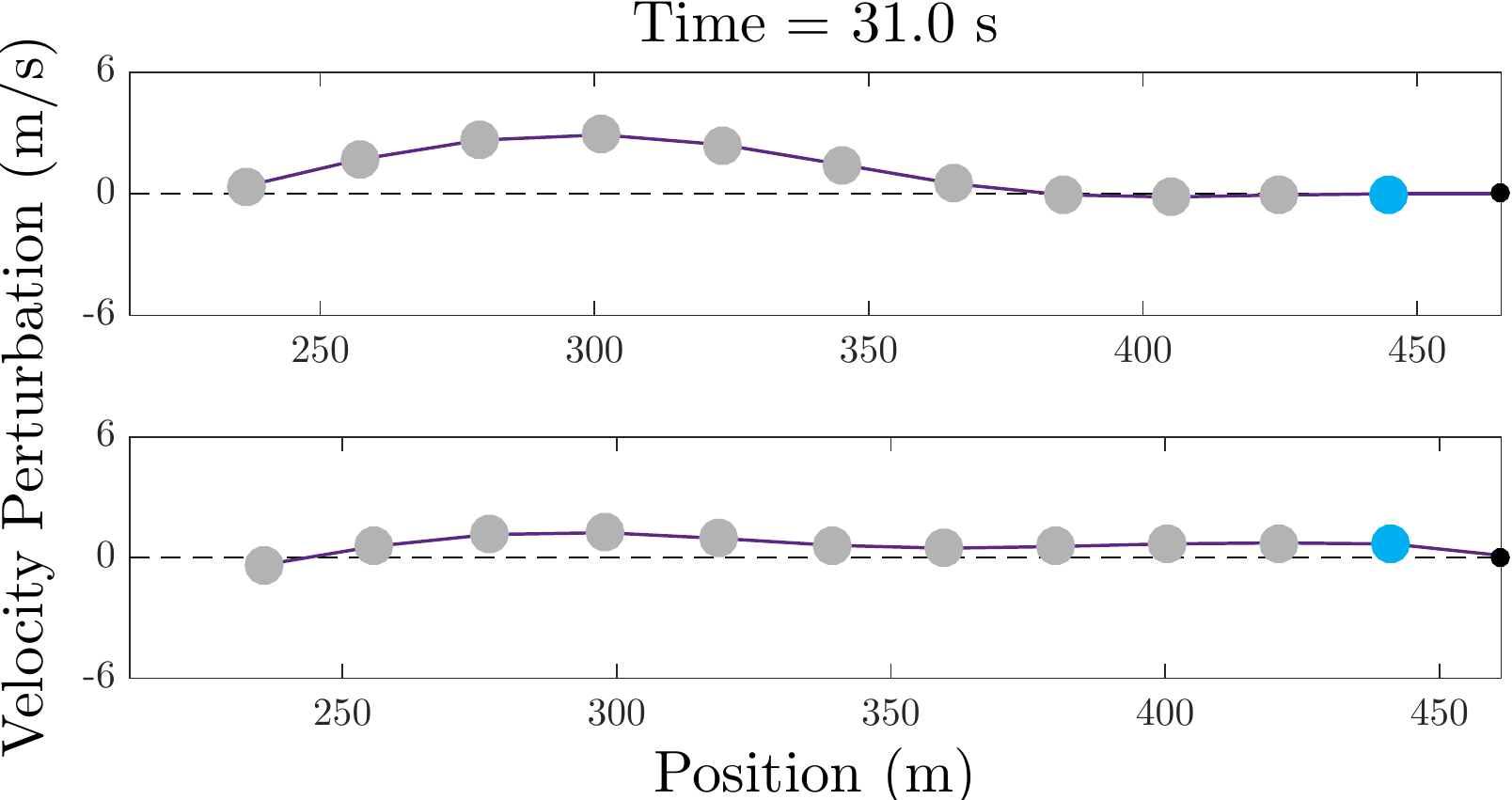}}
	\vspace{-1mm}
	\caption{Snapshots of the LCC system when a perturbation happens at the vehicle immediately behind the CAV, \ie, vehicle $1$. The gray nodes and the blue node represent the HDVs and the CAV, respectively. At each time moment, the upper panel demonstrates the result when the CAV makes no response to the perturbation behind, representing typical ``looking ahead" only strategies, while the lower panel demonstrates the result under the LCC framework. The deviation of each node from the middle dashed line represents the velocity perturbation of each vehicle from the equilibrium velocity $15 \, \mathrm{m/s}$. (a) The result of the FD-LCC system, where there is no vehicle ahead of the CAV. The control input is designed as $u(t)=-0.5 \tilde{v}_{0}(t)-0.2 \tilde{s}_{1}(t)+0.05 \tilde{v}_{1}(t)-0.1 \tilde{s}_{2}(t)+0.05 \tilde{v}_{2}(t)$. (b) The result of the CF-LCC system, where there exists one single preceding vehicle, \ie, the head vehicle, which is represented by the black dot. The control input is designed as $u(t) = 0.1 \tilde{s}_{0}(t)-0.5 \tilde{v}_{0}(t)-0.2 \tilde{s}_{1}(t)+0.05 \tilde{v}_{1}(t)-0.1 \tilde{s}_{2}(t)+0.05 \tilde{v}_{2}(t)$. A video demonstration can be found in \url{https://github.com/wangjw18/LCC}.}
	\label{Fig:Simulation_PerturbationBehind}
\end{figure*}

The previous experiment highlights the improvement of LCC in mitigating the perturbations that happen ahead. This is the prevailing perspective in many existing studies for ``looking ahead" strategies; see, \eg, \cite{jin2014dynamics,di2019cooperative}. As discussed before, LCC reveals the potential of CAVs in leading the motion of the following HDVs and improving the performance of the entire upstream traffic flow.  
Our second simulation aims to highlight this distinctive feature of LCC systems. Specifically, the two special LCC systems, FD-LCC and CF-LCC, are under consideration. We consider the scenario where there are $10$ HDVs behind the CAV, and the CAV responds to the motion of the two of them --- vehicle $1$ and vehicle $2$. We assume that the HDV immediately behind the CAV, \ie, vehicle $1$, is under a sudden strong perturbation and observe how the CAV in these two frameworks reacts to this perturbation that happens behind.

At the beginning of the simulation, all the vehicles are running at an initial equilibrium velocity of $15\,\mathrm{ m/s}$ ($v^* = 15$). At $t=20 \, \mathrm{s}$, the HDV immediately behind the CAV brakes at $-5 \, \mathrm {m/s^2}$ for one second. The snapshots of the whole system at consecutive time moments are demonstrated in Fig.~\ref{Fig:Simulation_PerturbationBehind}. As shown in the upper panel at each time moment in Fig.~\ref{Fig:Simulation_PerturbationBehind}, when the CAV makes no response to the perturbations behind, which is common in existing frameworks under ``looking ahead" only strategies, the perturbation naturally propagates upstream the traffic flow. Since there exist a finite number of HDVs behind, this traffic wave is gradually dissipated by the HDVs themselves.

In the two LCC systems, a static state feedback controller is designed respectively for the CAV with the motion of two HDVs behind under incorporation. Specifically, we design the control input for FD-LCC as $u(t)=-0.5 \tilde{v}_{0}(t)-0.2 \tilde{s}_{1}(t)+0.05 \tilde{v}_{1}(t)-0.1 \tilde{s}_{2}(t)+0.05 \tilde{v}_{2}(t)$, and the control input of CF-LCC as $u(t) = 0.1 \tilde{s}_{0}(t)-0.5 \tilde{v}_{0}(t)-0.2 \tilde{s}_{1}(t)+0.05 \tilde{v}_{1}(t)-0.1 \tilde{s}_{2}(t)+0.05 \tilde{v}_{2}(t)$. These choices are motivated by our previous results in ring-road 
mixed traffic systems in~\cite{wang2020controllability,zheng2020smoothing}. As shown in the lower panel at each time moment in Fig.~\ref{Fig:Simulation_PerturbationBehind}, in both systems, the CAV makes an active response to the perturbation behind. Precisely, when the HDV behind brakes hard, the CAV also makes a slight deceleration, letting the HDV behind easier to catch up with its own motion. Consequently, the HDV behind would not accelerate to a high velocity with a large deviation from the equilibrium (see, \eg, the moment at $25.0\,\mathrm{s}$), reducing its velocity fluctuations. Thus, the propagation of the traffic wave along the upstream traffic flow is evidently dampened.

We further introduce two performance indexes to make specific comparisons between the two LCC systems and ``looking ahead" only strategies. Precisely, an instantaneous fuel consumption model proposed in~\cite{bowyer1985guide} is utilized to calculate the fuel consumption rate $f_i$ ($\mathrm{mL/s}$) of the $i$-th vehicle, which is given by
	\begin{equation*}
	f_i = \begin{cases}
	0.444+0.090 R_i v_i + [0.054 a_i^2 v_i] _{a_i>0},& \text{if}\; R_i>0,\\
	0.444, & \text{if} \; R_i \le 0,
	\end{cases}
	\end{equation*}
where $R_i = 0.333+0.00108 v_i^2 + 1.200 a_i$ with $a_i$ denoting the acceleration of vehicle $i$. Then, we calculate the total fuel consumption (FC) of all the vehicles throughout the simulation from $t = 20 \,\mathrm{ s}$ to $t = 40 \,\mathrm{ s}$. In addition, we also introduce an index of average absolute velocity error (AAVE), which is calculated by $\sum_{0}^{10} \int_{20}^{40} | v_{i}(\tau)-v^{*}| d \tau/20/11$, to depict the degree of velocity fluctuations in the simulation. The results of the two indexes are presented in Table~\ref{Tb:Comparison}. As can be clearly observed, both the two LCC systems contribute to significant improvement in fuel consumption and traffic smoothness for the entire upstream traffic flow, compared with traditional ``looking ahead" only strategies, which typically make no response to the traffic perturbations behind.

\begin{table}[t]
	\begin{center}
		\caption{Comparison of Performance Indexes}\label{Tb:Comparison}
		\begin{tabular}{cccc} \toprule
			&``Looking ahead"  & FD-LCC & CF-LCC \\ \hline
			AAVE ($\mathrm{m/s}$) & $ 0.89$  & $0.58$ & $0.81$  \\
			Reduction rate of AAVE &  / & $34.97\%$ & $8.95\%$  \\ 
			FC ($\mathrm{mL}$) & $392.86$ & $321.94$ & $340.56$  \\
			Reduction rate of FC & / & $18.05\%$ & $13.31\%$  \\ 	\bottomrule
		\end{tabular}
	\end{center}
\end{table}

This result validates the effectiveness of LCC in incorporating the vehicles behind. Since the CAV could actively lead the motion of the vehicles behind and improve the performance of the upstream traffic flow, LCC further improves the CAV's capability in dampening traffic waves and smoothing traffic flow, compared with traditional ``looking ahead" only strategies. Note that in CF-LCC, the CAV also catches up with the head vehicle. Thus, CF-LCC allows the CAV to maintain the same equilibrium spacing from the preceding vehicle. 

\balance

\section{Conclusions}

\label{Sec:6}
In this paper, we have introduced the notion of Leading Cruise Control (LCC) for CAVs in mixed traffic flow. Based on the dynamical model of the general LCC system and three special cases, we have proved the controllability of the states of the HDVs behind through the CAV's active action, and the observability of the states of the HDVs ahead through direct measurement of the state of one individual vehicle. Moreover, we investigate the head-to-tail string stability of the proposed framework. 
These results reveal great potential of incorporating the vehicles behind into CAV's control. Our LCC frameworks take full advantage of V2V connectivity and~vehicular automation, bringing further improvements to mixed traffic systems where CAVs and HDVs~coexist.

This paper has mainly focused on establishing the concept of LCC and investigating its fundamental properties. There are indeed many exciting future directions. First, note that incorporating the information of upstream traffic might decrease the original weight for maintaining the spacing of the CAV from the preceding vehicle, and thus it is an important topic to design constrained control strategies for the CAV in LCC systems based on measurable output data to achieve optimal and safe control, as widely discussed in ACC~\cite{li2010model} or vehicle platooning~\cite{zheng2016distributed}. Considering that in real traffic flow HDVs might have heterogeneous car-following dynamics and reaction time, analyzing the influence of these practical factors on the LCC performance is worth further investigation. Some results has been discussed in CCC-related works~\cite{jin2017optimal,orosz2016connected}. In addition, extending the LCC frameworks to the scenarios with cooperation of multiple CAVs in mixed traffic flow deserves further investigation. Finally, validating the potential of LCC via large-scale traffic simulations or real-world experiments is another interesting research direction.



\ifCLASSOPTIONcaptionsoff
  \newpage
\fi



%

\bibliographystyle{IEEEtran}
\bibliography{IEEEabrv,mybibfile}

\appendices

\section*{Appendices}

In this appendix, we provide some auxiliary results to the main text, including derivation of the head-to-tail transfer function of LCC systems, spacing results of the nonlinear simulations in Section~\ref{Sec:SimulationAhead}, and an additional experiment with consideration heterogeneous HDV dynamics and reaction time, corresponding to the experiment in Section~\ref{Sec:SimulationBehind}.

\addcontentsline{toc}{section}{Appendices}
\renewcommand{\thesubsection}{\Alph{subsection}}

\subsection{Derivation of Head-to-Tail Transfer Function~\eqref{Eq:LCCTransferFunction}}

The specific derivation process of the head-to-tail transfer function~\eqref{Eq:LCCTransferFunction} is presented as follows. In the time domain, we have $\dot{\tilde{s}}_i(t)=\tilde{v}_{i-1}(t)-\tilde{v}_i(t)$ for each vehicle ($i \in \{0\} \cup \mathcal{F} \cup \mathcal{P}$). Recall that the definition of the control input of the CAV is given by~\eqref{Eq:ControlInputDefinition} and we have $u(t)=\dot{\tilde{v}}_0(t)$.
Thus, we obtain
\begin{equation}
		\widetilde{S}_i (s)= \frac{\widetilde{V}_{i-1} (s)-\widetilde{V}_{i} (s)}{s}, \; \forall i \in \{0\} \cup \mathcal{F} \cup \mathcal{P}, \label{Eq:FrequencySpacing} 
\end{equation}
and~\eqref{Eq:FrequencyInput} in the frequency domain after Laplace transform.
\begin{figure*}[!t]
\normalsize
\setcounter{equation}{33}
\begin{equation}
		s\widetilde{V}_0(s)= \alpha _{1}\widetilde{S}_{0}(s)- \alpha _{2}\widetilde{V}_{0}(s)+ \alpha _{3}\widetilde{V}_{-1}(s)
	+  \sum _{i\in \mathcal{F} \cup \mathcal{P}} \left(  \mu _{i}\widetilde{S}_{i} \left( s \right) +k_{i}\widetilde{V}_{i} \left( s \right)  \right). \label{Eq:FrequencyInput}
\end{equation}
\begin{equation} \label{Eq:TransferFunction_1}
	(s^2 + \alpha _{2} s + \alpha _{1}) \widetilde{V}_0 = (\alpha _{3} s + \alpha _{1}) \widetilde{V}_{-1} +   \sum _{i\in \mathcal{F} \cup \mathcal{P}} \left(  \mu _{i}\widetilde{V}_{i-1} +(k_{i}s - \mu_i) \widetilde{V}_{i}  \right).
\end{equation}
\setcounter{equation}{39}
\begin{equation}
\label{Eq:TransferFunction_2}
	\gamma(s) \left(\frac{ \varphi  \left( s \right) }{ \gamma  \left( s \right) }\right)^{-n} \widetilde{V}_{\mathrm{t}} = \varphi(s)  \left(\frac{ \varphi  \left( s \right) }{ \gamma  \left( s \right) }\right)^{m} \widetilde{V}_{\mathrm{h}} +   \sum _{i\in \mathcal{F} }  H_{i} \left( s \right)  \left(\frac{ \varphi  \left( s \right) }{ \gamma  \left( s \right) }\right)^{i-n} \widetilde{V}_{\mathrm{t}} + \sum _{i\in \mathcal{P} }  H_{i} \left( s \right)  \left(\frac{ \varphi  \left( s \right) }{ \gamma  \left( s \right) }\right)^{i+1+m} \widetilde{V}_{\mathrm{h}} .
\end{equation}
\hrulefill
\vspace*{4pt}
\end{figure*}

\setcounter{equation}{35}

For simplicity, we use $\widetilde{V}_i,\widetilde{S}_i$ to denote $\widetilde{V}_i(s),\widetilde{S}_i(s)$ in the following derivation. Substituting \eqref{Eq:FrequencySpacing} into \eqref{Eq:FrequencyInput} yields~\eqref{Eq:TransferFunction_1}.
Note that for HDVs, we have already obtained
\begin{equation} \label{Eq:HDVTransferFunction}
	 \frac{\widetilde{V}_{i}}{\widetilde{V}_{i-1}  }=\frac{ \alpha _{3}s+ \alpha _{1}}{s^{2}+ \alpha _{2}s+ \alpha _{1}}=\frac{ \varphi  \left( s \right) }{ \gamma  \left( s \right) }, \; i \in \mathcal{F} \cup \mathcal{P}.
\end{equation}
Substituting \eqref{Eq:HDVTransferFunction} into \eqref{Eq:TransferFunction_1} leads to
\begin{equation} \label{Eq:TransferFunction_2}
	\gamma(s) \widetilde{V}_0 = \varphi(s) \widetilde{V}_{-1} +   \sum _{i\in \mathcal{F} \cup \mathcal{P}}  H_{i} \left( s \right)  \widetilde{V}_{i} .
\end{equation}
where
\begin{equation*}
	H_{i} \left( s \right) = \mu _{i}\left(\frac{\gamma (  s )}{\varphi  ( s )} - 1\right)+k_{i}s,\,i \in \mathcal{F} \cup \mathcal{P}.
\end{equation*}
In addition, for the velocity error signals of the following HDVs and the CAV, their relationship with that of the vehicle at the tail can be given by (recall that the vehicle at the tail is indexed as vehicle $n$)
\begin{equation} \label{Eq:FollowingVehiclePerturbation}
	\frac{\widetilde{V}_{\mathrm{t}}}{\widetilde{V}_{i}} = \left(\frac{ \varphi  \left( s \right) }{ \gamma  \left( s \right) }\right)^{n-i}, \; i \in \{0\} \cup \mathcal{F} =\{0,1,2,\ldots,n\},
\end{equation}
while for velocity error signals of the preceding HDVs, their relationship with that of the vehicle at the head can be given by (recall that the vehicle at the head is the one ahead of vehicle $-m$)
\begin{equation} \label{Eq:PrecedingVehiclePerturbation}
	\frac{\widetilde{V}_{i}}{\widetilde{V}_{\mathrm{h}}} = \left(\frac{ \varphi  \left( s \right) }{ \gamma  \left( s \right) }\right)^{m+1+i}, \; i \in \mathcal{P}=\{-1,-2,\ldots,-m\}.
\end{equation}
Substituting \eqref{Eq:FollowingVehiclePerturbation} and \eqref{Eq:PrecedingVehiclePerturbation} into \eqref{Eq:TransferFunction_1} yields~\eqref{Eq:TransferFunction_2}.

Finally, after reorganizing the expression in~\eqref{Eq:TransferFunction_2}, we obtain the head-to-tail transfer function, which is given by
\setcounter{equation}{40}
\begin{equation}
	\Gamma (s) = \frac{\widetilde{V}_\mathrm{t} (s)  }{\widetilde{V}_\mathrm{h} (s) } = G(s) \cdot \left( \frac{ \varphi  \left( s \right) }{ \gamma  \left( s \right) } \right) ^{n+m},
\end{equation}
where
\begin{equation*}
	G(s)=\frac{ \varphi  \left( s \right) + \sum _{i\in \mathcal{P}}H_{i} \left( s \right)  ( \frac{ \varphi  \left( s \right) }{ \gamma  \left( s \right) } ) ^{i+1}}{ \gamma  \left( s \right) - \sum_{i \in \mathcal{F}} H_{i} \left( s \right)  ( \frac{ \varphi  \left( s \right) }{ \gamma  \left( s \right) } ) ^{i}}.
\end{equation*}

\subsection{Spacing Fluctuations in Simulations at Mitigating Traffic Perturbations Ahead}

\begin{figure}[t]
	\centering
	\subfigure[Case A]
	{\includegraphics[scale=0.36,height=3cm]{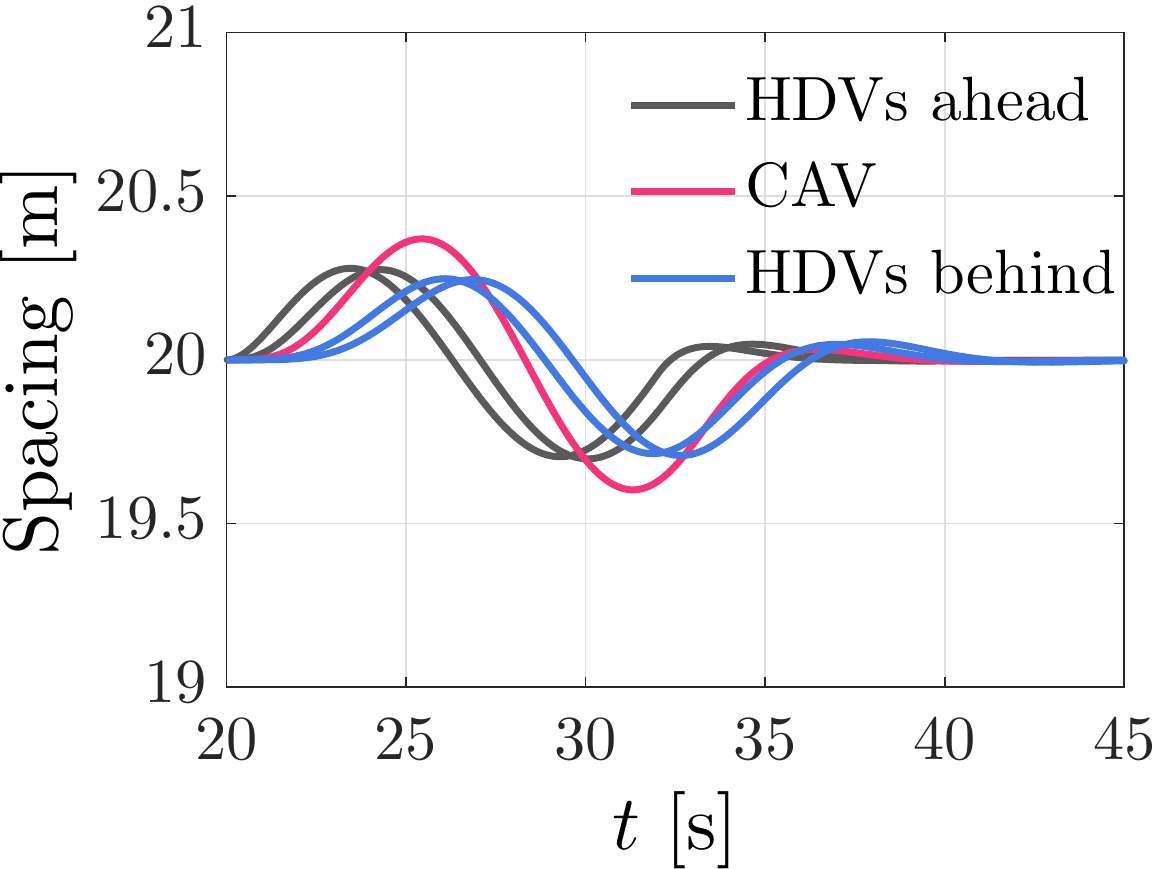}
		\label{Fig:Simulation_PerturbationAhead_1_Spacing}}
	\subfigure[Case B]
	{\includegraphics[scale=0.36,height=3cm]{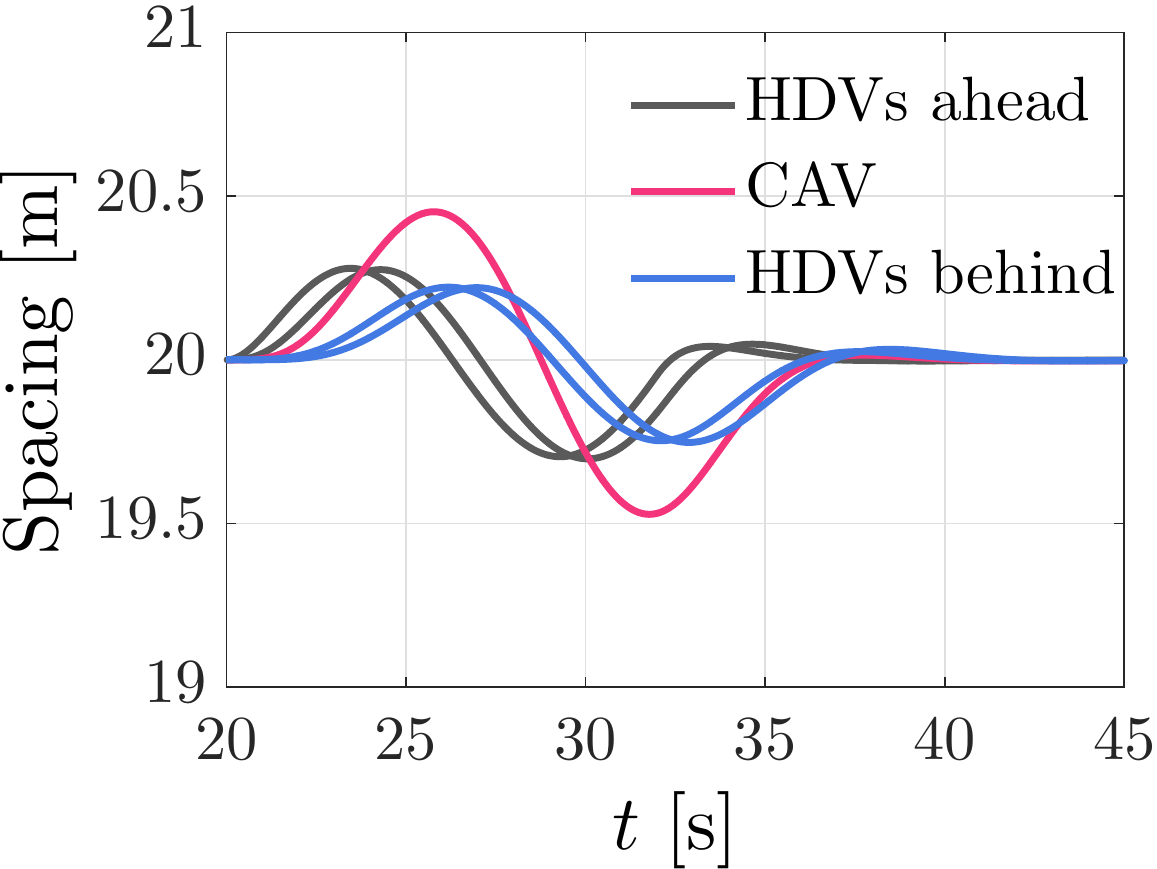}
		\label{Fig:Simulation_PerturbationAhead_2_Spacing}}
	\subfigure[Case C]
	{\includegraphics[scale=0.36,height=3cm]{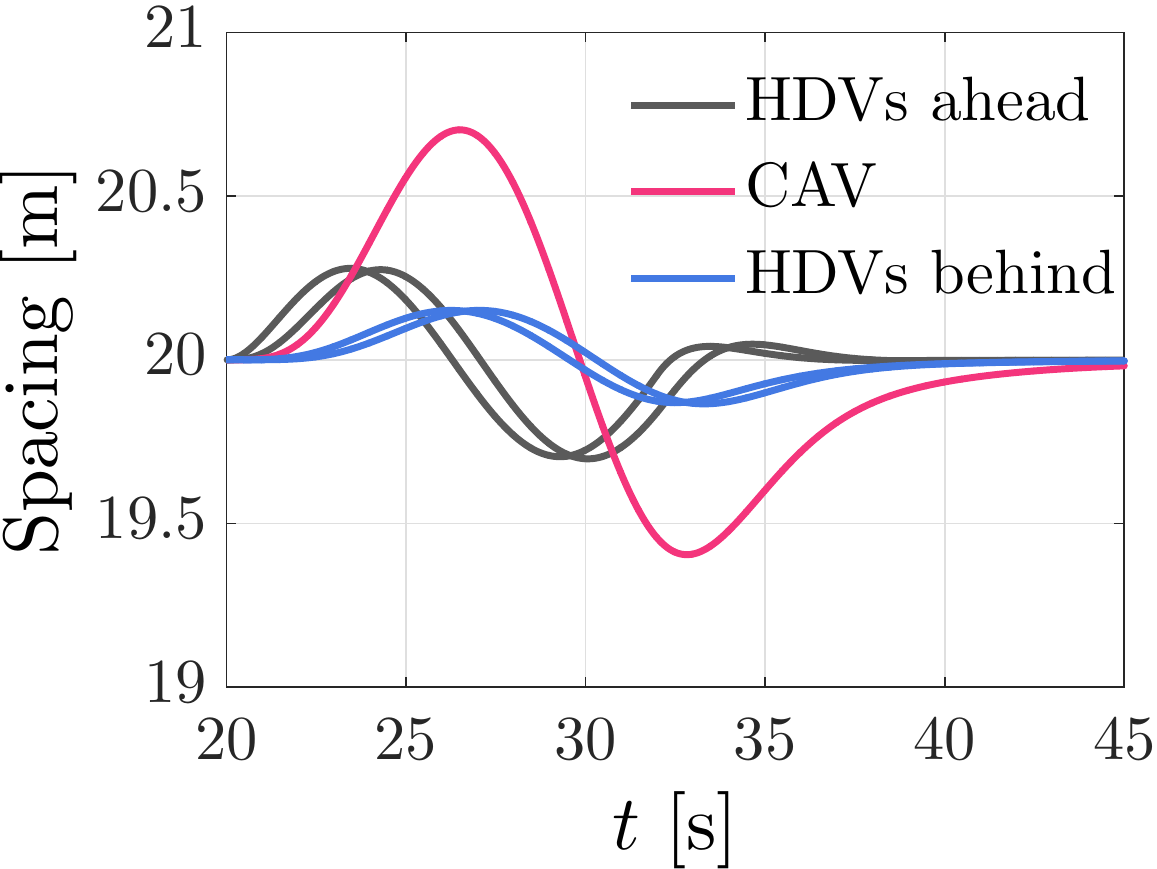}
		\label{Fig:Simulation_PerturbationAhead_3_Spacing}}
	\subfigure[Case D]
	{\includegraphics[scale=0.36,height=3cm]{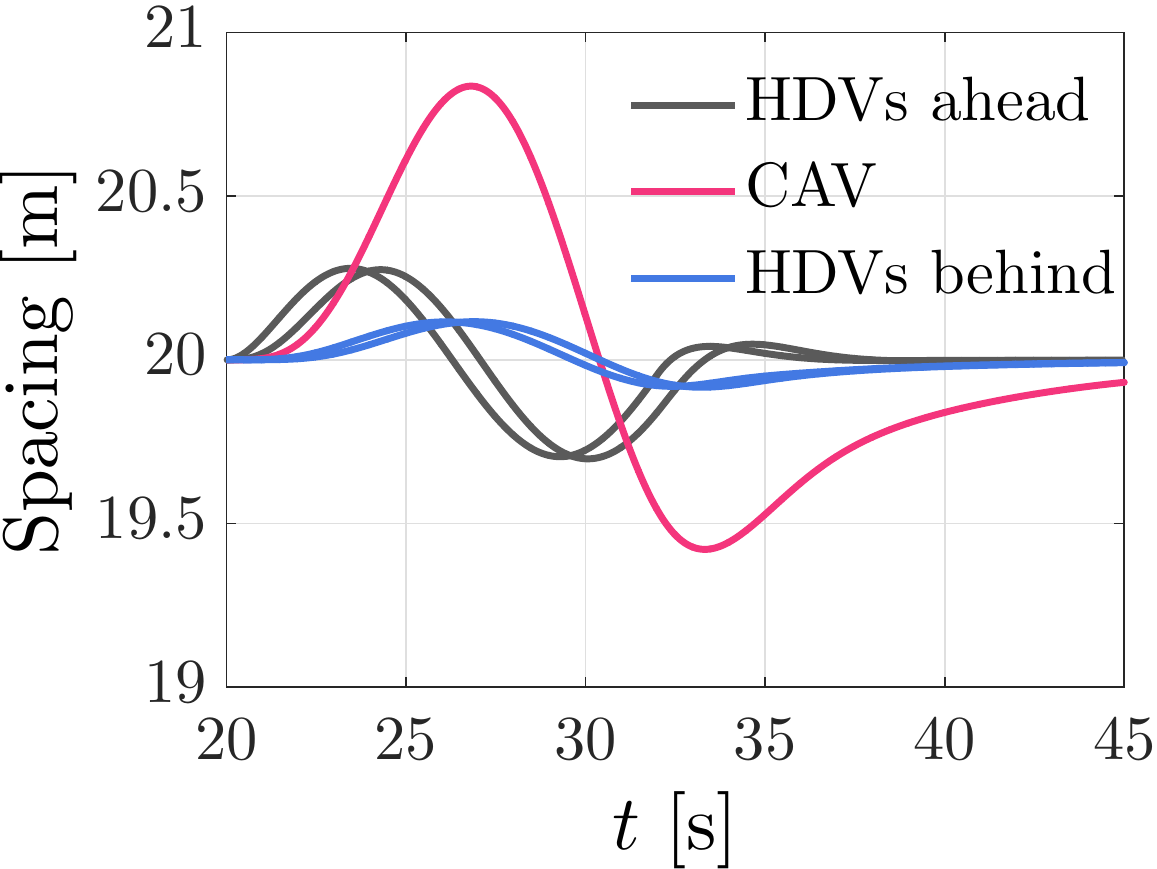}
		\label{Fig:Simulation_PerturbationAhead_4_Spacing}}
	\vspace{-1mm}
	\caption{Spacing (distance headway) profile of each vehicle in the nonlinear simulation a sinusoid perturbation is imposed on the head vehicle ($n=m=2$), corresponding to the velocity profiles in Figure~\ref{Fig:Simulation_PerturbationAhead}. (a) - (d) correspond to Case A to Case D in Table~\ref{Tb:FeedbackGainSetup}.}
	\label{Fig:Simulation_PerturbationAhead_Spacing}
\end{figure}

In Section~\ref{Sec:SimulationAhead}, we have presented the velocity profile of each vehicle in the simulations when a sinusoid perturbation is imposed on the head vehicle. Recall that the CAV only utilizes the information of preceding vehicles in Case A and Case B, while in Case C and Case D, the following vehicles are added into consideration. Our results of velocity perturbations in Fig.~\ref{Fig:Simulation_PerturbationAhead} have shown that the amplitude of the velocity fluctuations of the following vehicles becomes smaller from Case A to Case D, and validate the capability of LCC in mitigating traffic perturbations ahead.

\begin{figure*}[t]
		\vspace{1mm}
		\centering
		\subfigure{\includegraphics[scale=0.35]{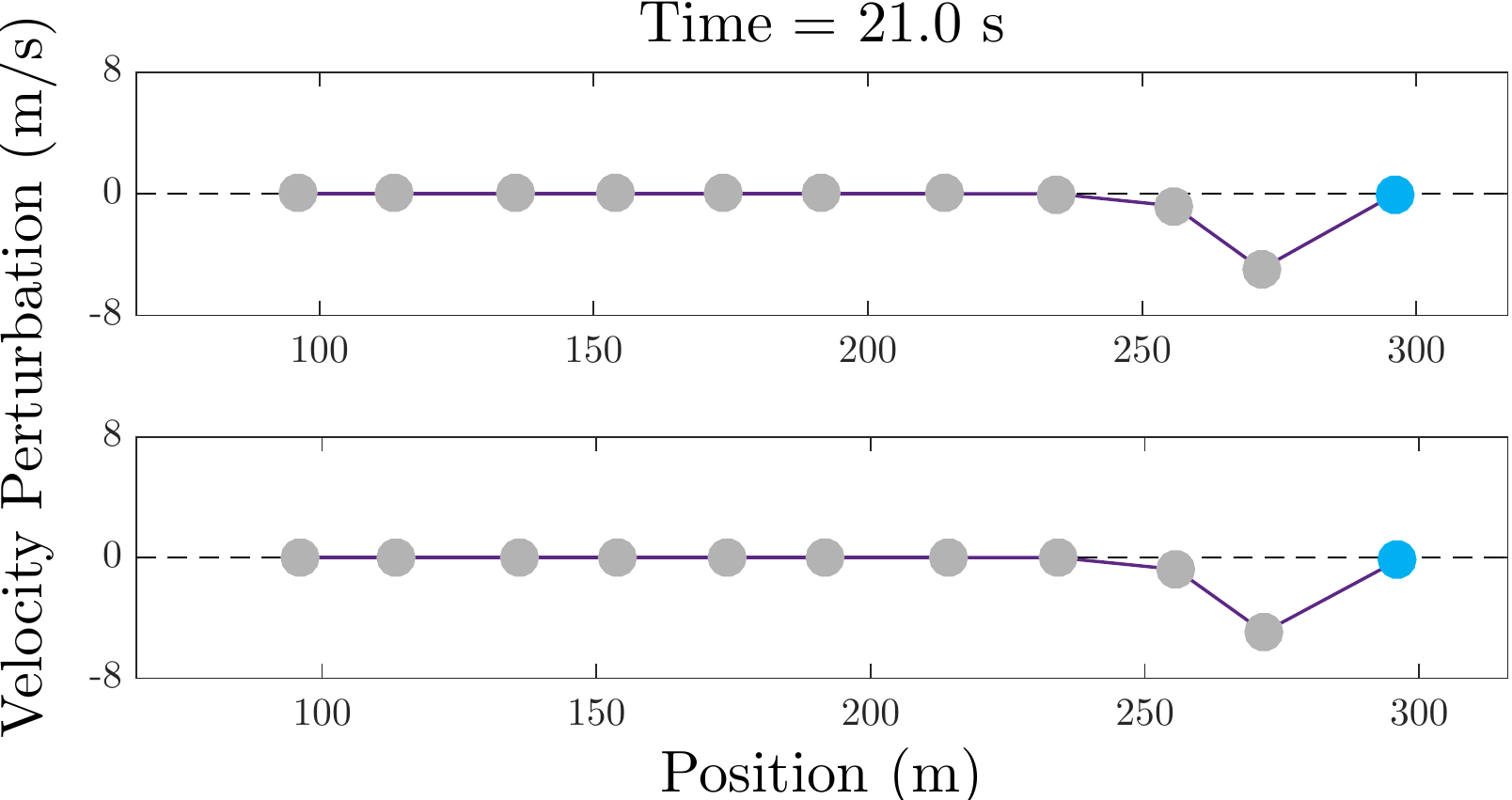}}
		\addtocounter{subfigure}{-1}
		\subfigure{\includegraphics[scale=0.35]{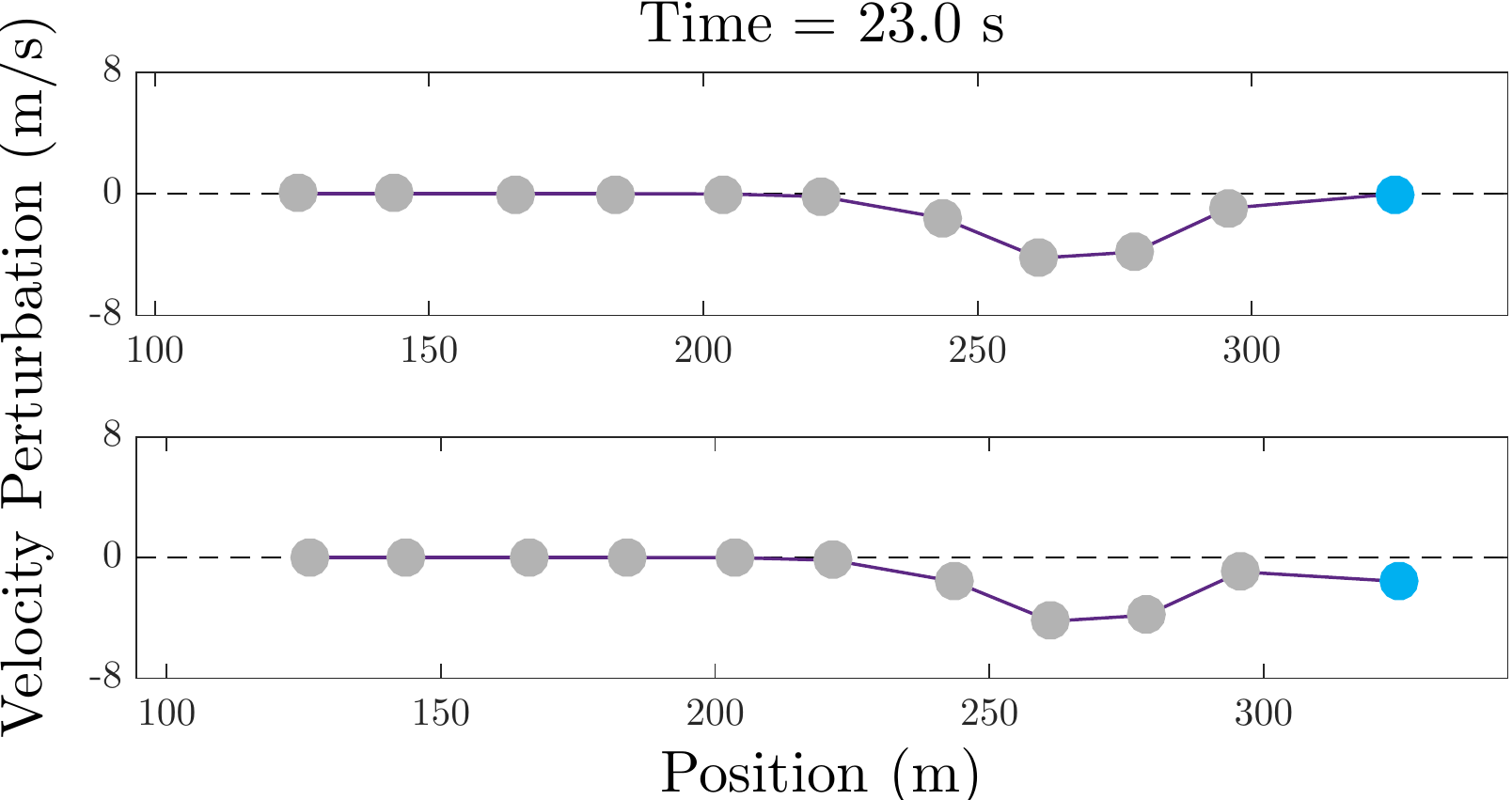}}
		\addtocounter{subfigure}{-1}
		\subfigure{\includegraphics[scale=0.35]{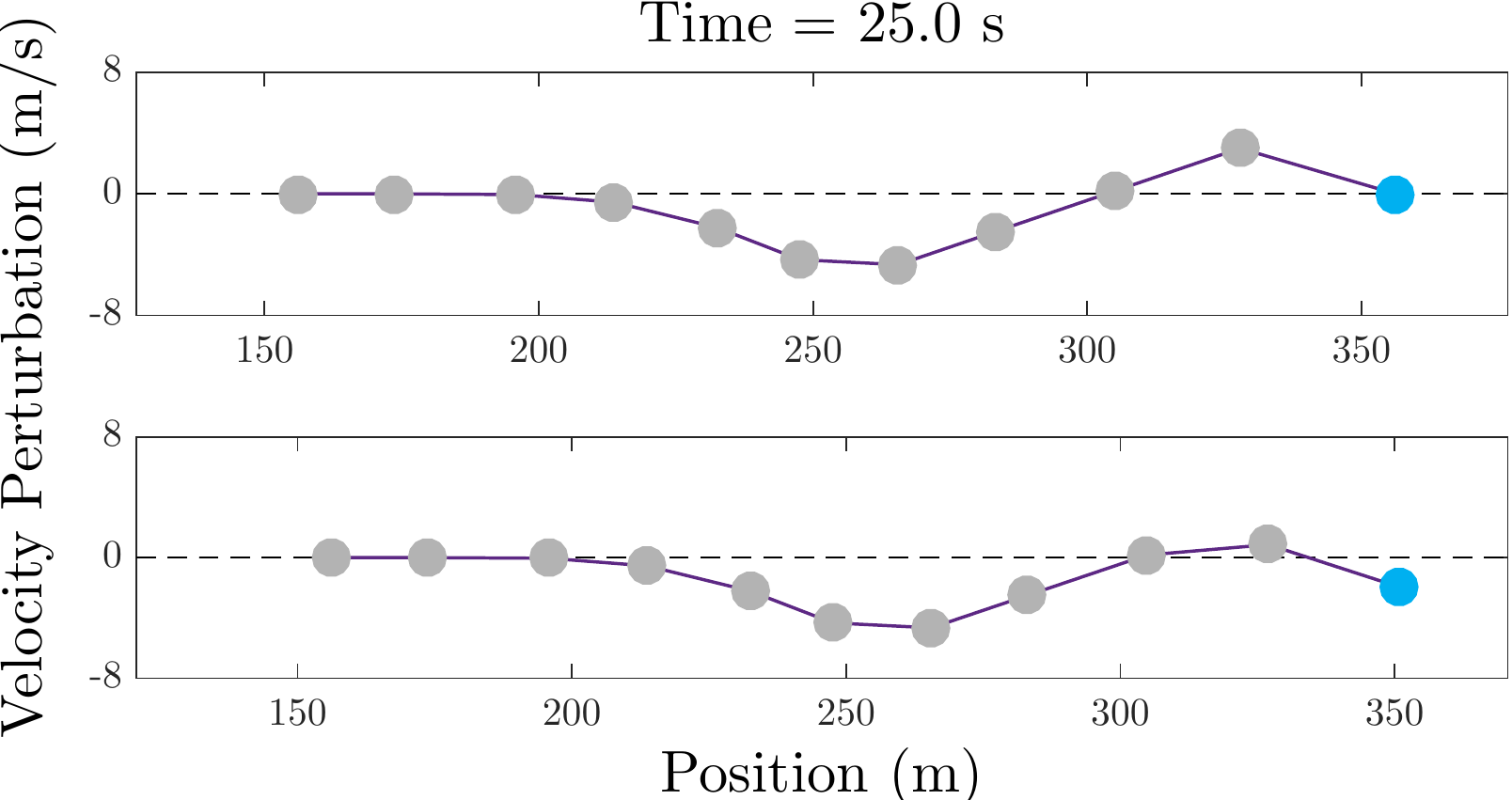}}
		\addtocounter{subfigure}{-1}
		\subfigure{\includegraphics[scale=0.35]{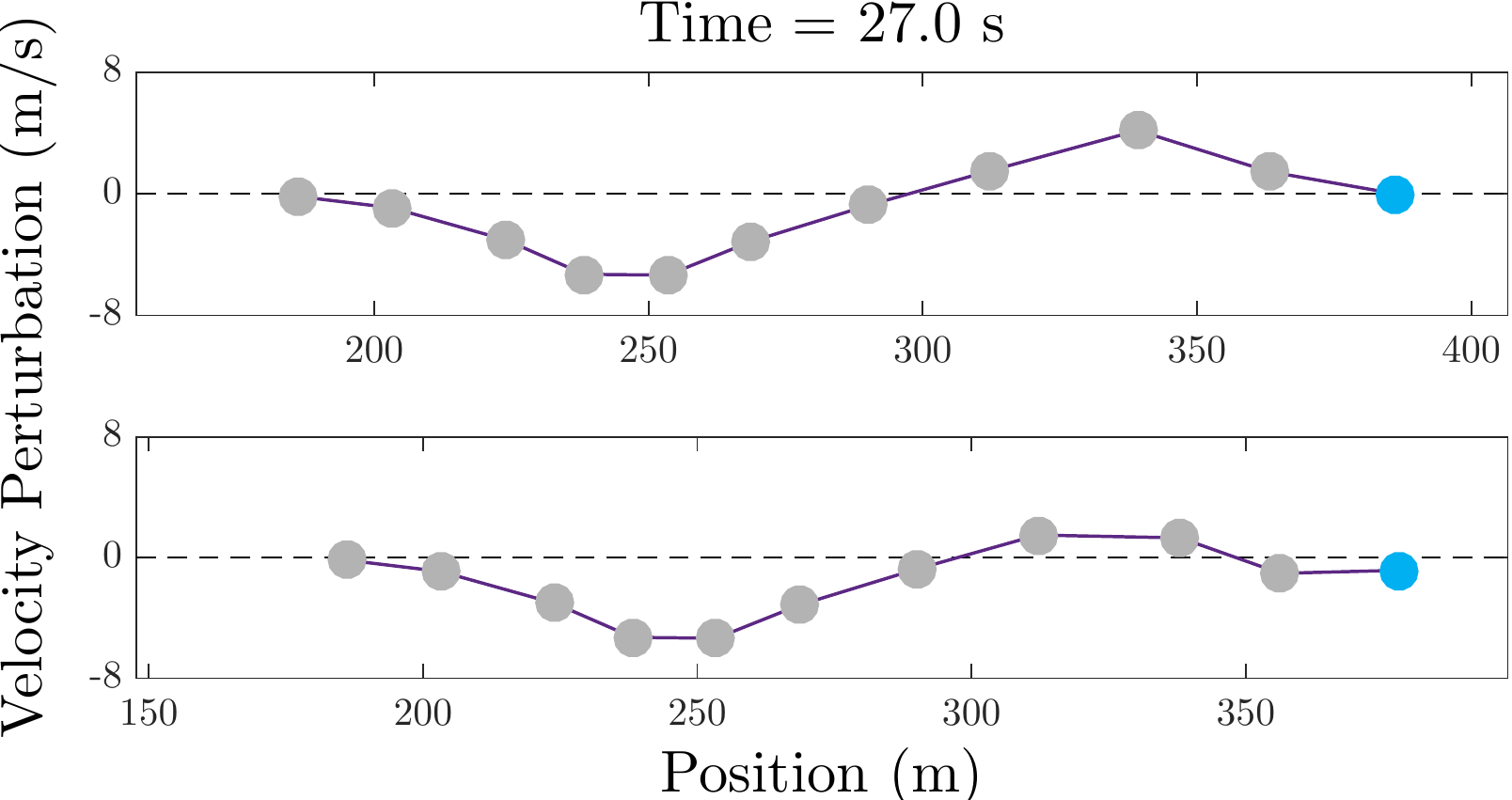}}
		\addtocounter{subfigure}{-1}
		\subfigure[FD-LCC]{\includegraphics[scale=0.35]{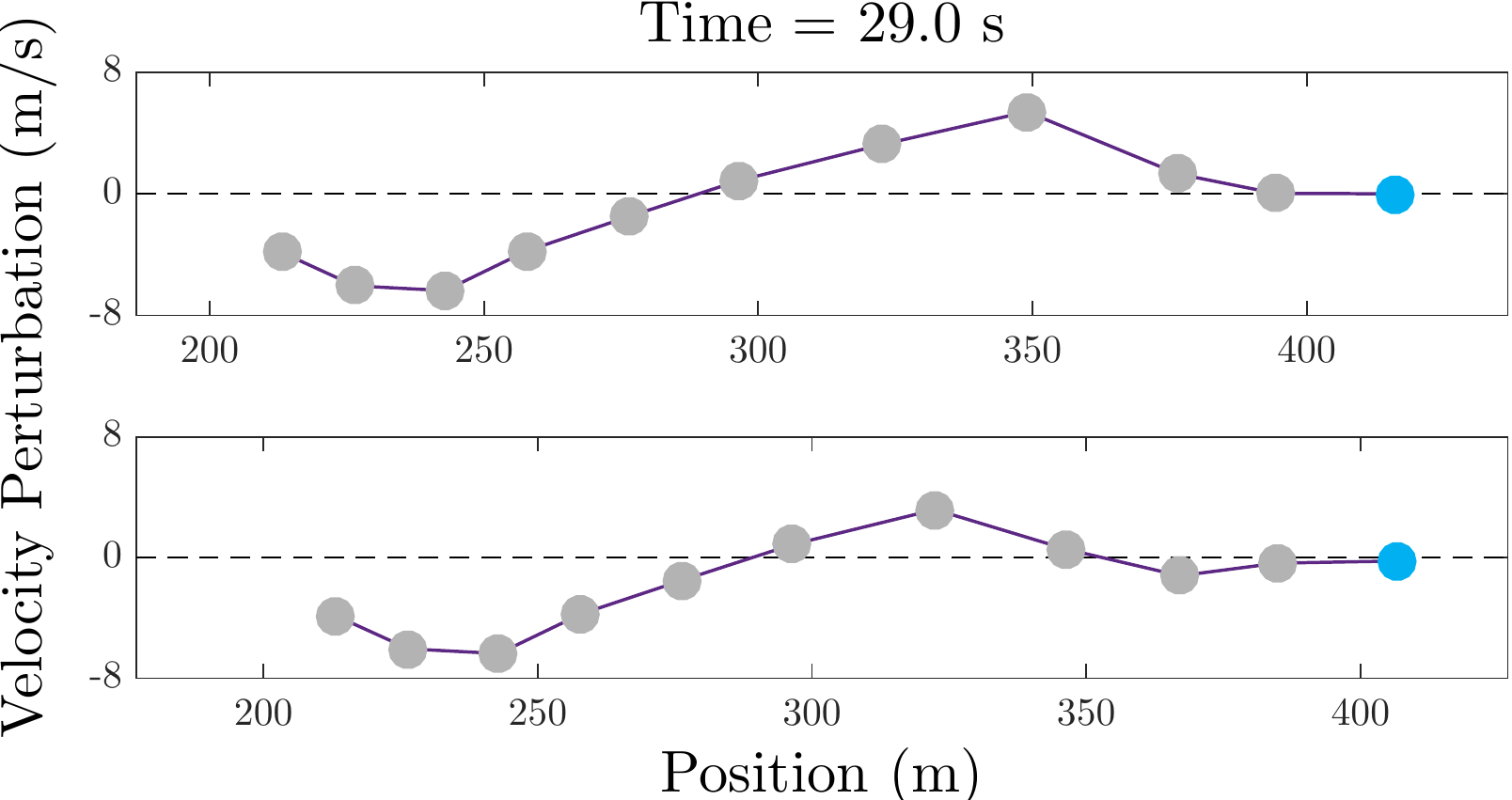}}
		\addtocounter{subfigure}{-1}
		\subfigure{\includegraphics[scale=0.35]{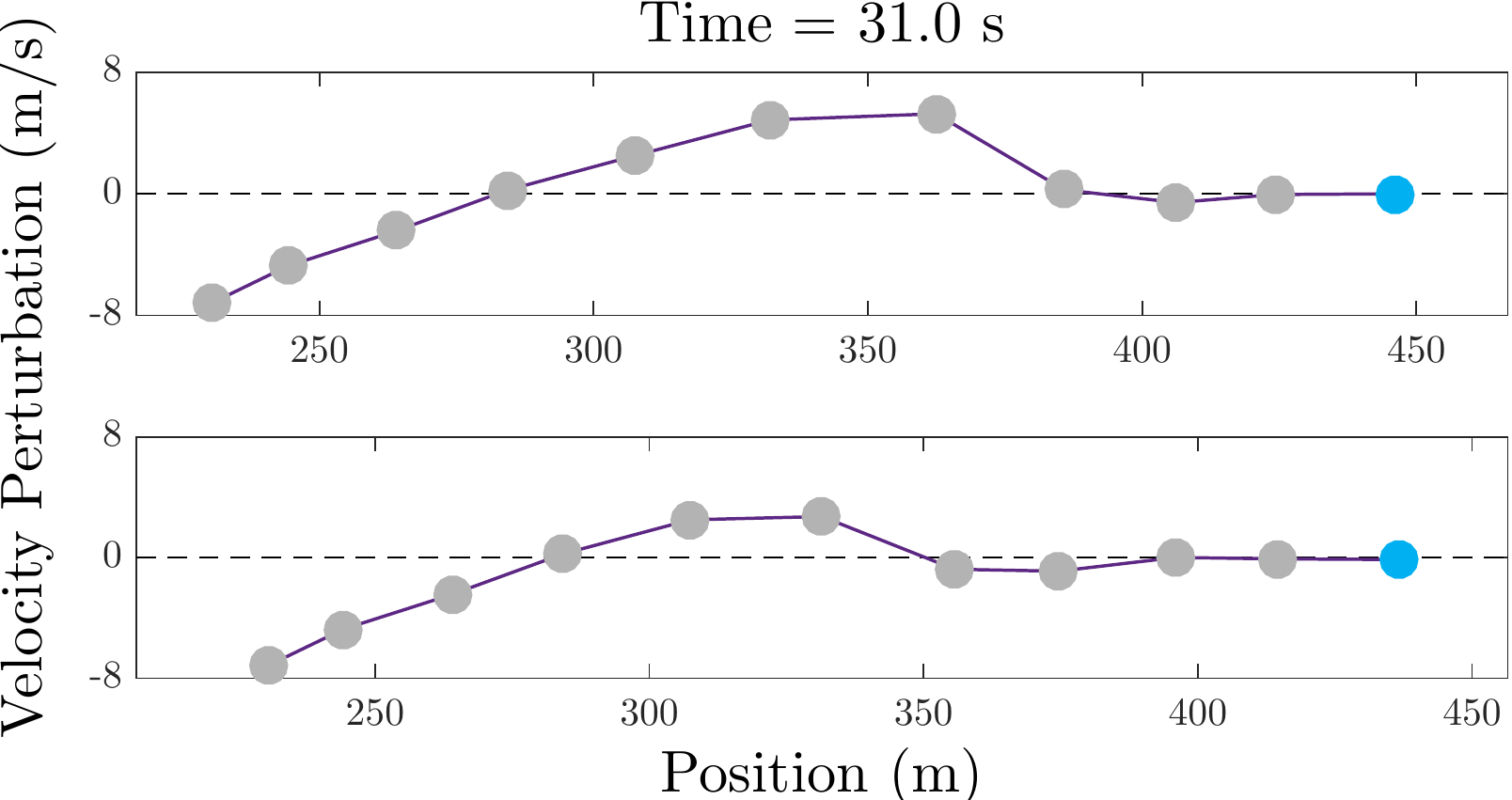}}
		\subfigure{\includegraphics[scale=0.35]{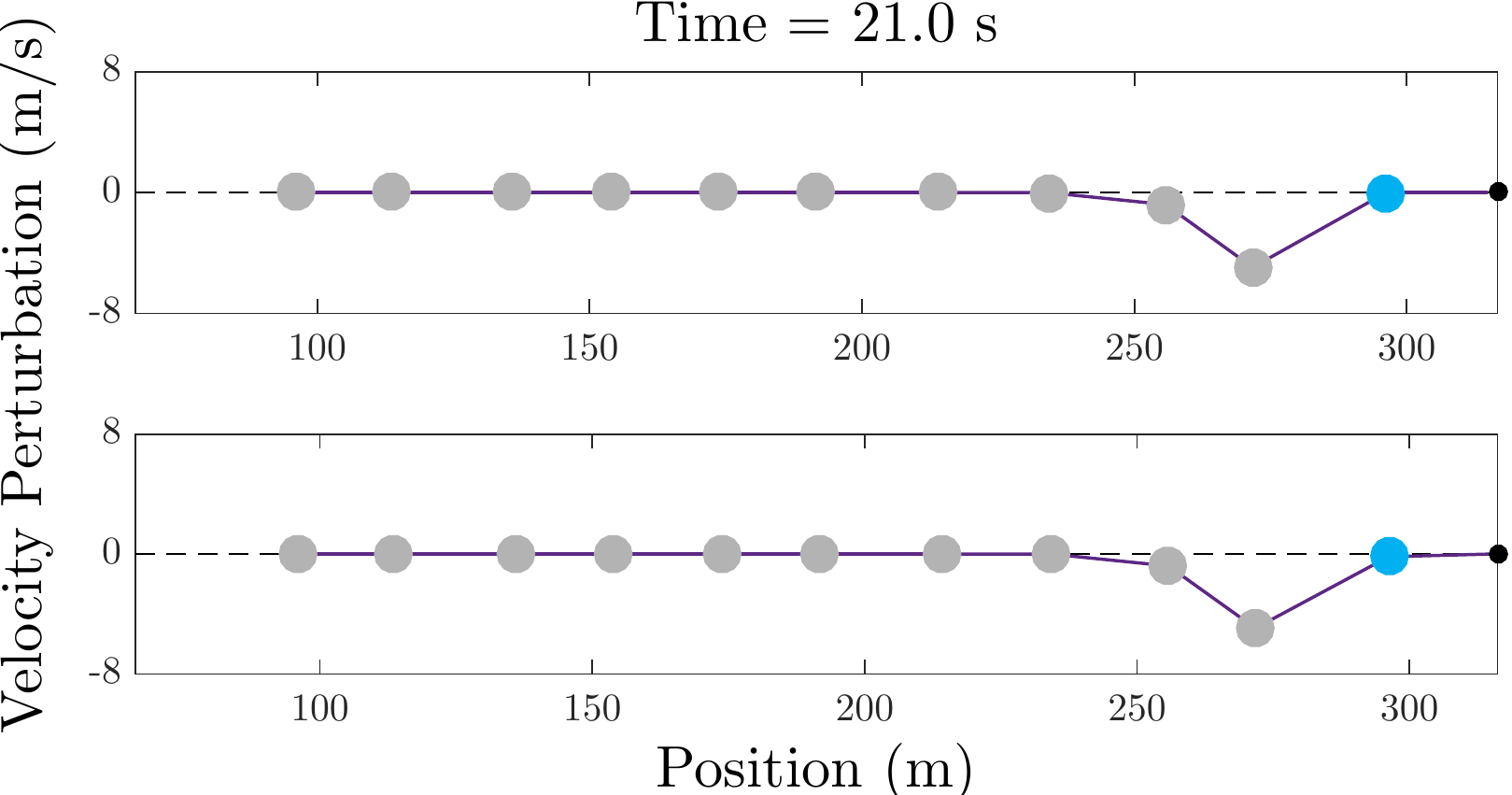}}
		\addtocounter{subfigure}{-1}
		\subfigure{\includegraphics[scale=0.35]{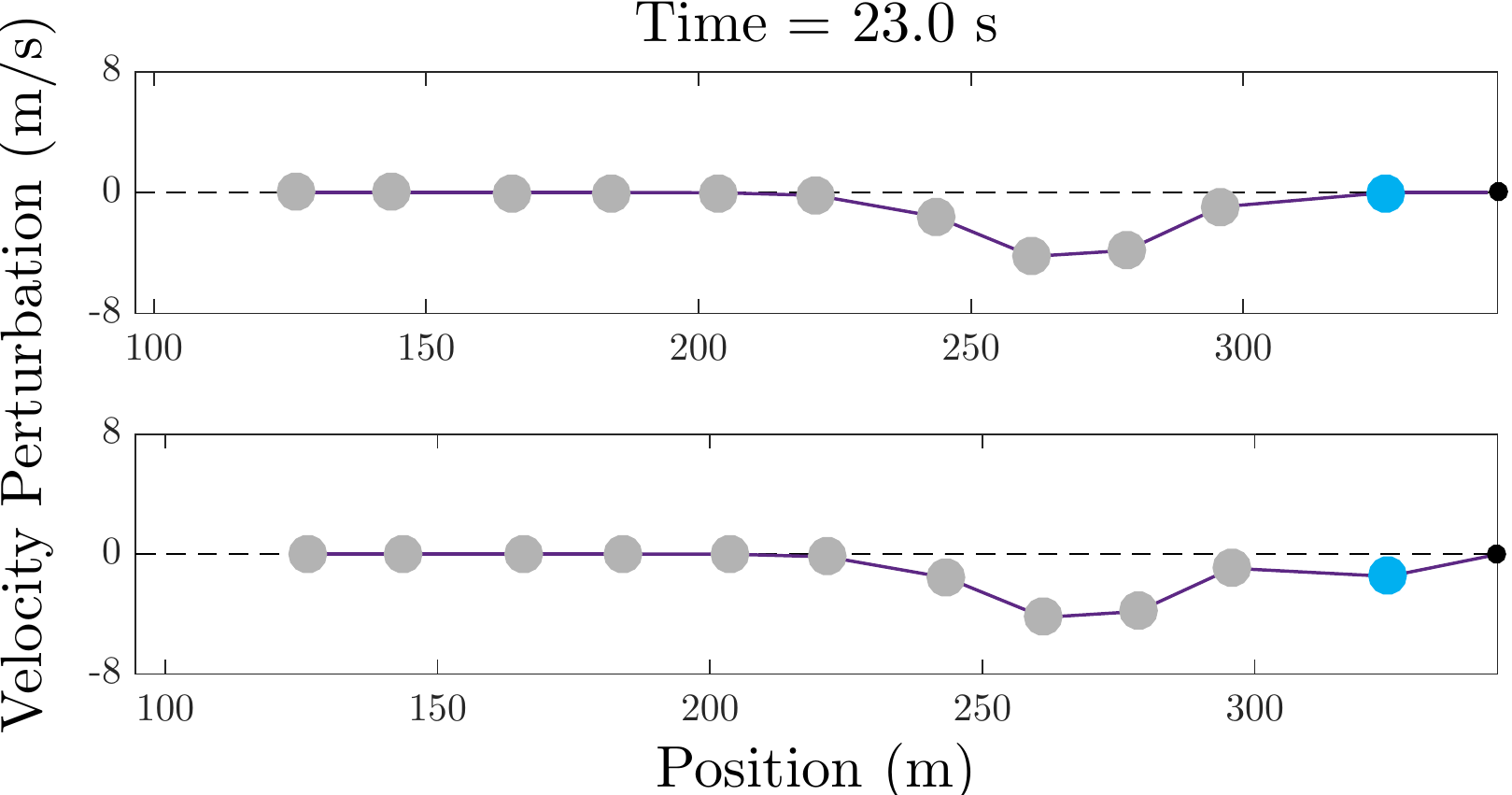}}
		\addtocounter{subfigure}{-1}
		\subfigure{\includegraphics[scale=0.35]{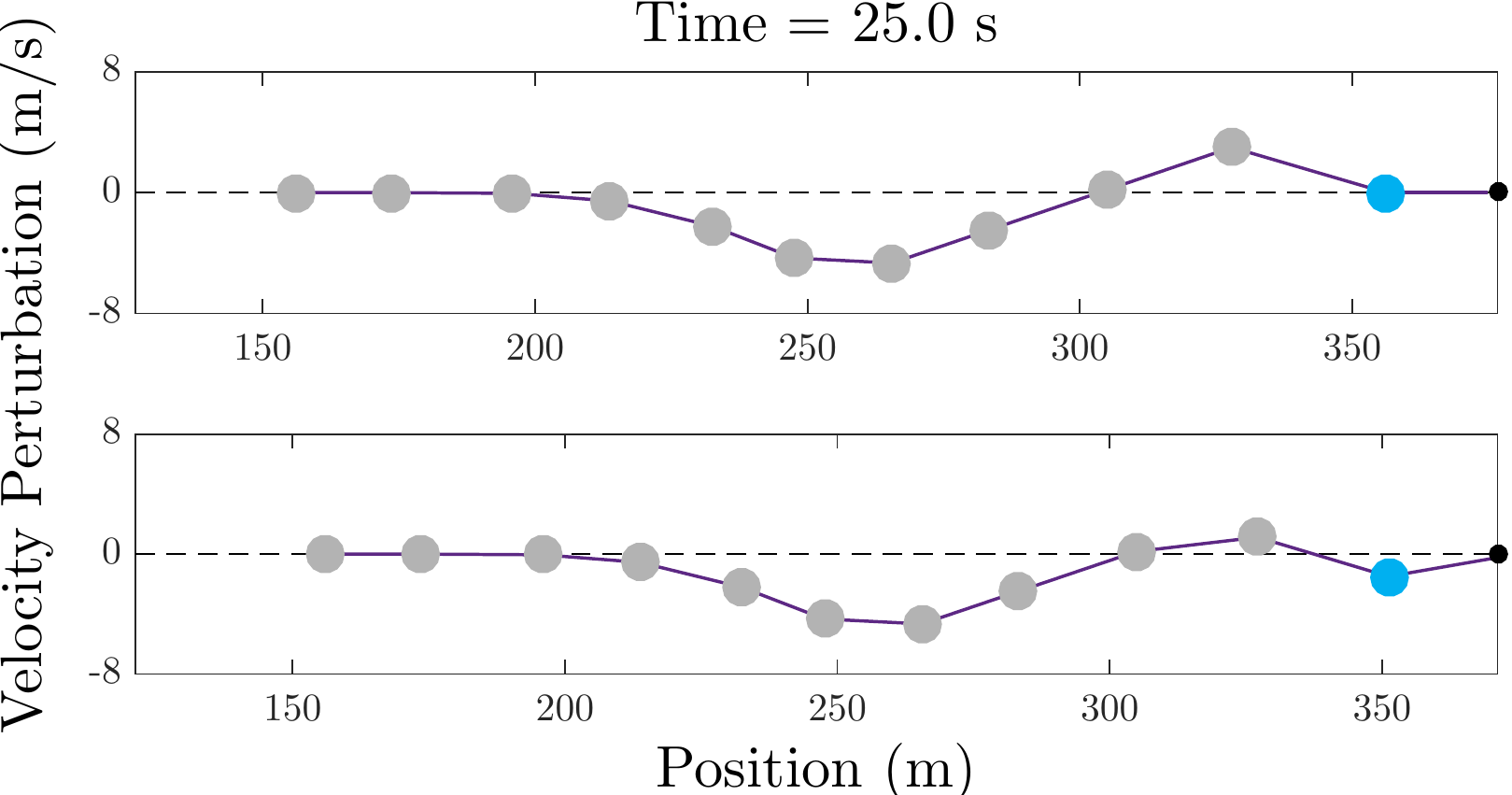}}
		\addtocounter{subfigure}{-1}
		\subfigure{\includegraphics[scale=0.35]{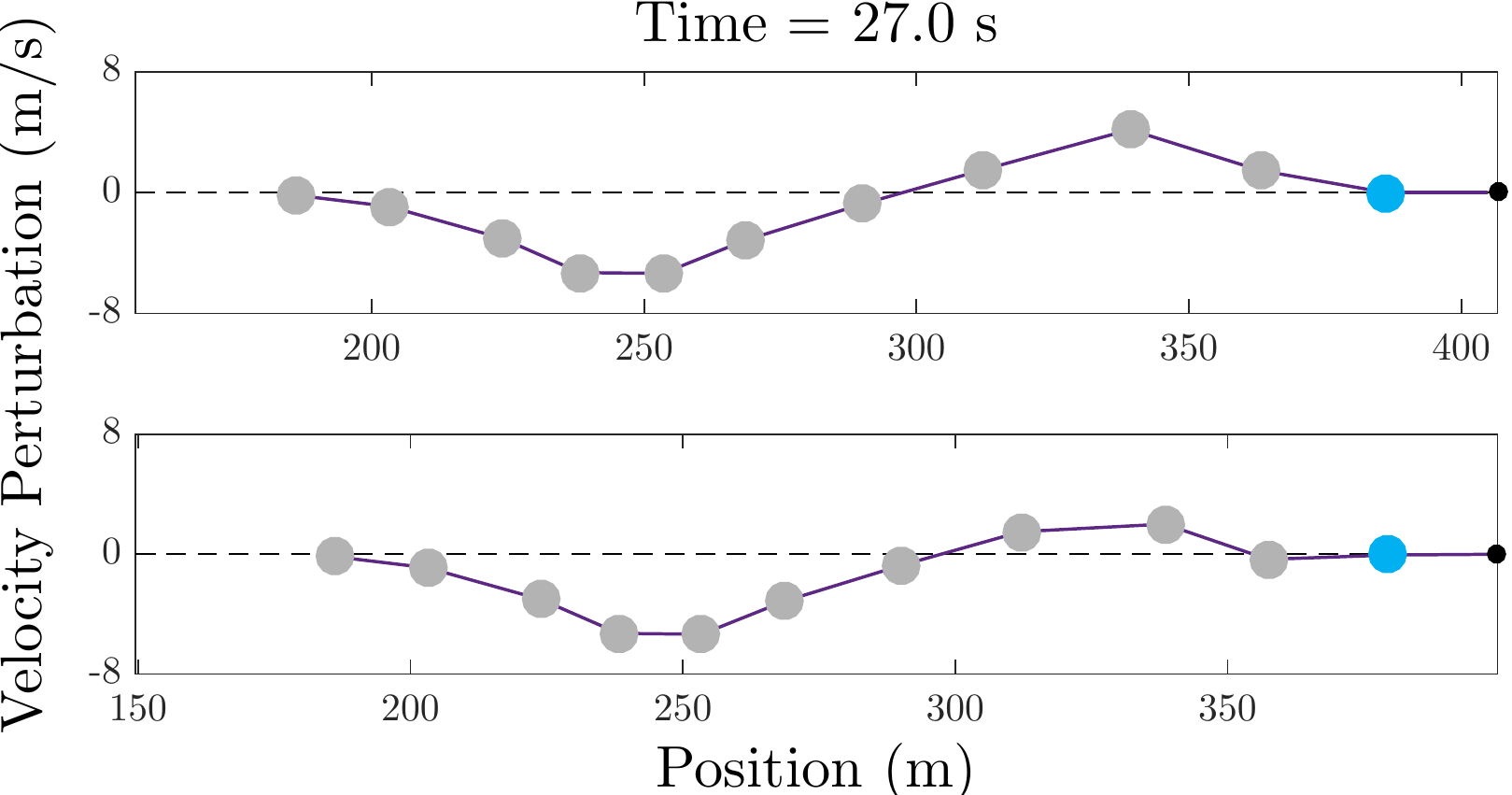}}
		\addtocounter{subfigure}{-1}
		\subfigure[CF-LCC]{\includegraphics[scale=0.35]{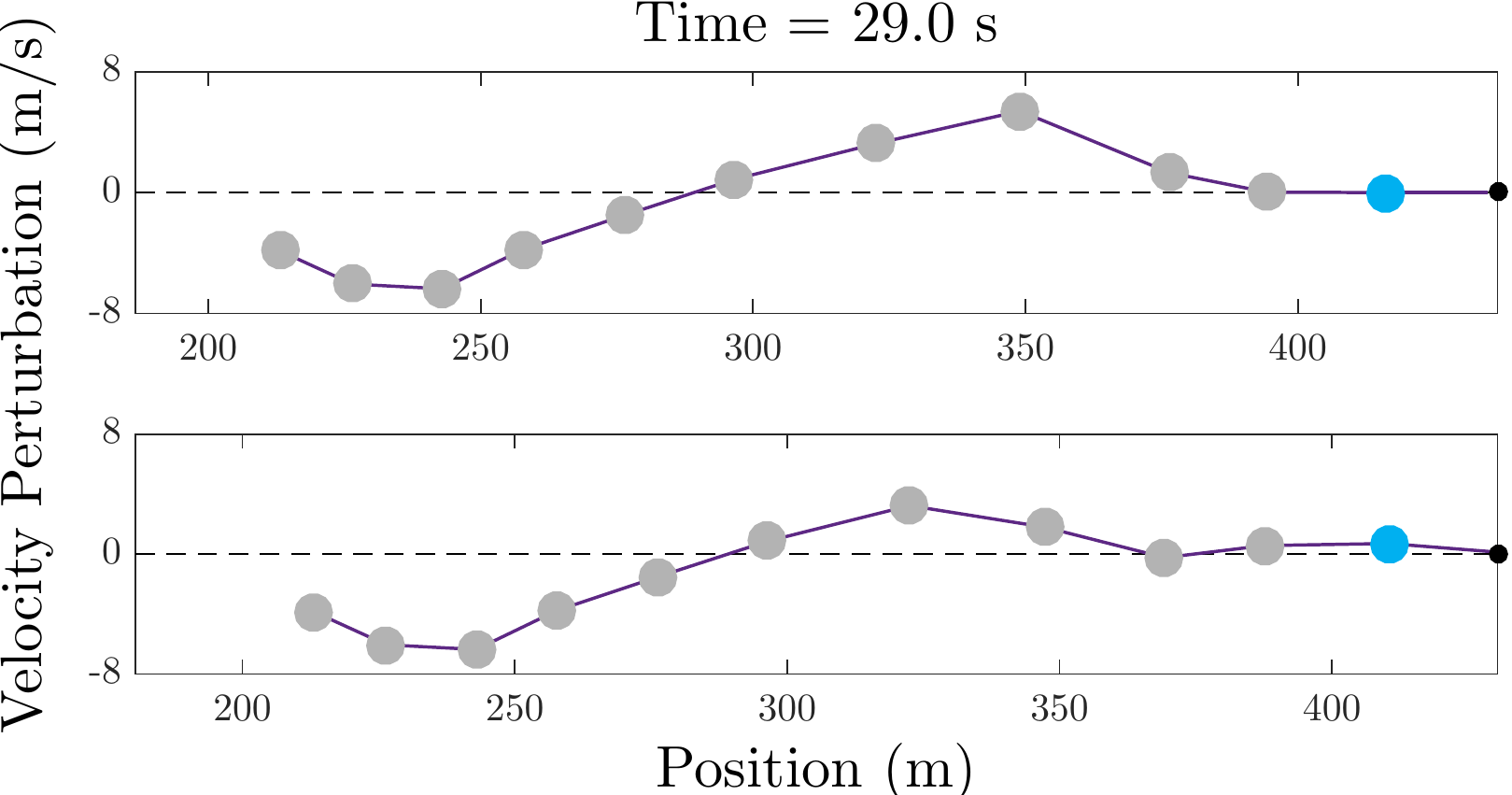}}
		\addtocounter{subfigure}{-1}
		\subfigure{\includegraphics[scale=0.35]{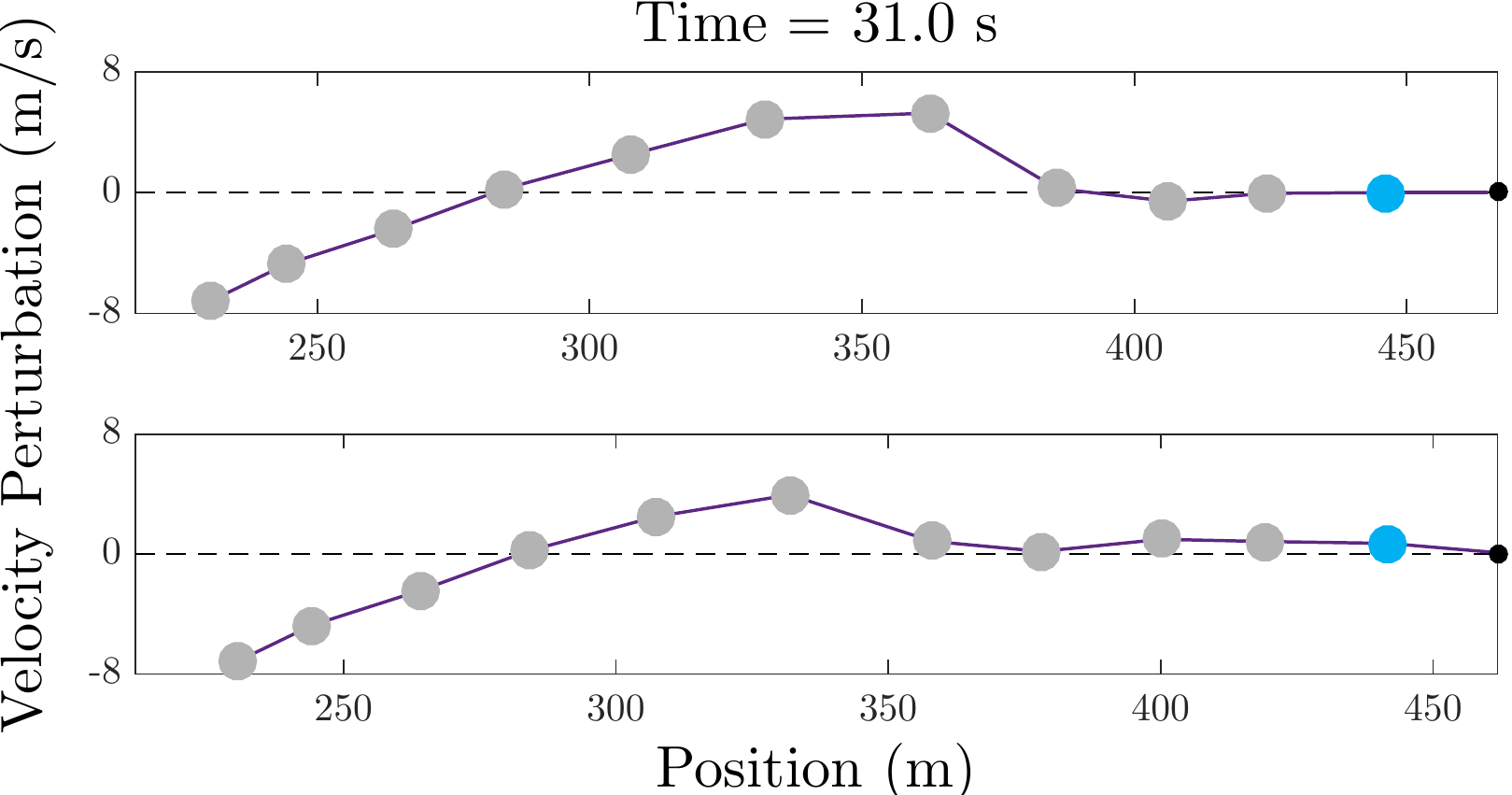}}
		\vspace{-1mm}
		\caption{Additional experiment: snapshots of the LCC system when a perturbation happens at the vehicle immediately behind the CAV, \ie, vehicle $1$. Heterogeneous dynamics and reaction delay of HDVs are under consideration, compared with the original experiment shown in Figure~\ref{Fig:Simulation_PerturbationBehind} in the main text. }
		\label{Fig:Simulation_PerturbationBehind_Additional}
\end{figure*}

However, there indeed exists a cost for reducing velocity perturbations. 
The spacing profile of each vehicle is presented in Figure~\ref{Fig:Simulation_PerturbationAhead_Spacing} in this appendix. It is observed that from Case A to Case D, the amplitude of the spacing fluctuations of the following HDVs decrease, similarly to their velocity fluctuations; for the CAV itself, by contrast, its spacing fluctuation becomes slightly larger. This phenomenon is somewhat expected, since the CAV has a smaller velocity fluctuation and then its spacing fluctuation grows up --- the CAV might leave a larger or smaller spacing with the preceding vehicle. These results are also consistent with the intuition that incorporating the vehicles behind into consideration would decrease the original weight of the CAV in maintaining the spacing from the preceding vehicle. In particular, such phenomenon might raise safety concerns. In all our simulations, we have added a low-level emergency braking system~\eqref{Eq:AEB} for the CAV to avoid collisions. Nevertheless, more detailed investigations on guaranteeing safety are very interesting for future work on LCC. For example, one can design a model predictive controller (MPC) with safety constraints under the LCC framework, which has seen great success in ACC~\cite{li2010model} and vehicle platoon~\cite{zheng2016distributed}, with the aim of achieving optimal and safe control.

\subsection{Simulations at Leading the Motion of the Vehicles Behind under Heterogeneous HDV Dynamics and Reaction Time }

In the following, we present the simulation results under the influence of HDVs' heterogeneous car-following dynamics and reaction time. The scenario is the same as that in Section V-B, where there are ten HDVs behind the CAV and the HDV immediately behind is under a sudden perturbation. Some critical human parameters are set as heterogeneous values: $\alpha = 0.6 + U[-0.1,0.1]$, $\beta = 0.9 + U[-0.1,0.1]$, $s_\mathrm{go} = 35 + U[-5,5]$, where $U[\cdot]$ denotes the uniform distribution, and the rest parameters remain the same as those in the main text. Additionally, a heterogeneous reaction delay is under consideration for each HDV with the value set as $0.4 + U[-0.1,0.1] $ seconds. The feedback controller and its specific feedback gains remain unchanged from Section~\ref{Sec:SimulationBehind} (recall that only the information of the two HDVs directly behind are utilized).

The new simulation results are shown in Figure~\ref{Fig:Simulation_PerturbationBehind_Additional} in this appendix. As can be clearly observed, the two LCC systems still show an apparently better performance in stabilizing upstream traffic flow than the traditional ``looking ahead" only strategies, where the CAV makes no response to the perturbation behind. Precisely, by calculating performance indexes with trajectory data from $t = 20\, \mathrm{ s}$ to $t = 40\, \mathrm{ s}$, it is observed that the average absolute velocity error (AAVE) is reduced by $23\%$ and $8.77\%$ for FD-LCC and CF-LCC, respectively, while the fuel consumption (FC) is reduced by $17.55\%$ and $13.72\%$, respectively. Note that we directly choose a linear feedback controller to carry out this research, and it is anticipated that the performance can be further improved by careful design of a controller, which explicitly addresses the influence of reaction delay and heterogeneity; see, \eg, previous discussions on CCC-related works~\cite{jin2017optimal,orosz2016connected}.
\end{document}